\documentclass[reqno,11pt]{amsart}
\usepackage[utf8]{inputenc}
\usepackage{graphicx}
\usepackage{slashed}
\usepackage{amscd}
\usepackage{amssymb}
\usepackage{esint}
\usepackage{enumitem}
\usepackage[arrow, matrix, curve]{xy}
\usepackage{csquotes}
\usepackage[mathscr]{eucal}
\numberwithin{equation}{section}
\allowdisplaybreaks[1]

\title[Introduction to Causal Fermion Systems]{Introduction to Causal Fermion Systems} %in Momentum Space

\author[C.\ Langer]{Christoph Langer \\ \\ November 2021}
\address{Fakult\"at f\"ur Mathematik \\ Universit\"at Regensburg \\ D-93040 Regensburg \\ Germany}
\email{%finster@ur.de, 
christoph.langer@ur.de}

\newtheorem{Def}{Definition}[section]
\newtheorem{Thm}[Def]{Theorem}
\newtheorem{Prp}[Def]{Proposition}

\newcommand{\Thanks}{\vspace*{.5em} \noindent \thanks}
\newcommand{\beq}{\begin{equation}}
\newcommand{\eeq}{\end{equation}}
\newcommand{\Proof}{\begin{proof}}
	\newcommand{\QED}{\end{proof} \noindent}

\newcommand{\la}{\langle}
\newcommand{\ra}{\rangle}
\newcommand{\bra}{\mathopen{<}}
\newcommand{\ket}{\mathclose{>}}

\newcommand{\C}{\mathbb{C}}
\newcommand{\R}{\mathbb{R}}
\newcommand{\Id}{\mbox{\rm 1 \hspace{-1.0505 em} 1}}
\newcommand{\Z}{\mathbb{Z}}
\newcommand{\N}{\mathbb{N}}
\newcommand{\Pdd}{\mbox{$\partial$ \hspace{-1.2 em} $/$}}

\DeclareMathOperator{\tr}{tr}

\DeclareMathOperator{\rank}{rank}
\newcommand{\GL}{\text{\rm{GL}}}

\renewcommand{\O}{{\mathscr{O}}}
\renewcommand{\L}{{\mathcal{L}}}
\newcommand{\LL}{{\text{\rm{L}}}}
\newcommand{\LDirac}{{\mathcal{L}}_{\text{\tiny{\rm{Dirac}}}}}
\newcommand{\LYM}{{\mathcal{L}}_{\text{\tiny{\rm{YM}}}}}
\newcommand{\LEH}{{\mathcal{L}}_{\text{\tiny{\rm{EH}}}}}
\newcommand{\Lint}{{\mathcal{L}}_{\text{\tiny{\rm{int}}}}}
\newcommand{\LQCD}{{\mathcal{L}}_{\text{\tiny{\rm{QCD}}}}}
\newcommand{\LQED}{{\mathcal{L}}_{\text{\tiny{\rm{QED}}}}}
\newcommand{\LMaxwell}{{\mathcal{L}}_{\text{\tiny{\rm{Maxwell}}}}}

\DeclareMathOperator{\ric}{Ric}
\newcommand{\Sact}{{\mathcal{S}}}
\newcommand{\T}{{\mathcal{T}}}

\newcommand\B{{\mathscr{B}}}
\newcommand{\U}{\text{\rm{U}}}
\newcommand{\SU}{\text{\rm{SU}}}

\newcommand{\np}{n_{\mathrm{p}}}
\newcommand{\na}{n_{\mathrm{a}}}

\newcommand{\Dir}{{\mathcal{D}}}
\renewcommand{\H}{\mathscr{H}}

\DeclareMathOperator{\diag}{diag}

\newcommand{\Lin}{\text{\rm{L}}}
\newcommand{\F}{{\mathscr{F}}}

\newcommand{\G}{{\mathcal{G}}}
\DeclareMathOperator{\re}{Re}
\DeclareMathOperator{\im}{Im}

\DeclareMathOperator{\Pexp}{Pexp}
\DeclareMathOperator{\Pe}{Pe}

\DeclareMathOperator{\supp}{supp}
\newcommand{\PP}{\text{\rm{PP}}}

\newcommand{\sea}{\text{\rm{sea}}}
\newcommand{\res}{\text{\rm{res}}}
\newcommand{\reg}{\text{\rm{reg}}}
\newcommand{\lec}{\text{\rm{le}}}
\newcommand{\hec}{\text{\rm{he}}}
\newcommand{\pseudo}{\Gamma}

\newcommand{\scrM}{\mycal M}
\newcommand{\hscrM}{\,\,\hat{\!\!\scrM}}
\newcommand{\scrN}{\mycal N}

\newcommand{\UMNS}{U_\text{\rm{\tiny{MNS}}}}
\newcommand{\UCKM}{U_\text{\rm{\tiny{CKM}}}}

\newcommand{\bitem}{\begin{itemize}[leftmargin=2.5em]}
	\newcommand{\eitem}{\end{itemize}}

\usepackage{scalerel, stackengine}
\stackMath
\newcommand\reallywidehat[1]{%
	\savestack{\tmpbox}{\stretchto{%
			\scaleto{%
				\scalerel*[\widthof{\ensuremath{#1}}]{\kern-.6pt\bigwedge\kern-.6pt}%
				{\rule[-\textheight/2]{1ex}{\textheight}}%WIDTH-LIMITED BIG WEDGE
			}{\textheight}% 
		}{0.5ex}}%
	\stackon[1pt]{#1}{\tmpbox}%
}

\DeclareFontFamily{OT1}{rsfso}{}
\DeclareFontShape{OT1}{rsfso}{m}{n}{ <-7> rsfso5 <7-10> rsfso7 <10-> rsfso10}{}
\DeclareMathAlphabet{\mycal}{OT1}{rsfso}{m}{n}

\begin{document}

\begin{abstract}
This paper is dedicated to give a concise introduction to the theory of causal fermion systems. After putting the theory of causal fermion systems into the historical context, we recall fundamental physical preliminaries. Afterwards, we enter the underlying ideas of the principle of the fermionic projector and clarify its connection to causal fermion systems. We finally outline the main structures of the theory of causal fermion systems. 
\end{abstract}
	
\maketitle
\tableofcontents

\thispagestyle{empty}
\newpage

%\begin{quote}
%	\smaller
%	\textsc{Abstract.} We give a short introduction to the theory of causal fermion systems. After putting the theory of causal fermion systems into the historical context, we recall fundamental physical preliminaries. Afterwards, we enter the underlying ideas of the principle of the fermionic projector and clarify its connection to causal fermion systems. We finally outline the main structures of the theory of causal fermion systems. 
%\end{quote}

%\vspace*{0.25cm}

\section{Introduction}\label{c:Section intro-cfs} 
The theory of causal fermion systems is a new approach to describe theoretical physics by means of a variational principle.  
It aims to provide a unified description of quantum mechanics and general relativity.  
Based on Dirac's original concept of describing the vacuum in terms of a completely filled ``Dirac sea,'' the theory of causal fermion systems evolved from the ``principle of the fermionic projector,'' an approach 
which originated by the attempt to resolve some of the shortcomings of relativistic quantum field theory. 
The variational principle at the heart of the theory of causal fermion systems is known as ``causal action principle.'' 
Spacetime and the structures therein are described by minimizers of the causal action principle. 
The underlying picture is to think of ``macroscopic'' spacetime as a certain limiting case (the so-called ``continuum limit'') of a more fundamental ``microscopic'' structure which 
is described mathematically in terms of a so-called ``regularization.''  
In the continuum limit (that is, when the regularization is removed), the resulting Euler-Lagrange equations give rise to 
an effective interaction 
which corresponds to  
classical gravity as well as the strong and electroweak gauge fields of the Standard Model.

Let us first recall some essential facts concerning modern physics.  
At present, there are four known fundamental forces in nature, which are \emph{gravitation, electromagnetic interaction, weak interaction} and \emph{strong interaction}. 
These four fundamental forces 
are described by \emph{Einstein's theory of general relativity} (gravitation) and the \emph{Standard Model of elementary particle physics} (electromagnetic, strong and weak interaction).  
All elementary particles known at present are either \emph{fermions} or \emph{bosons}. 
Fermions are the subatomic constituents of matter, comprising \emph{quarks} and \emph{leptons}. 
The fundamental bosons, on the other hand, are the mediating particles of the various interactions: \emph{photons} (electromagnetism), \emph{$W^{\pm}$ and~$Z$ bosons} (weak interaction) and \emph{gluons} (strong interaction). 
Furthermore, there might exist not yet detected \emph{gravitons} (gravitation). 
The Standard Model successfully incorporates all known properties of strong, weak and electromagnetic forces, whereas gravitational interactions are absent in the Standard Model. Thus it is considered one of the great physical problems of this century to unify general relativity and quantum mechanics~\cite{folland-qft, kaku, zeidler-gauge}.

In order to motivate the underlying ideas of the theory of causal fermion systems, let us point out the following drawback of the Standard Model of elementary particle physics. 
The description of the Standard Model rests upon relativistic quantum field theory. Unfortunately, one of the serious complications in quantum field theory is that the theory is naively divergent in the ultraviolet region. More precisely, it turns out that 
many physical quantities of interest are not integrable in momentum space due to unbounded contributions for large momenta, 
that is, for high energies. In order to bypass these divergences, physicists developed the so-called \emph{renormalization program}, which is a set of prescriptions for how to make sense of divergent integrals~\cite{kaku, peskin+schroeder}. To this end, one commonly introduces an ``ultraviolet cutoff'' in momentum space, thereby ``cutting off'' the region of high energies, a procedure known as ``regularization.'' Despite its considerable success, this method of ``regularization'' seems artificial from the physical point of view, and it is unjustified from the mathematical one. Or, to put it in Dirac's own words (cf.~\cite[p.~184]{kragh}),   
\begin{quote}
	``I must say that I am very dissatisfied with the situation because the so-called 'good theory' does involve neglecting infinities in its equations, ignoring them in an arbitrary way. This is just not sensible mathema\-tics. Sensible mathematics involves disregarding a quantity when it is small --- not neglecting it just because it is infinitely great and you do not want it!''
\end{quote}
The need for a cutoff in momentum space indicates that large momenta are not taken into account in the correct way in quantum field theory. More specifically, the cutoff in momentum space, introduced in order to render divergent integrals finite, is often associated to Planck energy~$E_P \approx 1.22 \cdot 10^{28}\;\text{eV}$.\footnote{For clarity, one \emph{electron volt} is approximately~$1 \; \text{eV} \approx 1.60 \cdot 10^{-19} \; \text{kg m}^2/\text{s}^2$.}  
In view of Heisenberg's uncertainty principle, physicists infer a correspondence between large momenta (high energies) and small distances in position space. Because of that,  
in quantum field theory 
one disregards distances which are smaller than the Planck length~$\ell_P \approx 1.6 \cdot 10^{-35}$~m.  
Consequently, the microstructure of spacetime is completely unknown. Unfortunately, at present there is no consensus on what the correct mathematical model for ``Planck scale physics'' should be. 

The simplest and maybe most natural approach is to assume that on the Planck scale, spacetime is no longer a continuum but becomes in some way ``discrete." This is the starting point in the monograph~\cite{pfp}, which later evolved to the theory of causal fermion systems~\cite{cfs}. The objective of this paper is to outline the main structure of the theory of causal fermion systems and to clarify its relation to physics. To begin with, let us first state the formal definition of causal fermion systems according to~\cite{cfs}. 
\begin{Def} \label{c:defparticle} %(causal fermion system) {\em{ \hspace*{1em}
	Let~$\H$ be a separable complex Hilbert space together with a scalar product~$\la . \, | \, . \ra_\H$, and denote by~$\LL(\H)$ the set of linear operators on~$\H$. 
	Moreover, given a parameter~$n \in \N$ (the so-called {\bf{``spin dimension''}}), we let~$\F \subset \Lin(\H)$ be the set of all
	self-adjoint operators on~$\H$ of finite rank, which (counting multiplicities) have
	at most~$n$ positive and at most~$n$ negative eigenvalues. Furthermore, let~$d\rho$ be a positive measure on~$\F$ (defined on a $\sigma$-algebra of subsets of~$\F$), the so-called
	{\bf{universal measure}}. We refer to~$(\H, \F, d\rho)$ as a {\bf{causal fermion system}}.
\end{Def} 

In order to single out the physically admissible
causal fermion systems, one must formulate physical equations. Following the principle of least action, we impose that
the universal measure should be a minimizer of the causal action principle, 
which we now introduce. For any~$x, y \in \F$, the product~$x y$ is an operator
of rank at most~$2n$.  
Denoting its eigenvalues (counting algebraic multiplicities)
by~$\lambda^{xy}_1, \ldots, \lambda^{xy}_{2n} \in \C$ 
and introducing the {\em{spectral weight}}~$| \,.\, |$
of an operator as the sum of the absolute values
of its eigenvalues, 
the spectral weight of the operator
products~$xy$ and~$(xy)^2$ are given by
\[ |xy| = \sum_{i=1}^{2n} \big| \lambda^{xy}_i \big|
\qquad \text{and} \qquad \big| (xy)^2 \big| = \sum_{i=1}^{2n} \big| \lambda^{xy}_i \big|^2 \:. \]
We introduce the \emph{Lagrangian} by
\begin{align*}
	\L : \F \times \F \to \R \:, \qquad \L(x,y) &= \big| (xy)^2 \big| - \frac{1}{2n}\: |xy|^2 % \label{c:cfs, (1.1.1)}
\end{align*}
and the \emph{causal action} by
\begin{align}
%	\text{\em{Lagrangian:}} && \L(x,y) &= \big| (xy)^2 \big| - \frac{1}{2n}\: |xy|^2 \label{c:cfs, (1.1.1)} \\
%	\text{\em{causal action:}} && 
	\Sact(\rho) = \iint_{\F \times \F} \L(x,y)\: d\rho(x)\, d\rho(y) \:. \label{c:cfs, (1.1.2)}
\end{align}
The {\em{causal action principle}} is to minimize~$\Sact$ by varying the universal measure~$d\rho$ 
under the following constraints:
\begin{align}
	\text{\em{volume constraint:}} && \rho(\F) = \text{const} \quad\;\; & \label{c:cfs, (1.1.3)} \\
	\text{\em{trace constraint:}} && \int_\F \tr(x)\: d\rho(x) = \text{const}& \label{c:cfs, (1.1.4)} \\
	\text{\em{boundedness constraint:}} && \T(\rho) := \iint_{\F \times \F} |xy|^2\: d\rho(x)\, d\rho(y) &\leq C \:, \label{c:cfs, (1.1.5)}
\end{align}
where~$C$ is a given parameter (and~$\tr$ denotes the trace of a linear operator on~$\H$).
%\sindex{constraint!volume}%
%\sindex{constraint!trace}%
%\sindex{constraint!boundedness}%

In order  
to specify the class of measures in which to vary~$d\rho$, 
on~$\F$ we consider the topology induced by the operator norm 
\begin{align*}
	\|A\| := \sup \big\{ \|A u \|_\H \text{ with } \| u \|_\H = 1 \big\} \:.
\end{align*}
The \emph{Borel $\sigma$-algebra}~$\B(\F)$ is generated by the open sets of~$\F$. Its elements are called \emph{Borel sets}.
A {\em{regular Borel measure}}
is a measure on the Borel sets with the property that
it is continuous under approximations by compact sets from inside and by open sets from outside
(for basics see for instance~\cite{gardner+pfeffer, halmosmt}). 
%or Appendix~\ref{Section Measure Theory}).
The right prescription is to vary the universal measure~$d\rho$ within the class of regular Borel measures on~$\F$. 
The constraints~\eqref{c:cfs, (1.1.3)}--\eqref{c:cfs, (1.1.5)} are needed to avoid trivial minimizers and in order for the variational principle to be well-posed. 

It is of central importance to observe that  
the causal action principle essentially depends on the eigenvalues of the operator product~$xy$ with~$x,y \in \F$. Instead of working with the operator product~$xy$, however, it is convenient to proceed as follows.  
For every~$x \in \F$, the corresponding \emph{spin space}~$S_x := x(\H)$ is a subspace of~$\H$ of dimension at most~$2n$. Denoting the orthogonal projection onto the subspace~$x(\H)$  
by~$\pi_x$,  
for all~$x,y \in \F$  one defines the \emph{kernel of the fermionic projector}~$P(x,y)$ by
\begin{align}\label{c:cfs, (1.1.13)}
	P(x,y) := \pi_x y|_{S_y} : S_y \to S_x \:. 
\end{align}
In order to express the eigenvalues of the operator~$xy$, one then introduces the \emph{closed chain}~$A_{xy}$ by
\begin{align}\label{c:cfs, (1.1.14)}
	A_{xy} := P(x,y)P(y,x) : S_x \to S_x \:. 
\end{align} 
Repeating the arguments after~\cite[eq.~(1.1.10)]{cfs}, it turns out that the eigenvalues of the closed chain~$A_{xy}$ coincide with the non-trivial eigenvalues~$\lambda_1^{xy}, \ldots, \lambda_{2n}^{xy}$ of~$xy$. 

At this stage, however, the reader who is unfamiliar with the theory of causal fermion systems might wonder  
how the causal action principle comes about and in which way it is connected to physics. This introduction is intended to  
hopefully settle these issues; 
more precisely, the goal of this paper is to put the theory of causal fermion systems into the physical context, to give a terse summary of its main concepts and to clarify the specific form of the causal action principle. Following~\cite{vester}, we shall restrict attention to the crucial constructions in a non-rigorous way without entering most of the details.

The paper is organized as follows. In order to put the theory of causal fermion systems into the historical context, in Section~\ref{c:Section History} we give a succinct overview of the history of physics, thereby summarizing particularly significant physical achievements. In Section~\ref{c:Section Physical Preliminaries} we recall the fundamental physical preliminaries required for the theory of causal fermion systems: classical mechanics (\S \ref{c:S classical mechanics}), electrodynamics (\S \ref{c:S electrodynamics}), special relativity (\S \ref{c:S special relativity}), general relativity (\S \ref{c:S general relativity}) as well as quantum mechanics (\S \ref{c:S qm}). After recalling some aspects of group theory in physics (\S \ref{c:S group theory}), we give a brief introduction to elementary particles (\S \ref{c:S elementary particles}) and outline the Standard Model of elementary particle physics (\S \ref{c:S SM}). We conclude this section with a glimpse of quantum field theory (\S \ref{c:S qft}). After these preparations, in Section~\ref{c:Section Principle} we outline the principle of the fermionic projector, from which the theory of causal fermion systems eventually evolved. To this end, we first sketch the derivation of local gauge freedom (\S \ref{c:S gauge freedom}) and then enter the principle of the fermionic projector (\S \ref{c:S principle}). Afterwards, the connection to causal fermion systems is established (\S \ref{c:S connection CFS}). In Section~\ref{c:Section CFS}, we finally introduce the theory of causal fermion systems. We first present the general strategy (\S \ref{c:S general strategy}) and clarify the connection to Dirac's original concept of a Dirac sea (\S \ref{c:S kernel}). We then recall the external field problem (\S \ref{c:S external field problem}) and describe the construction of the ``auxiliary'' fermionic projector (\S \ref{c:S auxiliary}). In order to explain the connection to modern physics, we outline the so-called ``light-cone expansion'' (\S \ref{c:S light-cone expansion}) and describe the formalism of the continuum limit in a few words (\S \ref{c:S continuum limit}). This allows us to formulate the Euler-Lagrange equations (\S \ref{c:S EL equations}) in a concise way, from which finally the classical field equations can be derived (\S \ref{c:S field equations}) and the connection to the Standard Model of particle physics may be established (\S \ref{c:S connection SM}).

\section{A Brief History of Physics}\label{c:Section History}
The notion of ``physics'' was coined by one of \textsc{Aristotle's} (384--322 BC) major works~\cite{russell-history}. 
Back to the ancient Greeks, the idea of an atomic structure of matter was postulated by \textsc{Democritus} (460--371 BC), and the Athenian philosopher \textsc{Plato} (428/427--348/347 BC) developed a first approach towards cosmology~\cite{russell-history}. At the beginning of modern ages, \textsc{Nikolaus Kopernikus} (1473--1543) described in his principal work \emph{De revolutionibus orbium coelestium} (1543) a heliocentric view of the world~\cite{born-relativitatstheorie}, which was affirmed by astronomical observations by \textsc{Galileo Galilei} (1564--1641) and calculations of his contemporary, \textsc{Johannes Kepler} (1571--1630). The foundations of modern mechanics were developed and formulated in \textsc{Isaac Newton}'s magnum opus \emph{Philosophiae Naturalis Principia Mathematica} (1687) in order to explain Kepler's laws by a universal law of gravitation~\cite{newton-1999}.
The principle of least action goes back to \textsc{Pierre Louis Moreau de Maupertuis} (1698--1759). Employing variational principles, \textsc{Leonhard Euler} (1707--1783) and \textsc{Joseph-Louis Lagrange} (1736--1813) made fundamental contributions to the further development of mechanics.

At the end of the 18th century, \textsc{Charles-Augustin de Coulomb} (1736--1806) formulated the laws of electrostatics. Observations concerning heat conduction by \textsc{Joseph Fourier} (1768--1830) led to the fundamental method of \emph{Fourier analysis}, whereas the foundations of electrodynamics were laid by \textsc{Andr\'e-Marie Amp\`ere} (1775--1836) and \textsc{Michael Faraday} (1791--1867).  
The further development of
electrodynamics as well as the theory of thermodynamics in the 19th century is mainly due to \textsc{James Prescott Joule} (1818--1889), \textsc{Lord Kelvin} (1824--1907), \textsc{Hermann von Helmholtz} (1821--1894), \textsc{James Clerk Maxwell} (1831--1879) and \textsc{Ludwig Boltzmann} (1844--1906),
reaching its peak in \emph{Maxwell's equations}. In honor of \textsc{William Rowan Hamilton} (1805--1865), the principle of least action is also referred to as \emph{Hamilton's principle}. Moreover, \textsc{Oliver Heaviside} (1850--1925) and \textsc{Heinrich Hertz} (1857--1894) made important contributions for developing the theory of electromagnetism~\cite{born-relativitatstheorie, gribbin, griffiths, pal}.

At the beginning of the 20th century, the two fundamental pillars of modern physics were discovered: quantum theory by \textsc{Max Planck} (1858--1947) as well as the theory of relativity by \textsc{Albert Einstein} (1879--1955). Both theories substantially changed the physical view of the world~\cite{born-relativitatstheorie, veltman}. Preparatory works by \textsc{Hendrik Antoon Lorentz} (1853--1928) and \textsc{Henri Poincar\'e} (1854--1912) led to Einstein's special relativity (1905), a theory which is solely based on two postulates, the principle of relativity and constancy of the speed of light. Subsequently, \textsc{Hermann Minkowski} (1864--1909) developed the formal basis of four-dimensional spacetime in special relativity theory. The foundations of general relativity go back to Albert Einstein as well. Originating from the principle of equivalence, he applied differential geometry as developed by \textsc{Carl Friedrich Gau\ss} (1777--1855), \textsc{Bernhard Riemann} (1826--1866), \textsc{Elwin Bruno Christoffel} (1829--1900), \textsc{Gregorio Ricci-Curbastro} (1853--1925) and \textsc{Tullio Levi-Civita} (1873--1941) in order to formulate gravitation as a geometric property of curved four-dimensional spacetime~\cite{simon}. The \emph{Einstein field equations} form the fundamental equations of general relativity. \textsc{David Hilbert} (1862--1943) was the first one to recognize a particularly elegant derivation of the Einstein equations by varying the \emph{Einstein-Hilbert action}, thereby employing the principle of least action. The \emph{Schwarzschild metric}, discovered in 1916 by \textsc{Karl Schwarzschild} (1873--1916) was the first known exact solution of the Einstein equations~\cite{schwarzschild-1916}.

Quantum theory owes its origins to Planck's quantum hypothesis~\cite{planck-1900}, according to which radiation is not emitted continuously, but in form of discrete ``quanta.'' Based on Planck's hypothesis, Einstein deduced the particle character of light (photons). In order to resolve the resulting contrast to the wave theory of light, the French physicist \textsc{Louis de Broglie} (1892--1987) postulated \emph{wave-particle duality} to be the fundamental principle of nature. In the mid-1920's,  
\textsc{Werner Heisenberg} (1901--1976), \textsc{Max Born} (1882--1970), \textsc{Pascual Jordan} (1902--1980), \textsc{Wolfgang Pauli} (1900--1958) and \textsc{Erwin Schr\"odinger} (1887--1961) developed quantum mechanics~\cite{weinberg-I}. In 1928, the British physicist \textsc{Paul Dirac} (1902--1984) formulated the \emph{Dirac equation}~\cite{dirac-1928}, thereby postulating the existence of anti-matter. Together with the \emph{Klein-Gordon equation} (due to \textsc{Oskar Klein} (1894--1977) and \textsc{Walter Gordon} (1893--1939)), the Dirac equation admits a relativistic formulation of quantum theory.  
The discovery of spin, a fundamental property of all particles, led to a subdivision of matter into \emph{bosons}\footnote{Particles with integer spin, in honor of the Indian physicist \textsc{Satyendra Nath Bose} (1894--1974).} and \emph{fermions}\footnote{Particles with half-integer spin, named after the Italian physicist \textsc{Enrico Fermi} (1901--1954).}. Max Born introduced the statistical interpretation of wave functions, which was expanded by the \emph{Copenhagen interpretation of quantum theory} by \textsc{Niels Bohr} (1885--1962) and Werner Heisenberg in 1928. By eliminating determinism, quantum physics drastically changed the paradigms of physics~\cite{born-relativitatstheorie}. 

The 1930's were characterized by developments in nuclear physics, supported by the invention of particle accelerators. 
For instance, the neutron was discovered, which in the sequel led to the discovery of weak interaction.  
The further development of particle accelerators after the Second World War led to the discovery of a whole ``zoo'' of elementary particles. At the end of the 1940's, quantum field theory (QFT) emerged~\cite{schweber-qed, schweber, weinberg-I}. The development of quantum electrodynamics (QED) is mostly due to \textsc{Shinichiro Tomonaga} (1906--1979), 
\textsc{Julian Schwinger} (1918--1994) as well as \textsc{Richard Feynman} (1918--1988) and \textsc{Freeman Dyson} (1923--2020).  
Originally introduced by \textsc{Hermann Weyl} (1885--1955), \emph{gauge theories} turned out to be of central importance for the emerging \emph{Standard Model of elementary particles}, in particular \emph{Yang-Mills theories}~\cite{yang+mills} (due to \textsc{Chen-Ning Yang} (1922--) and \textsc{Robert Mills} (1927--1999)).  
Unifying the electromagnetic and weak interaction to the electroweak theory was accomplished by \textsc{Abdus Salam} (1926--1996), \textsc{Sheldon Glashow} \mbox{(1932--)} and \textsc{Steven Weinberg} {(1933--2021)} in the 1960's. The roots of quantum chromodynamics (QCD), a Yang-Mills theory for the strong interaction, go back to the 1970's.  
Important contributions concerning the strong interaction are due to \textsc{Murray Gell-Mann} (1929--2019).  
Since the end of the 1970's, elementary particle physics was governed by string theory~\cite{yau+nadis}, aiming for unifying gravity and quantum theory, which also is the main objective of other approaches like quantum gravity~\cite{rovelli}. 
The origins of what later evolved to the theory of causal fermion systems go back to the 1990's~\cite{finster1996ableitung, gauge, local}.

\section{Physical Preliminaries}\label{c:Section Physical Preliminaries}
The aim of this subsection is to outline fundamental physical preliminaries which are of essential importance for the theory of causal fermion systems. For an introduction to concepts of modern physics, the interested reader is referred to~\cite{beiser}. For a non-technical introduction to physics we refer to~\cite{feynman-1, feynman-2, feynman-3} as well as~\cite{gribbin}. The explanations in this subsection mostly follow the non-rigorous style common in physics textbooks, and we usually shall employ the physicists' conventions. Concerning the physicists' notation, we highly recommend~\cite{folland-qft}. For instance, physicists like to indicate the dimensionality of their integrals explicitly by writing the volume element on~$\R^n$ as~$d^nx$ (see~\cite{folland-qft}). 
Moreover, vectors in Euclidean space~$\R^3$ are usually written as~$\mathbf{x}$ or~$\vec{x}$. In order to symbolically distinguish between conventional physics and and the theory of causal fermion systems, we shall first employ the bold-face notation and switch to the vector notation afterwards.

Let us finally address the subject of physical units. The fundamental aspects of the universe to which all physical measurements relate are mass, length or distance and time (cf.~\cite{folland-qft}). All quantities in physics come with ``dimensions'' that can be expressed in terms of these three basic ones. 
Very often is is customary and convenient to use so-called \emph{natural units}~$\hbar = c = 1$ (where~$\hbar$ is reduced Planck's constant and~$c$ denotes the speed of light). In natural units,  
we have~$[\textup{length}] = [\textup{time}] = [\textup{energy}]^{-1} = [\textup{mass}]^{-1}$, 
and  the mass of a particle is equivalent to its rest energy~$mc^2$, and also to its inverse Compton wavelength~$mc/\hbar$ (see~\cite{kane, peskin+schroeder}). Factors of~$\hbar$ or~$c$, which are not displayed, can be reinserted by ``getting the dimension right.''

\subsection{Classical Mechanics}\label{c:S classical mechanics}
All physical theories are based on certain fundamental laws, which are referred to as \emph{laws of nature}. In classical mechanics, \emph{Newton's axioms} are regarded as being these fundamental laws. 
Based on Newton's axioms, classical mechanics provides the foundations of theoretical physics, and its methods, including the Hamilton and Lagrangian formalism, can be regarded as fundamental principles of theoretical physics. An alternative approach is based on the principle of least action, in which case the laws of nature in classical mechanics are formulated in terms of an action principle (cf.~\cite{fliessbach}). In preparation for the theory of causal fermion systems, let us enter the Lagrangian formalism in classical mechanics in some more detail.  
The subsequent explanations are due to~\cite{folland-qft} and~\cite{landau+lifshitz-mechanics}. 

\subsubsection*{The Equations of Motion}
The main subject of classical mechanics is to study time-dependent motions in three-dimensional Euclidean space~$\R^3$. 
One of the fundamental concepts of classical mechanics is that of a \emph{particle}.  
The position of a particle in~$\R^3$ 
is indicated by its vector~$\mathbf{r} = \mathbf{r}(t)$, depending on time~$t$, whose components are its Cartesian coordinates~$x = x(t),y = y (t), z = z(t)$. The derivative~$d\mathbf{r}/dt$ of~$\mathbf{r}$ with respect to time~$t$ is called \emph{velocity} of the particle, and the second derivative~$d^2 \mathbf{r}/dt^2$ is referred to as \emph{acceleration}. It is customary to denote differentiation with respect to time by~$\mathbf{v} = \dot{\mathbf{r}}$.  

To define the position of a system of~$N$ particles in space, one needs to specify~$N$ radius vectors, thus giving rise to~$3N$ coordinates. Accordingly, if no constraints are involved, the number of \emph{degrees of freedom} of such a system is~$3N$, which in general is defined as the number of independent quantities required in order to uniquely specify the position of any system. In general,
any~$s$ quantities~$q_1, \ldots, q_s$ which completely define the position of a system with~$s$ degrees of freedom are called \emph{generalized coordinates} of the system, and the derivatives~$\dot{q}_i$ are called its \emph{generalized velocities}. The manifold described by generalized coordinates is known as \emph{configuration space}. The relations between the coordinates, velocities and accelerations are referred to as \emph{equations of motion}. 

The \emph{principle of least action} (or \emph{Hamilton's principle}), according to which every mechanical system is characterized by a definite function~$L(q_1, \ldots, q_s, \dot{q}_1, \ldots, \dot{q}_s, t)$, or briefly~$L({q}, \dot{{q}}, t)$, is
the most general formulation of the law governing the motion of mechanical systems. 
Given a path~$t \mapsto {q}(t)$ in configuration space with~$t_1 \le t \le t_2$, the \emph{action} is introduced by  
\begin{align*}%\label{c:(LL,2.1)}
	S = \int_{t_1}^{t_2} L({q}, \dot{{q}}, t) \: dt \:,
\end{align*}
and the problem is to \emph{minimize the action over all paths~${q}(t)$ that begin at~${q}(t_1)$ and end at~${q}(t_2)$}.  
The function~$L$ is called the \emph{Lagrangian} of the system concerned. 
The necessary condition for~$S$ to have an extremum is that its \emph{first variation} equals zero. Thus the principle of the least action may be expressed by imposing that 
\begin{align*}%\label{c:(LL, 2.4)}
	\delta S = \delta \int_{t_1}^{t_2} L({q}, \dot{{q}}, t) \: dt = 0 \:. 
\end{align*}
We then obtain~$s$ equations of the form 
\begin{align}\label{c:(11.2')}
	\frac{d}{dt} \left(\frac{\partial L}{\partial \dot{q}_i} \right) - \frac{\partial L}{\partial q_i} = 0 \qquad \text{for all~$i=1, \ldots, s$} \:.
\end{align}
These are the equations of motions,  
also known as \emph{Euler-Lagrange equations}.\footnote{Introducing the \emph{momentum~$p$ conjugate to~$q$} by~$p \equiv \partial L/\partial \dot{q}$, 
	%\begin{align}\label{c:BD2, (11.3)}
	%p \equiv \frac{\partial L}{\partial \dot{q}} \:,
	%\end{align}
	and the \emph{Hamiltonian} by the Legendre transformation~$H(p,q) = p\dot{q} - L(q, \dot{q})$, 
	the equations of motion~\eqref{c:(11.2')} become
	\begin{align*}%\label{c:BD2, (11.5)}
		\frac{\partial H}{\partial p} = \dot{q} \qquad \text{and} \qquad - \frac{\partial H}{\partial q} = \dot{p} \:. 
\end{align*}}

Let us now consider a system of~$N$ particles which may interact with one another, but are not subject to external interactions. This is called a \emph{closed system}. In this case, the Lagrangian can be written as
\begin{align}\label{c:LL1, (5.1)}
	L = \sum_{a=1}^N \frac{1}{2} m_a v_a^2 - U(\mathbf{r}_1, \ldots, \mathbf{r}_N) \:, 
\end{align}
where~$\mathbf{r}_a$ is the radius vector of the $a$th particle ($a=1, \ldots, N$). This is the general form of the Lagrangian for a closed system in classical mechanics, where~$T \equiv \sum_{a=1}^N \frac{1}{2} m_a v_a^2$ is called the \emph{kinetic energy}, and~$U$ is referred to as the \emph{potential energy}.

\subsubsection*{Conservation Laws} 
In order to describe mechanical phenomena mathematically one needs to choose a \emph{reference frame}.  
In classical mechanics, it is sensible to assume 
that space is \emph{homogeneous} as well as \emph{isotropic} and time is \emph{homogeneous} (where homogeneity means that no point   
in space and time is distinguished, and isotropy means that no spatial direction is distinguished, see e.g.~\cite{fliessbach} and~\cite[Chapter~52]{feynman-1}).

We now consider a mechanical system with~$s$ degrees of freedom whose dynamics 
respects the symmetries of space and time; more precisely, we assume its Lagrangian to be  
invariant under translations in space and time as well as rotations in space.  
To such symmetries correspond the following conservation laws. 
The first conservation law resulting from the \emph{homogeneity of time} asserts that the quantity
\begin{align*}%\label{c:(LL1, 6.1)}
	E \equiv \sum_{i=1}^{s} \dot{q}_i \frac{\partial L}{\partial \dot{q}_i} - L
\end{align*}
remains constant during the motion of a closed system; it is called the \emph{energy} of the system. Mechanical systems whose energy is conserved are said to be \emph{conservative}.   
A second conservation law follows from the \emph{homogeneity of space}. More explicitly, in a closed mechanical system the vector
\begin{align*}%\label{c:LL1, (7.2)}
	\mathbf{p} \equiv \sum_{a=1}^N \frac{\partial L}{\partial \mathbf{v}_a} \:, 
\end{align*}
known as the \emph{momentum} of the system, 
remains constant in time. Differentiating the Lagrangian~\eqref{c:LL1, (5.1)},  
in terms of the velocities of the particles the momentum reads
\begin{align*}%\label{c:LL, (7.3)}
	\mathbf{p} = \sum_{a=1}^N m_a \mathbf{v}_a \:,
\end{align*}
where~$m_a$ denotes the \emph{mass} of the~$a$th particle ($a = 1, \ldots, N$).  
A third conservation law follows from the \emph{isotropy of space}. 
In this case, the vector
\begin{align*}%\label{c:LL1, (9.3)}
	\mathbf{M} \equiv \sum_{a=1}^N \mathbf{r}_a \times \mathbf{p}_a \:, 
\end{align*}
which is called the \emph{angular momentum} of the system, is a conserved quantity of a closed system. 
In greater generality, Noether's theorem states that symmetries of a system give rise to conserved quantities (see e.g.~\cite[Chapter~4]{griffiths} or~\cite[Section~1.5]{jost-calculus}).

\subsection{Electrodynamics}\label{c:S electrodynamics} 
The interaction of particles in classical electrodynamics is described by means of \emph{fields}~\cite{landau+lifshitz-fields}.\footnote{According to~\cite[Section~1-2]{feynman-2}, a ``field'' is any physical quantity which takes on different values at different points in space. It is precisely because~$\mathbf{E}$ and~$\mathbf{B}$ can be specified at every point in space that they are called ``fields.''} Electromagnetic phenomena are governed by Maxwell's equations~\cite{jackson}, specifying the behavior of an electric field~$\mathbf{E}$ and a magnetic field~$\mathbf{B}$. 
Following~\cite[Section~2.4]{folland-qft}, the electric field~$\mathbf{E}$ and the magnetic field~$\mathbf{B}$ are vector-valued functions of position~$\mathbf{x} \in \R^3$ and time~$t$ that enter the \emph{Lorentz force law}: The force on a particle with charge~$q$ moving with velocity~$\mathbf{v}$ is
\begin{align*}%\label{c:folland-qft, (2.17)}
	\mathbf{F} = q \mathbf{E} + \frac{q}{c} \mathbf{v} \times \mathbf{B} \:,
\end{align*}
where~$c$ is the speed of light, and the fields~$\mathbf{E}$ and~$\mathbf{B}$ are evaluated at the location of the particle 
(further details can be found in~\cite[Section~7.4]{griffiths}).

The behavior of~$\mathbf{E}$ and~$\mathbf{B}$ is governed by \emph{Maxwell's equations}.  
Using the Heaviside-Lorentz convention  
(see~\cite[\S 1.1]{folland-qft}), they read
\begin{align}
	\nabla \cdot \mathbf{E} &= \rho \label{c:folland-qft, (2.18)} \\
	\nabla \cdot \mathbf{B} &= 0 \label{c:folland-qft, (2.19)} \\
	\nabla \times \mathbf{E} + \frac{1}{c} \frac{\partial \mathbf{B}}{\partial t} &= 0 \label{c:folland-qft, (2.20)} \\
	\nabla \times \mathbf{B} - \frac{1}{c} \frac{\partial \mathbf{E}}{\partial t} &= \frac{1}{c} \mathbf{j} \:,  \label{c:folland-qft, (2.21)} 
\end{align}
where~$\rho$ is the charge density and~$\mathbf{j}$ is the current density, both of which are functions of position and time.  
The quantities~$\rho$ and~$\mathbf{j}$ are not independent; indeed, they satisfy the so-called \emph{continuity equation}
\begin{align}\label{c:folland-qft, (2.22)}
	\frac{\partial \rho}{\partial t} + \nabla \cdot \mathbf{j} = 0 \:,
\end{align}
which describes the conservation of electric charge. 

For the homogeneous Maxwell equations,  
formula~\eqref{c:folland-qft, (2.19)} is equivalent to the fact that~$\mathbf{B}$ can be written as the curl of a \emph{vector potential}~$\mathbf{A}$, 
\begin{align}\label{c:Griffiths, (7.75)}
	\mathbf{B} = \nabla \times \mathbf{A} \:. 
\end{align}
With this, formula~\eqref{c:folland-qft, (2.20)} becomes
\begin{align*}%\label{c:Griffiths, (7.76)}
	\nabla \times \left(\mathbf{E} + \frac{1}{c} \frac{\partial \mathbf{A}}{\partial t} \right) = 0 \:,
\end{align*}
which is equivalent to the statement that the expression in brackets,~$\mathbf{E} + (1/c)(\partial \mathbf{A}/\partial t)$,  can be written as the gradient of a \emph{scalar potential}~$\phi$, 
\begin{align}\label{c:Griffiths, (7.77)}
	\mathbf{E} = - \nabla \phi - \frac{1}{c} \frac{\partial \mathbf{A}}{\partial t} \:. 
\end{align} 
Note that~$\mathbf{A}$ and~$\phi$ are not uniquely determined; 
adjustments of~$\mathbf{A}$ and~$\phi$ which do not affect~$\mathbf{E}$ and~$\mathbf{B}$ are called \emph{gauge transformations} (for details see~\cite[Chapter~9]{folland-qft}). A frequently imposed gauge condition is the \emph{Landau gauge} (or \emph{Lorentz gauge}), 
\begin{align}\label{c:folland-qft, (2.23)}
	\nabla \cdot \mathbf{A} + \frac{1}{c} \frac{\partial \phi}{\partial t} = 0 \:.
\end{align} 
Denoting the \emph{wave operator} or \emph{d'Alembertian} by~$\Box$, 
\begin{align*}
	\Box \equiv \frac{1}{c^2} \frac{\partial^2}{\partial t^2} - \nabla^2 \:, 
\end{align*}
in the presence of the Lorentz gauge condition~\eqref{c:folland-qft, (2.23)}
Maxwell's equations read\footnote{For a relativistic formulation of electrodynamics we refer to 
	the end of 
	the next subsection.}
\begin{align*}%\label{c:folland-qft, (2.24)}
	\Box \phi = \rho \:, \qquad \Box \mathbf{A} = \mathbf{j} \:. 
\end{align*} 

In case that~$\rho$ and~$\mathbf{j}$ vanish, %both the electromagnetic potentials~$A$ and~$\phi$ as well 
the Maxwell equations imply that the electromagnetic fields~$\mathbf{E}$ and~$\mathbf{B}$ satisfy the \emph{wave equation}  
\begin{align*}
	\nabla^2 \psi = \frac{1}{c^2} \frac{\partial^2 \psi}{\partial t^2} \:,
\end{align*}
which describes waves propagating at the speed of light~$c$. 
Of special interest are waves that are periodic in space and time, 
which are 
referred to as \emph{plane waves}. Periodicity in space is characterized by the \emph{wave length}~$\lambda$, while periodicity in time is represented by the \emph{time period}~$T$. The related quantities~$k = 2\pi /\lambda$ and~$\omega = 2\pi/T$ are called \emph{wave number} and \emph{angular frequency}, respectively. 
Such waves travels at a speed~$c = \omega/k$. 
In three dimensions, a plane wave~$\psi$ at time~$t$ and position~$\mathbf{r}$, moving in~$\mathbf{k}$-direction (where~$\mathbf{k}$ is called \emph{wave vector}), can be written as
\begin{align*}
	\psi(t, \mathbf{r}) = a e^{i(\mathbf{k} \cdot \mathbf{r} - \omega t)}
\end{align*}
with (real) \emph{amplitude}~$a$ and~$\omega = |\mathbf{k}|c$.  
The quantity~$(\mathbf{k} \cdot \mathbf{r} - \omega t)$ is known as \emph{phase}.  
For details see~\cite[\S 6]{landau+lifshitz-qm} and~\cite[Section~3.1]{shankar}. 

\subsection{Special Relativity}\label{c:S special relativity}
Special relativity is based on two fundamental postulates: the principle of relativity and the constancy of the speed of light. The basic difficulty is to merge both facts, each of them being confirmed experimentally. Einstein's brilliant contribution~\cite{einstein-1905} amounts to the insight that to this end, the common \emph{notion of simultaneity} has to be abandoned. In simple terms, Einstein claims that \emph{there is no absolute simultaneity}, thereby radically changing our understanding of space and time~\cite{born-relativitatstheorie}.

In order to describe processes taking place in nature, one considers a \emph{reference frame}, that is, 
a system of coordinates in order to indicate the position of particles in space, together with a clock fixed in this system in order to indicate time. A reference frame, in which a particle which is not acted upon by external forces proceeds with constant velocity, is called an \emph{inertial frame}. 
Classical mechanics is governed by the so-called \emph{principle of relativity}, according to which all laws of nature are identical in all inertial systems of reference~\cite{feynman-1, landau+lifshitz-fields}. 
On the other hand, it is found that the speed of light, usually denoted by~$c$, is the \emph{same} in \emph{all} inertial systems of reference.\footnote{Concerning the connection to the famous Michelson-Morley experiment, carried out by \textsc{Albert Michelson} (1852--1931) and \textsc{Edward Morley} (1838--1923) in~1887, we refer to~\cite{sexl+urbantke} and~\cite{feynman-1}.}

Based on preparatory works by Lorentz~\cite{lorentz-1892}, Einstein's solution to the problem raised is to claim that space and time become \emph{relative}. More specifically,  
assume that~$S$ and~$S'$ are inertial frames, where~$S'$ moves with uniform velocity of magnitude~$v$ in~$x$-direction with respect to~$S$. 
Denoting the coordinates of space and time in~$S$ and~$S'$ by~$x,y,z,t$ and~$x', y', z', t'$, respectively, the so-called \emph{Lorentz transformations} establish the following connection, 
\begin{align}\label{c:Griffiths, (3.1)}
	x' = \gamma (x - vt) \:, \qquad y' = y \:, \qquad z' = z \:, \qquad t' = \gamma (t- \frac{v}{c^2}x) \:, 
\end{align}
where~$\gamma \equiv 1/\sqrt{1 - v^2/c^2}$. 
The Lorentz transformations have a number of immediate consequences: the \emph{relativity of simultaneity}, \emph{Lorentz contraction}, \emph{time dilation} and the \emph{addition of velocity} (for details we refer to~\cite[Chapter~3]{griffiths}).

It is well-known that the velocity of light has the numerical value
\begin{align*}
	c = 2.998 \times 10^{8} \: \textup{m/s} \:. 
\end{align*} 
It is precisely the large value of~$c$ which explains why in most cases of practical interest classical mechanics appears to be sufficiently accurate. In other words, classical mechanics may be regarded as the limiting case of relativistic mechanics: Whenever the velocity~$v$ is small compared to the speed of light, i.e.~$v \ll c$,    
the Lorentz transformations~\eqref{c:Griffiths, (3.1)} go over to the \emph{Galilean transformations}
\begin{align*}
	x' = x - vt \:, \qquad y' = y \:, \qquad z' = z \:, \qquad t' =  t 
\end{align*}
of classical mechanics (see e.g.~\cite[eq.~(4.1)]{landau+lifshitz-fields}). 

\subsubsection*{Four-Vectors}
In order to obtain a convenient framework, it is customary to introduce so-called \emph{four-vectors}~$x \in \R^4$ by 
\begin{align*}%\label{c:Griffiths, (3.7)}
	x^0 = ct \:, \qquad x^1 = x \:, \qquad x^2 = y \:, \qquad x^3 = z \:. 
\end{align*}
Introducing the abbreviation~$\beta \equiv v/c$, the Lorentz transformations take the form
\begin{align}\label{c:Griffiths, (3.8)}
	{x^0}' = \gamma (x^0 - \beta x^1) \:, \qquad {x^1}' = \gamma (x^1 - \beta x^0) \:, \qquad {x^2}' = x^2 \:, \qquad {x^3}' = x^3 \:. 
\end{align}
Moreover, introducing the matrix~$\Lambda$ by
\begin{align*}%\label{c:Griffiths, (3.11)}
	\Lambda = \begin{pmatrix}
		\gamma & - \gamma \beta & 0 & 0 \\ 
		-\gamma \beta & \gamma & 0 & 0 \\
		0 & 0 & 1 & 0 \\
		0 & 0 & 0 & 1
	\end{pmatrix} \:, 
\end{align*}
the Lorentz transformations~\eqref{c:Griffiths, (3.8)} of a four-vector~$x$ can be written as~$x' = \Lambda x$. It is common, however, in the physics literature to denote the vector whose components are~$x^0, \ldots, x^3$ by~$x^{\mu}$ rather than~$x$ (see~\cite{folland-qft}); thus in usual physics notation, this transformation of a four-vector~$x^{\mu}$ is written as
\begin{align*}%\label{c:Griffiths, (3.12)}
	{x^{\mu}}' = \Lambda^{\mu}_{\nu} x^{\nu} \:,
\end{align*} 
where in the last expression we employed the \emph{Einstein summation convention}, stating that in any product of vectors and tensors\footnote{Concerning tensor calculus, we refer the interested reader to~\cite{lichnerowicz}.} in which an index appears once as a subscript and once as a superscript, that index is to be summed from~$0$ to~$3$ (cf.~\cite{folland-qft}). 

In other words, the matrix~$\Lambda$ allows us to describe transformations from~$S$ to~$S'$.  
It is found that for any Lorentz transformation the following expression holds,  
\begin{align*}%\label{c:Griffiths, (3.13)}
	(x^0)^2 - (x^1)^2 - (x^2)^2 - (x^3)^2 = ({x^0}')^2 - ({x^1}')^2 - ({x^2}')^2 - ({x^3}')^2 \:. 
\end{align*} 
Introducing the \emph{Minkowski metric}~$\eta = \eta_{\mu \nu}$ by
\begin{align*}%\label{c:Griffiths, (3.14)}
	\eta_{\mu\nu} = \diag(1, -1, -1, -1) \:, 
\end{align*}
the \emph{Minkowski inner product} given by
\begin{align*}%\label{c:pfp, (1.1.1)}
	\langle x, y \rangle \equiv \eta_{\mu \nu} x^{\mu} y^{\nu}
\end{align*}
is Lorentz invariant. 
The Euclidean space~$\R^4$ endowed with~$\langle . , . \rangle$ is called \emph{Minkowski space}; in the sequel, we usually denote Minkowski space by~$\scrM$. Defining the \emph{covariant} four-vector~$x_{\mu}$ by
\begin{align*}%\label{c:Griffiths, (3.16)}
	x_{\mu} \equiv \eta_{\mu\nu} x^{\nu}
\end{align*}
(the ``original'' four-vector~$x^{\mu}$ is said to be \emph{contravariant}),   
the Minkowski inner product of two four-vectors~$x^{\mu}$ and~$y^{\mu}$ can be written as
\begin{align*}%\label{c:Griffiths, (3.22)}
	x \cdot y \equiv x_{\mu} y^{\mu} = x^0 y^0 - \mathbf{x} \cdot \mathbf{y} \:.
\end{align*}
For simplicity, the dot is sometimes omitted, i.e.~$xy \equiv x \cdot y$. Moreover, 
\begin{align*}%\label{c:Griffiths, (3.24)}
	x^2 \equiv x \cdot x = (x^0)^2 - \mathbf{x}^2 \:. 
\end{align*}
It is important to observe that~$x^2$ need not be positive. Indeed, a four-vector~$x$ is said to be
\begin{align*}%\label{c:Griffiths, (3.25)}
	\begin{split}
		\emph{timelike} \qquad &\text{if~$x^2 > 0$} \\ 
		\emph{spacelike} \qquad &\text{if~$x^2 < 0$} \\ 
		\emph{lightlike} \qquad &\text{if~$x^2 = 0$} \:. 
	\end{split}
\end{align*}
Lightlike vectors are also referred to as \emph{null} vectors. 
The subsets 
\begin{align*}
	L := \{x \in \scrM : x^2 = 0 \} \:, \qquad I := \{x \in \scrM : x^2 > 0 \} \:, \qquad J := \{x \in \scrM : x^2 \ge 0 \}
\end{align*} 
are called \emph{light cone}, \emph{interior light cone} and \emph{closed light cone}, respectively. They give rise to the following decomposition of Minkowski space~$\scrM$, 
\begin{align*}
	\scrM = L  \: {\dot{\cup}} \: I \: {\dot{\cup}} \: \scrM \setminus J \:. 
\end{align*}
For an illustration of the light cone we refer to~\cite[Figure~8]{hawking+ellis} or~\cite[Figure~1.3.1]{naber}. 
In this framework, the notation of derivatives on~$\R^4$ is
\begin{align*}
	\partial_{\mu} = \frac{\partial}{\partial x^{\mu}} \qquad \text{for all~$\mu = 0, \ldots, 3$} \:,
\end{align*}
so that, with respect to traditional coordinates~$\mathbf{x}$ and~$t$, 
\begin{align}\label{c:folland-qft, (1.3)}
	(\partial_0, \ldots, \partial_3) = (c^{-1} \partial_t, \nabla_{\mathbf{x}}) \:, \qquad (\partial^{0}, \ldots, \partial^{3}) = (c^{-1} \partial_t, - \nabla_{\mathbf{x}}) \:. 
\end{align}
Further details can be found in~\cite{pfp, folland-qft, griffiths, landau+lifshitz-fields}. 
We point out that the light cone plays a central role in the theory of causal fermion systems.

\subsubsection*{Energy and Momentum}
In what follows, we consider a particle of rest mass~$m$, which is described by the  four-vector~$x^{\mu}$. Then its \emph{four-velocity}~$u^{\mu}$ is introduced by
\begin{align*}%\label{c:Griffiths, (3.33)}
	u^{\mu} \equiv \frac{dx^{\mu}}{ds} \:, 
\end{align*}
where~$ds = c\gamma \: dt$ (with~$\gamma$ defined above), 
and its \emph{four-momentum}~$p = p^{\mu}$ is defined by
\begin{align*}%\label{c:Griffiths, (3.38)}
	p^{\mu} \equiv m u^{\mu} \:. 
\end{align*} 
Making use of the fact that~$u^{\mu} = (\gamma c, \mathbf{v})$, the four-momentum~$p^{\mu}$ takes the form
\begin{align*}
	p^{\mu} = (p^0, \mathbf{p}) = \gamma ( mc, m\mathbf{v}) \:. 
\end{align*}
Defining the \emph{relativistic energy}~$E$ by (cf.~\cite[eq.~(3.41)]{griffiths})
\begin{align*}%\label{c:Griffiths, (3.41)}
	E \equiv \gamma mc^2 = \frac{mc^2}{\sqrt{1- v^2/c^2}} \:, 
\end{align*}
we conclude that~$p^0 = E/c$ and~$p_{\mu}p^{\mu} = m^2c^2$.  
From this we deduce that
\begin{align}\label{c:(Energie-Impuls-Gleichung)}
	E^2 - \mathbf{p}^2c^2 = m^2c^4 \:, 
\end{align}
implying that~$E = mc^2$ is the \emph{rest energy} of the particle under consideration. The set of admissible four-momenta~$p^{\mu}$ satisfying~\eqref{c:(Energie-Impuls-Gleichung)} forms a \emph{hyperboloid} in~$\R^4$.  

The space of four-momenta~$p^{\mu} \in \R^4$, which for clarity we denote by~$\hscrM$, is known as \emph{momentum space}. Identifying momentum space~$\hscrM$ with Minkowski space~$\scrM$, the Minkowski inner product can be considered as a mapping  
\begin{align*}
	\langle . , . \rangle : \hscrM \times \scrM \to \R \:, \qquad (p,x) \mapsto \langle p, x \rangle = g_{\mu\nu} p^{\mu} x^{\nu} \:. 
\end{align*} 
Given~$m > 0$, one introduces the so-called \emph{mass shell} by
\begin{align*}
	C_m := \{p \in \hscrM : p^2=m^2 \} \:. 
\end{align*}
Furthermore, one often distinguishes its subsets
\begin{align*}
	C_m^+ := \{p \in \hscrM : \text{$p^2=m^2$, $p_0 > 0$} \}  \qquad \text{and} \qquad 
	C_m^- := \{p \in \hscrM : \text{$p^2=m^2$, $p_0 < 0$} \} \:, 
\end{align*}
which are known as \emph{upper} and \emph{lower mass shell}, respectively. 
The sets~$C_m^{\pm}$ form the two parts of a hyperbola in momentum space. In the case~$m= 0$, the corresponding sets~$C_0^+$ and~$C_0^-$ are called \emph{forward} and \emph{backward light cone}, respectively (cf.~\cite[Section~1.3]{folland-qft}). These constructions will be of crucial importance in the theory of causal fermion systems.

\subsubsection*{Relativistic Formulation of Electrodynamics}
Applying the notation of special relativity to electrodynamics gives rise to particularly concise formulas. 
More specifically, introducing the 
\emph{electromagnetic 4-potential} by
\begin{align*}%\label{c:folland-qft, (2.25)}
	A^{\mu} = (\phi, \mathbf{A}) \qquad \text{or} \qquad A_{\mu} = (\phi, - \mathbf{A}) \:,
\end{align*}
the electric and magnetic fields are combined into the \emph{electromagnetic field tensor}
\begin{align*}%\label{c:folland-qft, (2.26)}
	F_{\mu \nu} = \partial_{\mu} A_{\nu} - \partial_{\nu} A_{\mu} \:, \qquad \text{or} \qquad F^{\mu \nu} = \partial^{\mu} A^{\nu} - \partial^{\nu} A^{\mu} \:. 
\end{align*}
Thus the Maxwell equations~\eqref{c:folland-qft, (2.19)} and~\eqref{c:folland-qft, (2.20)} become
\begin{align*}%\label{c:folland-qft, (2.27)}
	\partial_{\kappa} F_{\mu \nu} + \partial_{\mu} F_{\nu \kappa} + \partial_{\nu} F_{\kappa \mu} = 0 \:,
\end{align*}
and the equations~\eqref{c:folland-qft, (2.18)} and~\eqref{c:folland-qft, (2.21)} take the form
\begin{align*}%\label{c:folland-qft, (2.28)}
	\partial_{\mu} F^{\mu \nu} = j^{\nu} \:, 
\end{align*}
where the \emph{4-current density~$j^{\mu}$} is defined by
\begin{align*}
	j^{\mu} \equiv (\rho, \mathbf{j}) \:. 
\end{align*}
The Lorentz gauge condition~\eqref{c:folland-qft, (2.23)} reads 
\begin{align*}
	\partial_{\mu} A^{\mu} = 0 \:, 
\end{align*} 
and the continuity equation~\eqref{c:folland-qft, (2.22)} can be written as 
\begin{align*}%\label{c:folland-qft, (2.29)}
	\partial_{\mu} j^{\mu} = 0 \:. 
\end{align*}
The previous relations may also be expressed in terms of differential forms, making use of the Hodge star operator; for details we refer to~\cite[Section~2.4]{folland-qft} or~\cite{zeidler-gauge}. 

Let us finally point out that the Maxwell equations~\eqref{c:folland-qft, (2.18)}--\eqref{c:folland-qft, (2.21)} can be obtained from a variational principle. To this end, one considers the action (cf.~\cite[eq.~(28.6)]{landau+lifshitz-fields})
\begin{align*}%\label{c:LL2, (28.6)}
	S = - \frac{1}{c^2} \int A_{\mu} j^{\mu} \: d\Omega - \frac{1}{16\pi c} \int F_{\mu\nu} F^{\mu\nu} \: d\Omega \:, 
\end{align*}
where~$d\Omega = c \: dt \: dx \: dy \: dz$ (see~\cite[eq.~(27.4)]{landau+lifshitz-fields}). Following~\cite[\S 30]{landau+lifshitz-fields}, by varying the electromagnetic potential one obtains Maxwell's equations~\eqref{c:folland-qft, (2.18)} and~\eqref{c:folland-qft, (2.21)}, whereas the equations~\eqref{c:folland-qft, (2.19)} and~\eqref{c:folland-qft, (2.20)} are a consequence of~\eqref{c:Griffiths, (7.75)} and~\eqref{c:Griffiths, (7.77)} by~\cite[\S 26]{landau+lifshitz-fields}.

\subsection{General Relativity}\label{c:S general relativity}
The main reason for the fact that solely Einstein's name is associated with the theory of relativity is due to the further development: 
So far, the principle of relativity is restricted to inertial frames; influenced by the Austrian physicist and philosopher \textsc{Ernst Mach} (1838--1916), however, Einstein was guided by the postulate that the ``true'' laws of physics should hold in any reference frame. The starting point for his considerations was the insight that \emph{no physical experiment can distinguish between gravitational and inertial forces}. This is known as \emph{Einstein's principle of equivalence}~\cite{born-relativitatstheorie, simon}.  
The resulting ``general theory of relativity,''  
including a new theory of gravity,  
changed our ideas of the universe in a fundamental way. 

In short, Einstein's results can be summarized as follows: In the theory of general relativity, spacetime is described mathematically by a four-dimensional Lorentzian manifold~$(M,g)$, which is a semi-Riemannian manifold either of signature~$(-,+,+,+)$ or~$(+, -, -,-)$.\footnote{The two sign conventions are related by considering the metric~$g$ or~$-g$, respectively.}   
At each point~$p \in M$, the tangent space~$T_pM$ can be identified with Minkowski space, and special relativity can be regarded as general relativity of Minkowski spacetime (for details see~\cite{beem, oneill}).  
The famous \emph{Einstein equations} read
\begin{align}\label{c:Gron, (8.3)}
	G_{\mu\nu} + \Lambda g_{\mu\nu} = \kappa T_{\mu\nu} \:.
\end{align} 
where~$G_{\mu\nu}$ denotes the \emph{Einstein gravitational tensor},~$T_{\mu\nu}$ is known as \emph{stress-energy tensor} (or \emph{energy-momentum tensor}), and~$\kappa$ denotes Newton's gravitational constant.  
The Einstein field equations~\eqref{c:Gron, (8.3)} describe the behavior of curvature of spacetime in the presence of matter. 
For clarity we point out that~$G_{\mu\nu}$ crucially depends on the metric tensor~$g$, whereas~$T_{\mu\nu}$ is to be determined from physical observations dealing with the distribution of matter and energy (cf.~\cite[Chapter~3]{hawking+ellis} or~\cite[Chapter~5]{misner+thorne+wheeler}). In simple terms~\cite{wheeler, oneill}, the Einstein equations often are summarized by saying that 
\begin{quote}\centering
	``Spacetime tells matter how to move, \\ 
	matter tells spacetime how to curve.''
\end{quote} 

As being of relevance for the theory of causal fermion systems,  
let us consider basic objects on Lorentzian manifolds in some more detail.\footnote{Our explanations are due to~\cite{beem} with signature~$(-,+,+,+)$, as is common in many textbooks on general relativity; concerning the sign conventions we also refer to~\cite{oneill}.} 
More generally, let~$(M,g)$ be a semi-Riemannian manifold. Denoting the set of all smooth vector fields on~$M$ by~$\mathfrak{X}(M)$ and the set of all smooth real-valued functions on~$M$ by~$\mathfrak{F}(M)$, a \emph{connection} is a function
\begin{align*}
	\nabla : \mathfrak{X}(M) \times \mathfrak{X}(M) \to \mathfrak{X}(M)
\end{align*}
with the properties that
\begin{align*}
	\nabla_V(X + Y) &= \nabla_V X + \nabla_V Y \:, \\
	\nabla_{fV + hW}(X) &= f \nabla_V X + h \nabla_W X \:, \\
	\nabla_V (fX) &= f \nabla_V X + V(f) X 
\end{align*}
for all~$f,h \in \mathfrak{F}(M)$ and all~$X,Y,V,W \in \mathfrak{X}(M)$. The vector field~$\nabla_X Y$ is said to be the \emph{covariant derivative} of~$Y$ with respect to~$X$. 
The \emph{Lie bracket} of the ordered pair of vector fields~$X$ and~$Y$ is a vector field~$[X,Y]$ which acts on a smooth function~$f$ by
\begin{align*}
	[X,Y](f) = X(Y(f)) - Y(X(f)) \:. 
\end{align*}
Next, the \emph{torsion tensor}~$T$ of~$\nabla$ is the function~$T : \mathfrak{X}(M) \times \mathfrak{X}(M) \to \mathfrak{X}(M)$ given by
\begin{align*}
	T(X,Y) = \nabla_X Y - \nabla_Y X - [X, Y] \:. 
\end{align*}
The \emph{curvature}~$R(., .)$ of~$\nabla$ is a function which assigns to each pair~$X,Y \in \mathfrak{X}(M)$ the $f$-linear map~$R(X,Y) : \mathfrak{X}(M) \to \mathfrak{X}(M)$ given by
\begin{align*}
	R(X,Y) Z = \nabla_X \nabla_Y Z - \nabla_Y \nabla_X Z - \nabla_{[X,Y]} Z \:. 
\end{align*}
Thus curvature provides a measure of the non-commutativity of~$\nabla_X$ and~$\nabla_Y$.  
Given a semi-Riemannian manifold~$(M,g)$, there is a unique connection~$\nabla$ on~$M$ which is \emph{metric compatible}, i.e.
\begin{align*}
	Z(g(X,Y)) = g(\nabla_ZX, Y) + g(X, \nabla_ZY) \:,
\end{align*}
as well as \emph{torsion free}, i.e.
\begin{align*}
	[X,Y] = \nabla_X Y - \nabla_Y X
\end{align*}
for all~${X,Y,Z \in \mathfrak{X}(M)}$. This connection~$\nabla$ is called \emph{Levi-Civita connection}. The trace of the curvature tensor is the \emph{Ricci curvature}; for each~$p \in M$, the Ricci curvature may be interpreted as the bilinear map~$\ric_p : T_pM \times T_pM \to \R$. In terms of the \emph{Ricci tensor}~$R_{\mu\nu}$, Ricci curvature may be represented as
\begin{align*}
	\ric_p(v,w) = R_{\mu\nu} v^{\mu}w^{\nu} \:, 
\end{align*} 
and the trace of the Ricci curvature is the \emph{scalar curvature}~$R$, i.e.~$R = {R^{\mu}}_{\mu}$.  
Then the Einstein tensor~$G_{\mu\nu}$ is given by (also see~\cite[eq.~(4.29)]{carroll}
\begin{align*}%\label{c:Carroll, (4.29)}
	G_{\mu\nu} = R_{\mu\nu} - \frac{1}{2} R g_{\mu\nu} \:,
\end{align*}
and it satisfies~$\nabla^{\mu} G_{\mu\nu} = 0$. Just as Maxwell's equations govern how the electric and magnetic fields respond to charges and currents, Einstein's field equations determine how the metric responds to energy and momentum. 

In his paper~\cite{einstein-1917}, Einstein suggests to modify his field equations by introducing an additional constant (which is referred to as \emph{cosmological constant}~$\Lambda$).  
In this case, the \emph{vacuum field equations} read
\begin{align}\label{c:Gron, (8.27)}
	G_{\mu\nu} + \Lambda g_{\mu\nu} = 0 \:. 
\end{align}
In 1915, Hilbert~\cite{hilbert-1915} observed that the Einstein equations can be recovered by means of a variational principle.  
More precisely, varying the so-called \emph{Einstein-Hilbert action} 
\begin{align*}%\label{c:Carroll, (4.102)}
	S_{\text{\tiny{\rm{EH}}}} = \int_M \left[\frac{1}{2 \kappa} (R - 2 \Lambda) \right] \: d\mu
\end{align*}
with respect to the metric~$g$ yields~\eqref{c:Gron, (8.27)}, where~$d\mu := \sqrt{- \det g} \: d^4x$ 
(for details we refer to~\cite{carroll, gron+hervik}). Actually, introducing an action~$S$ in the fashion of~\cite[eq.~(1.1.9)]{pfp}, it is possible to recover the equations of motion, the Maxwell equations as well as the Einstein equations by varying the action~$S$ with respect to the spacetime curve, the electromagnetic potential and the metric, respectively (cf.~\cite[Section~1.1]{pfp}).  
The non-quantized field equations derived so far are also referred to as ``classical.''

\subsection{Quantum Mechanics}\label{c:S qm} 
From the conceptual point of view, it is instructive to motivate quantum mechanics by illustrating the famous \emph{double-slit experiment} (for details see~\cite[Chapter~3]{shankar}) in order to expose the inadequacy of classical physics.   
More precisely, performing the double-slit experiment for single {photons}, one observes that each photon carries the same energy~$E$ and the same momentum~$\mathbf{p}$, and its energy is given by~$E = |\mathbf{p}|c$. Applying the relativistic equation~$E^2 = \mathbf{p}^2c^2 + m^2c^4$ (see~\eqref{c:(Energie-Impuls-Gleichung)}), one infers that each photon is a particle of mass~$m= 0$. Moreover,  
varying the (angular) frequency~$\omega$ of the light source, one discovers that~$E = \hbar \omega$ and~$p = \hbar k$  
(where~$k = |\mathbf{k}|$ is the absolute value of the wave vector~$\mathbf{k}$ and~$\hbar$ is \emph{reduced Planck's constant}), implying that energy and momentum are \emph{quantized}~\cite{planck-1900}.  

From these facts Born drew the following conclusion: Each particle has associated with it a wave function~$\psi(t, \mathbf{x})$ such that~$|\psi(t,\mathbf{x})|^2$ gives the probability of finding it at a point~$\mathbf{x}$ at time~$t$. This is known as \emph{wave-particle duality}. The dynamics of the particle is described in terms of its wave function~$\psi(t, \mathbf{x})$, which in non-relativistic quantum mechanics is determined by \emph{Schr\"odinger's equation}. 
The resulting quantum theory is then based on a few fundamental postulates, which take into account these experimental findings (see~\cite[Chapter~4]{shankar}). 
As~$|\psi(t,\mathbf{x})|^2$ is interpreted as probability of finding a particle with wave function~$\psi$ at time~$t$ at a point~$\mathbf{x}$, for each time~$t$ the integral over the whole space must be normalized, 
\begin{align*}
	\int_{\R^3} |\psi(t, \mathbf{x})|^2 \: d^3x = 1 \:. 
\end{align*}
Thus for every~$t$, a wave function~$\psi(t, \cdot)$ is contained in the Hilbert space of square-integrable functions~$L^2(\R^3, \C)$, endowed with the inner product~$\langle \, . \, | \, . \,\rangle_{L^2(\R^3, \C)}$. 
This property is reflected in the first postulate of quantum mechanics, according to which the \emph{state} of a particle is represented by a normalized vector in a complex Hilbert space, and physically measurable quantities (referred to as \emph{observables}) correspond to self-adjoint operators on the Hilbert space (thereby establishing a connection to spectral calculus, for details see~\cite{dimock, thaller}). Instead of entering the postulates of quantum mechanics in more detail, let us stick to the most important consequences concerning the theory of causal fermion systems.

Besides the classical ``orbital'' angular momentum, 
a specific feature of quantum mechanics is that 
every elementary particle comes with a certain ``intrinsic'' angular momentum which is independent of its motion in space~\cite{landau+lifshitz-qm}. This intrinsic angular momentum is called \emph{spin}.  
Denoting the spin of a particle by~$s$, it is remarkable that~$s$ only takes integer or half-integer values, i.e.~$s \in \{0, 1/2, 1, 3/2, \ldots \}$. Particles with integer spin are called \emph{bosons}, whereas particles with half-integer spin are known as \emph{fermions}~\cite{griffiths, weinberg-I}. 
Concerning fermions with spin~$s=1/2$, there are~$(2s+1) = 2$ possible spin orientations (see~\cite[Section~4.2]{griffiths}); they are usually referred to as ``up'' and ``down'' (for details see~\cite{sakurai-qm}).

As mentioned before, the dynamics in non-relativistic quantum mechanics obeys Schr\"odinger's equation, in which case relativistic effects are not taken into consideration. However, as the principles of special relativity are generally accepted, a correct quantum theory should satisfy the requirement of relativity: laws of motion valid in one inertial system must hold in all inertial systems. Stated mathematically, relativistic quantum theory must be formulated in a Lorentz covariant form~\cite{bjorken+drell}.  
In other words, Schr\"odinger's equation is unsatisfactory from the relativistic point of view, as it treats space and time on a different footing. In relativistic mechanics, momentum~$\mathbf{p}$ and energy~$E$ of a free particle combine to a four-vector~$p^{\mu} = (E, \mathbf{p})$ (in ``natural units,'' in which Planck's constant~$\hbar$ and the speed of light~$c$ both equal one, $\hbar = c = 1$).  
The recipe for quantizing energy and momentum amounts to replacing~$p^{\mu} \to i \partial^{\mu}$, 
where~$\partial^{\mu}$ is given by~\eqref{c:folland-qft, (1.3)}. Applying this replacement rule to the free-particle energy-momentum relation~$p^2 = m^2$ yields the relativistic equation 
\begin{align*}
	(\Box - m^2) \phi = 0 \:, 
\end{align*} 
known as \emph{Klein-Gordon equation} (where~$\Box \equiv \partial_{\mu} \partial^{\mu}$ is the wave operator). 
Complex-valued solutions of this equation describe a scalar particle (that is, a particle without spin) of mass~$m$. 
Unfortunately, the scalar component~$\rho$ which appears in the corresponding continuity equation is not positive definite and therefore cannot be understood as a probability density. 
For further details we refer to~\cite{pfp, folland-qft, griffiths, schweber, thaller}.

Dirac's aim was to avoid this difficulty of negative probability densities, taking into account relativistic covariance. Dirac's strategy was to ``factorize'' the relativistic energy-momentum relation~$p^2 = m^2c^2$ (cf.~\cite{griffiths}). To this end, he introduced the so-called \emph{Dirac matrices}~$\gamma^{\mu} \in \C^{4 \times 4}$ ($\mu= 0, \ldots, 3$), satisfying the anti-commutation relations
\begin{align*}%\label{c:Griffiths, (7.15)}
	\{\gamma^{\mu}, \gamma^{\nu} \} \equiv \gamma^{\mu} \gamma^{\nu} + \gamma^{\nu} \gamma^{\mu} = 2 \eta^{\mu\nu} \qquad \text{for all~$\mu, \nu = 0, \ldots, 3$}
\end{align*}
(with Minkowski metric~$\eta^{\mu\nu}$). This approach finally led him to set up the \emph{Dirac equation} in Minkowski space, which in natural units takes the form 
\begin{align}\label{c:Griffiths, (7.20)}
	(i \slashed{\partial} - m) \psi = 0 \:,
\end{align}
where~$\slashed{\partial} \equiv \gamma^{\mu} \partial_{\mu}$ is known as \emph{Dirac slash} (cf.~\cite{griffiths, weinberg-I}). The Dirac equation is Lorentz covariant~\cite{bjorken+drell}.  
The solutions~$\psi$ of the Dirac equation are referred to as \emph{Dirac spinors}, which are mappings~$\psi : \scrM \to \C^4$ from Minkowski space~$\scrM$ to spinor space~$\C^4$; they describe particles of spin~$1/2$.  
Introducing the \emph{adjoint spinor}~$\overline{\psi} \equiv \psi^{\dagger} \gamma^0$ (where by~$\psi^{\dagger}$ we denote the adjoint of~$\psi$ with respect to the Euclidean inner product on~$\C^4$) gives rise to an indefinite inner product~$\overline{\psi}\phi$ on~$\C^4$ of signature~$(2,2)$, which will be of central relevance in the theory of causal fermion systems. 
The success of Dirac's intuition lies in the fact that the scalar component~$\rho$ of the corresponding continuity equation is given by~$\psi^{\dagger}\psi = |\psi|^2$, and thus non-negative. 
For a probabilistic interpretation of the wave functions, one requires the \emph{normalization condition} 
\begin{align}\label{c:pfp, (1.2.15)}
	\int_{\R^3} (\psi^{\dagger} \psi)(t, \mathbf{x}) \: d^3 \mathbf{x} = 1 \:, 
\end{align}
thus giving rise to the Hilbert space~$\mathcal{H} = L^2(\R^3)^4$,  
endowed with the scalar product
\begin{align}\label{c:pfp, (1.2.17)}
	(\psi \mid \phi) := \int_{\R^3} (\overline{\psi} \gamma^0 \phi)(t, \mathbf{x}) \: d^3 \mathbf{x} \qquad \text{for all~$\psi, \phi \in H$}
\end{align}
for any time~$t$ 
(cf.~\cite[Section~1.2]{pfp},~\cite[Section~4.1]{folland-qft} and~\cite[Section~1.3]{thaller}).

In what follows we make use of the \emph{energy function}~$\omega$, defined by 
\begin{align*}
	\R^3 \ni \mathbf{k} \mapsto \omega(\mathbf{k}) := \sqrt{\mathbf{k}^2 + m^2} \in \R \:. 
\end{align*}
Introducing for any~$\mathbf{k} \in \R^3$ the orthogonal projections~$p_{\pm}(\mathbf{k}) : \C^4 \to \C^4$ by
\begin{align*}%\label{c:oppio, (2.13)}
	p_{\pm}(\mathbf{k}) := \frac{\slashed{k} + m}{2k^0}\gamma^0\Big|_{k^0 = \pm \omega(\mathbf{k})} \:, 
\end{align*}
the spinor space~$\C^4$ decomposes into two orthogonal subspaces, 
\begin{align*}
	\C^4 = \mathcal{W}_{\mathbf{k}}^+ \oplus \mathcal{W}_{\mathbf{k}}^- \:,
\end{align*}
where~$\mathcal{W}_{\mathbf{k}}^{\pm} = p_{\pm}(\mathbf{k})(\C^4)$ for any~$\mathbf{k} \in \R^3$, and a general solution of the Dirac equation takes the form (cf.~\cite[eq.~(9.24)]{dimock})
\begin{align*}%\label{c:Dimock, (9.24)}
	\psi(t, \mathbf{x}) = \int_{\R^3} \frac{d\mathbf{p}}{(2\pi)^3} e^{i \mathbf{p} \cdot \mathbf{x}} \left(e^{-i\omega(\mathbf{p})t} \psi_+(\mathbf{p}) + e^{i\omega(\mathbf{p})t} \psi_-(\mathbf{p}) \right)
\end{align*}
with~$\psi_{\pm}(\mathbf{p}) \in \mathcal{W}_{\mathbf{p}}^{\pm}$ for all~$\mathbf{p} \in \R^3$. 
It is an important consequence that the solution space~$\mathcal{H}$ of the Dirac equation splits into a positive and negative energy subspaces, 
\begin{align*}%\label{c:Dimock, (9.26)}
	\mathcal{H} = \mathcal{H}^+ \oplus \mathcal{H}^- \:, 
\end{align*} 
where~$\mathcal{H}^{\pm} = \{\psi \in \mathcal{H} : \psi_{\mp} = 0\}$ (for details see~\cite{dimock, oppiocfs}). Solutions of \emph{positive} energy are interpreted as \emph{particles}; the physical interpretation of negative energy solutions, on the other hand, is due to Dirac~\cite{dirac1930theory}. Based on the \emph{Pauli exclusion principle}, according to which no two fermions can occupy the very same state, Dirac proposed that ``all the states of negative energy are occupied except perhaps a few of small velocity'' in order to prevent particles of positive energy to fall down into negative energy states. In accordance with Dirac's interpretation~\cite{dirac1930theory}, these few vacant states, regarded as ``holes'' in the sea of negative energy particles, were later discovered as \emph{anti-particles} (cf.~\cite{weinberg-I}). This concept, also referred to as a ``Dirac sea,'' became unpopular in quantum field theory;\footnote{To quote {Julian Schwinger} (cf.~\cite[p.~14]{weinberg-I}), ``the picture of an infinite sea of negative energy electrons is now best regarded as a historical curiosity, and forgotten.''} however,  
Dirac's original idea of a sea of particles was revived in the theory of causal fermion systems (see~\cite{finster2011formulation}).  

Whenever a wave function~$\psi = \psi(t, \mathbf{x})$ is given, the four-vector~$x = (t, \mathbf{x}) \in \R^4$ is said to be in ``position space.'' Denoting position space by~$\scrM \simeq \R^4$ and assuming that~$\psi \in L^1(\scrM)$, its Fourier transform is given by (for details see~\cite{folland-qft}) 
\begin{align*}
	\widehat{\psi}(k) = \int_{\scrM} e^{ikx} \psi(x) \: d^4x \qquad \text{for all~$k \in \R^4$} \:,
\end{align*}
where~$k \in \R^4$ is said to be in ``momentum space.'' Accordingly, we refer to~$\hscrM \simeq \R^4$ as \emph{momentum space}. For mathematical details we refer to~\cite{folland-harmonic}.

Let us finally deal with the famous uncertainty principle, one of the fundamental principles of quantum mechanics, discovered by Heisenberg in 1927 (cf.~\cite{landau+lifshitz-qm}). 
Due to the axioms of quantum mechanics, to each quantum mechanical system is associated a Hilbert space, and every measurable quantity (referred to as an ``observable,'' such as energy, position or momentum) is represented by a self-adjoint operator~\cite{thaller}.  
Given a self-adjoint operator~$A$ on a Hilbert space~$(H, \langle . , . \rangle)$, for any unit vector~$\psi \in H$ the \emph{expectation value} of~$A$ in the state~$\psi$ is defined by~$\langle A \rangle_{\psi} = \langle \psi , A \psi \rangle$, and the \emph{uncertainty of~$A$}, denoted by~$\Delta_{\psi}A$, is given by~$(\Delta_{\psi} A)^2 = \langle A^2 \rangle_{\psi} - (\langle A \rangle_{\psi})^2$ (cf.~\cite[Section~3.6]{hall}).  
Introducing the \emph{position operators}~$(X^{i})_{i=1,2,3}$ by
\begin{align*}
	(X^{i}\psi)(\mathbf{x}) = {x^{i}}\psi(\mathbf{x}) \qquad \text{for all~$i=1,2,3$}\:, 
\end{align*}
which amounts to a ``multiplication by~${x^{i}}$,'' the expectation value of~$X^{i}$ in the state~$\psi$ is then~$\langle X^{i} \rangle_{\psi} = \langle \psi, X^{i} \psi \rangle$ for all~$i=1,2,3$. Moreover, one can introduce the \emph{momentum operators}~$(P^{i})_{i=1,2,3}$ such that~$[X^{i},P^{j}] = \delta_{ij} i\hbar$ for all~$i,j =1,2,3$ (cf.~\cite[Chapter~3]{hall}). 
Making use of the fact that the position and momentum operators~$X^{i}$ and~$P^{i}$ do not commute for all~$i=1,2,3$, for any state~$\psi$ the \emph{Heisenberg uncertainty principle} reads (for details see~\cite[Chapter~12]{hall} or~\cite{folland-qft, griffiths-qm, landau+lifshitz-qm, shankar}) 
\begin{align*}%\label{c:Griffiths, [3.63]}
	(\Delta_{\psi} X^{i}) (\Delta_{\psi} P^{i}) \ge \frac{\hbar}{2\pi} \qquad{\text{for all~$i=1,2,3$}} \:,
\end{align*}
implying that position and momentum cannot be measured simultaneously.  
In other words, in order to probe \emph{small distances} one requires \emph{high energies}~\cite{griffiths}.

\subsection{Group Theory in Physics}\label{c:S group theory}
The importance of group theory in physics was recognized soon after the discovery of quantum mechanics~\cite{sternberg}, based on fundamental works by Weyl~\cite{weyl-1927, weyl-1977}. In the sequel, group theory evolved to a substantial part of modern theoretical physics. For this reason, 
let us briefly compile some basics from group theory required in the following.  
To begin with, we recall that, for any~$n \in \N$, the \emph{unitary group}~$\U(n)$ is given by
\begin{align*}
	\U(n) = \{A \in \GL(n, \C) : A^{\dagger} = A^{-1} \} \:,
\end{align*}
where~$\GL(n, \C) = \{A \in \C^{n \times n} : \det(A) \not= 0 \}$. 
The \emph{special unitary group} is defined by
\begin{align*}
	\SU(n) = \{A \in \U(n) : \det(A) = 1 \} \:. 
\end{align*}
Clearly, the norm of a spinor~$\psi(x) \in \C^{n}$ is invariant under unitary transformations
\begin{align*}%\label{c:Pal, (3.10)}
	\psi'(x) = U\psi(x) \qquad \text{with~$U \in \U(n)$} \:.
\end{align*} 

We recall that in general, a \emph{Lie group}~$G$ is a differentiable manifold which is also a group such that the mapping 
\begin{align*}
	G \times G \to G \:, \qquad (\sigma, \tau) \mapsto \sigma \tau^{-1} 
\end{align*}
is smooth (see e.g.~\cite{warner}). Whenever a Lie group~$G$ is a subgroup of~$\GL(n,\C)$, it is also referred to as \emph{matrix Lie group}~\cite{hall-lie}. Both~$\U(n)$ and~$\SU(n)$ are (compact) matrix Lie groups. In physical applications one is typically interested in a fixed Lie group~$G$ (the ``gauge group'') which represents a symmetry  
of the theory (cf.~\cite[Section~6.3]{marathe+martucci}).   

As~$\U(n)$ and~$\SU(n)$ are the most important Lie groups in particle physics, let us restrict attention to matrix Lie groups in what follows. According to~\cite[Chapter~2]{hall-lie}, the \emph{exponential map} is defined by 
\begin{align*}
	\exp : \C^{n \times n} \to \C^{n \times n} \:, \qquad X \mapsto e^{X} := \sum_{n=0}^{\infty} \frac{X^n}{n!} \:. 
\end{align*} 
Whenever~$G$ is a matrix Lie group, the \emph{Lie algebra of~$G$}, denoted by~$\mathfrak{g}$, is defined as the set of all matrices~$X$ such that~$e^{tX} \in G$ for all~$t \in \R$. For clarity we point out that the \emph{physicists' convention} is to consider the map~$X \mapsto e^{iX}$ instead of~$X \mapsto e^X$. Accordingly, physicists are accustomed to  
think of the Lie algebra of~$G$ as the set of all matrices~$X$ such that~$e^{itX} \in G$ for all~$t \in \R$.  
Moreover, physics literature does not always distinguish clearly between a Lie group and its Lie algebra. 

In general, a \emph{Lie algebra}~$\mathfrak{g}$ over~$\R$ is a real vector space~$\mathfrak{g}$ together with a bilinear operator~$[.,.] : \mathfrak{g} \times \mathfrak{g} \to \mathfrak{g}$ (called the \emph{Lie bracket}) such that for all~$X,Y,Z \in \mathfrak{g}$, 
\begin{itemize}[leftmargin=2em]
	\item[(i)] $[X,X] = 0$ for all~$X \in \mathfrak{g}$, and 
	\item[(ii)] $[[X,Y],Z] + [[Y,Z], X] + [[Z,X],Y] = 0$ for all~$X,Y,Z \in \mathfrak{g}$ (\emph{Jacobi identity}). 
\end{itemize} 
Note that each Lie group is closely related to a corresponding Lie algebra  
(cf.~\cite{warner}); indeed, 
whenever~$G$ is a Lie group with unit element~$e$, the vector space~$\mathfrak{g} := T_eG$ is called the \emph{Lie algebra} of~$G$ (cf.~\cite[Definition~(2.4)]{broecker+tomdieck}). It becomes a Lie algebra with the \emph{Lie product} (cf.~\cite[eq.~(2.13)]{broecker+tomdieck})
\begin{align*}%\label{c:Broecker, (2.13)} 
	[X,Y] = XY - YX \:. 
\end{align*}
Whenever~$G$ is a matrix Lie group with Lie algebra~$\mathfrak{g}$, and~$X_1, \ldots, X_n$ is a basis for~$\mathfrak{g}$ (as a vector space), then for all~$i,j \in \{1, \ldots, n\}$ the Lie bracket~$[X_i, X_j]$ can be written uniquely in the form
\begin{align*}
	[X_i, X_j] = \sum_{k=1}^{n} c_{ijk} X_k \:.
\end{align*}
The constants~$c_{ijk}$ are called the \emph{structure constants} of~$\mathfrak{g}$ (see~\cite[\S 2.8.1]{hall-lie}). In physics literature, the basis vectors~$X_1, \ldots, X_n$ are known as \emph{generators} (cf.~\cite{aitchison+hey-II, pal, weinberg-I}).

\subsection{Introduction to Elementary Particles}\label{c:S elementary particles}
To begin with, elementary particles are thought of as objects which are not made up of more fundamental ones; 
one thus can say that elementary particles are the fundamental constituents of all objects in the universe.

For a historical introduction to elementary particles we refer the interested reader to~\cite[Chapter~1]{griffiths} and~\cite[Chapter~9]{pal}.  
Nevertheless,  
it might be instructive to motivate our present understanding of elementary particles as follows. The idea that matter is composed of ``indivisible'' atoms goes back to the ancient Greeks~\cite{russell-history}. However, our knowledge of the detailed structure of atoms originates from scattering experiments due to \textsc{Ernest Rutherford} (1871--1937), according to which the positive charge as well as most of the mass are concentrated in the center of the atom, the \emph{nucleus}. In the 20th century it turned out that the nucleus itself is made up of \emph{neutrons} and \emph{protons}. 
The electrons, already discovered in~1897 by \textsc{Joseph John Thompson} (1856-1940), were supposed to ``orbit'' around the nucleus. This picture summarizes the period of classical elementary particle physics~\cite{griffiths}. 

In the mean time, further experimental findings revealed that neutrons and protons themselves consist of even more fundamental constituents, so-called \emph{quarks}, which are denoted by the first letters of their names. More precisely, there are six kinds of quarks, known as \emph{flavors}, referred to as up (u), down (d), strange (s), charm (c), bottom (b) and top (t), and each quark comes in three \emph{colors} (red, green, blue).  
They  
fall into doublets (called ``families'' or ``generations'');   
a first generation~(u, d), a second generation~(c, s) and a third generation~(t, b). It should be noticed that free quarks have never been observed (referred to as \emph{quark confinement}). Electrons, on the other hand, belong to the class of \emph{leptons}. There are six types of leptons, called \emph{flavors}, comprising three charged leptons (\emph{electron}~e, \emph{muon}~$\mu$ and \emph{tauon}~$\tau$) and three neutral leptons (\emph{electron neutrino}~$\nu_{\text{e}}$, \emph{muon neutrino}~$\nu_{\mu}$ and \emph{tau neutrino}~$\nu_{\tau}$). By contrast to quarks, leptons can exist as free particles. Similarly as quarks, leptons come in three generations: a first generation ($\nu_{\text{e}}$, e), a second generation ($\nu_{\mu}, \mu$) and a third generation ($\nu_{\tau}, \tau$).  
This comprises all known elementary particles,  
which therefore can be summarized in the following table (where the arrows~$\uparrow$ and~$\downarrow$ refer to ``isospin up'' and ``isospin down,'' respectively;  
for details concerning weak isospin see below):
\begin{align}\label{c:table} 
	\begin{tabular}{c|c|c|c|c}
		& 1$^{\tiny\textup{st}}$ generation & 2$^{\tiny\textup{nd}}$ generation & 3$^{\tiny\textup{rd}}$ generation & $I_3$ \\ %\hline
			$\text{quarks} \left\{\begin{array}{c} \text{red} \\ \text{\:\:green\:\:} \\ \text{blue} \\ \text{red} \\ \text{green} \\ \text{blue} \end{array} \right. $ & $\begin{array}{c} \text{u} \\ \text{u} \\ \text{u} \\ \text{d} \\ \text{d} \\ \text{d} \end{array} $ & $\begin{array}{c} \text{c} \\ \text{c} \\ \text{c} \\ \text{s} \\ \text{s} \\ \text{s} \end{array} $ & $\begin{array}{c} \text{t} \\ \text{t} \\ \text{t} \\ \text{b} \\ \text{b} \\ \text{b} \end{array} $  
			& $\begin{array}{c}
			\text{$\uparrow$} \\ \text{$\uparrow$} \\ \text{$\uparrow$} \\ \text{$\downarrow$} \\ \text{$\downarrow$} \\ \text{$\downarrow$} \\
			\end{array} $ \\
			$\text{leptons} \left\{\begin{array}{c} \text{neutral} \\ \text{charged} \end{array} \right. $ & $\begin{array}{c} \text{$\nu_{\text{e}}$} \\ \text{e} \end{array}$& $\begin{array}{c} \text{$\nu_{\mu}$} \\ \text{$\mu$} \end{array}$& $\begin{array}{c} \text{$\nu_{\tau}$} \\ \text{$\tau$} \end{array}$  
			& $\begin{array}{c}
			\text{$\uparrow$} \\ \text{$\downarrow$} 
			\end{array}$
	\end{tabular}  
\end{align}

Bearing in mind that to each particle there is a corresponding \emph{anti-particle}, there are~48 elementary particles in total, which are all Dirac particles (that is, particles of spin~1/2 which obey the Dirac equation). 
Thus one can say that on the fundamental level, all matter is described by the Dirac equation.
In the theory of causal fermion systems, Table~\eqref{c:table} will play an important role in the construction of the fermionic projector~\cite{pfp}. The fact that all particles in Table~\eqref{c:table} are fermions motivates to speak of ``fermionic'' projector and causal ``fermion'' systems.

Concerning the terminology of subatomic particles~\cite{folland-qft}, there are actually two basic dichotomies: all particles are either bosons, with integer spin, or fermions, with half-integer spin; and all particles are either \emph{hadrons}, which participate in the strong interaction, or non-hadrons. The fundamental fermions comprise quarks, which are hadrons, and leptons, which are not. Quarks combine in triplets to make \emph{baryons}, and in quark/
anti-quark pairs to make \emph{mesons}. Baryons are fermions, whereas mesons are bosons. The most familiar baryons are the proton and the neutron.

Elementary particles are characterized by several \emph{quantum numbers} which are often said to be ``intrinsic'' to them: \emph{electric charge} ($Q$), \emph{weak isospin} ($I$), \emph{baryon number} ($B$) and \emph{lepton number} ($L$).  
Note that, in natural units, electric charge is dimensionless, and the unit of charge in Heaviside-Lorentz units is simply the number~1. The basic physical constant is then the fundamental unit of charge, the absolute value of the charge of an electron, which usually is denoted by~$e$ (cf.~\cite{folland-qft}). The electron, muon and tauon have electric charge~$-e$, while each neutrino has electric charge zero~\cite{griffiths, kane}. 
Down, strange and bottom quarks carry electric charge~$-1/3e$, while
up, charm and top quarks have electric charge~$2/3e$. 

As mentioned before, elementary particles are either \emph{hadrons} or \emph{leptons}, depending on whether they respond to the strong interaction (hadrons) or not (leptons).
Now to each lepton is associated the \emph{lepton number}~$L=1$, and~$L=-1$ to all anti-leptons; all other particles have~$L = 0$. In a similar fashion, to each baryon is assigned 
the \emph{baryon number}~$B = 1$, and~$B = -1$ to all anti-baryons, while all other particles have~$B = 0$. Taking into account that baryons are composed of three quarks, the baryon number of a quark is~$B = 1/3$ (see~\cite{beiser}).  
Next, there are particles which in experiments behaved so unexpectedly that they were called ``strange particles.'' Accordingly, the so-called \emph{strangeness number}~$S$ was introduced; in particular, the strange quark has strangeness~$S = -1$, whereas all other quarks have strangeness~$S = 0$ (cf.~\cite[Table~13.4]{beiser}). A related quantity, called \emph{hypercharge}, is just the sum of strangeness~$S$ and baryon number~$B$, i.e.~$Y = S + B$ (cf.~\cite{eisberg+resnick}). There is one more property, \emph{(weak) isospin} (short for ``isotopic spin''), which has mathematical properties similar to those of spin, but which has no direct physical relationship to spin 
(cf.~\cite[Section~17-3]{eisberg+resnick} and~\cite{pal} for its connection to group theory).  
Of particular interest is its third component, usually denoted by~$I_3$, which satisfies the relation~$Y = 2(Q - I_3)$ (cf.~\cite{folland-qft} and~\cite[eq.~(9.139)]{griffiths}).  
Similarly to spin, isospin either takes integer or half-integer values, i.e.~$I_3 \in \{0, 1/2, 1, \ldots \}$; accordingly, a representation of isospin~$I_3 = 1/2$ contains~$(2I_3+1) = 2$ states, referred to as ``isospin up'' and ``isospin down,'' thus giving rise to \emph{isospin doublets} which coincide with the above generations of fermions (cf.~\cite[Section~18-8]{eisberg+resnick} and~\cite[Chapter~12]{aitchison+hey-II}). For a group theoretical treatment we refer the interested reader to~\cite[Chapter~5]{sternberg}.  
Moreover, a comprehensive overview of quantum numbers can be found in~\cite[Table~17-1]{eisberg+resnick}.

The reason for introducing the adjoint spinor~$\overline{\psi} = \psi^{\dagger}\gamma^0$ is that~$\overline{\psi} \psi$ is a relativistic invariant. Moreover, considering the \emph{parity transformation}
\begin{align*}
	P : \R^3 \to \R^3 \:, \qquad \mathbf{x} = (x,y,z) \mapsto (-x,-y,-z) = - \mathbf{x} \:, 
\end{align*}
the quantity~$\overline{\psi}\psi$ is invariant under~$P$. For this reason, it is called a \emph{scalar}. 
Introducing the so-called \emph{pseudoscalar matrix} (cf.~\cite[eq.~(A.2.18)]{quigg}) 
\begin{align*}
	\gamma^5 \equiv i\gamma^0\gamma^1\gamma^2\gamma^3 \:, 
\end{align*}
the quantity~$\overline{\psi}\gamma^5\psi$ changes sign under~$P$, i.e.~$P(\overline{\psi}\gamma^5\psi) = - \overline{\psi} \gamma^5 \psi$; for this reason, it is referred to as \emph{pseudoscalar}.  
In a similar fashion, for all~$\mu=0, \ldots, 3$ one refers to the quantity~$\overline{\psi} \gamma^{\mu} \psi$ as a \emph{vector}, whereas~$\overline{\psi} \gamma^{\mu} \gamma^5 \psi$ is said to be a \emph{pseudovector} (or \emph{axial vector}). Introducing the \emph{bilinear covariants}~$\sigma^{\mu\nu} \equiv \frac{i}{2}[\gamma^{\mu}, \gamma^{\nu}]$ gives rise to the \emph{anti-symmetric tensor}~$\overline{\psi}\sigma^{\mu\nu}\psi$. Making use of the fact that~$\Id, \gamma^5, \gamma^{\mu}, \gamma^{\mu}\gamma^5$ and~$\sigma^{\mu\nu}$ constitute a basis of~$\C^{4 \times 4}$, any~$4\times 4$-matrix can be written as a linear combination of the resulting~16 terms (for details see~\cite[Section~4.4 and Section~7.3]{griffiths} or~\cite{peskin+schroeder}). From the vector and pseudovector matrices, one can form the \emph{(Dirac) current}~$j^{\mu}$ and the \emph{axial vector current}~$j^{\mu 5}$ by (cf.~\cite[eq.~(3.73)]{peskin+schroeder}) 
\begin{align*}%\label{c:PS, (3.73)}
	j^{\mu} = \overline{\psi}\gamma^{\mu}\psi \qquad \text{and} \qquad \gamma^{\mu 5} = \overline{\psi} \gamma^{\mu} \gamma^5 \psi \:,
\end{align*}
respectively. 
Whenever~$\psi$ satisfies the Dirac equation, the current~$j^{\mu}$ is a conserved quantity, that is,~$\partial_{\mu}j^{\mu} = 0$. In the case~$m=0$,  
the axial vector current~$j^{\mu 5}$ is also conserved, that is,~$\partial_{\mu} j^{\mu 5} = 0$ (cf.~\cite[Section~3.4]{peskin+schroeder}).

Prior to~1956, it was taken for granted that the laws of physics are invariant with respect to parity transformations in the sense that the mirror image of any physical process also represents a perfectly possible physical process. However, as experiments carried out by \textsc{Chien-Shiung Wu} (1912--1997) revealed, weak interactions are \emph{not} invariant under parity~\cite{wu-1957}. 
Accordingly, 
neutrinos observed in nature are \emph{left-handed}, whereas the observed anti-neutrinos are \emph{right-handed} (see~\cite{griffiths}). 
These properties are summarized by the notion of \emph{chirality}. 
For a mathematical description of chirality (cf.~\cite[Section~1.2]{pfp} and~\cite[Chapter~5]{kane}), 
one introduces 
so-called \emph{chiral projectors}, 
\begin{align*}%\label{c:Kane, (5.49)--(5.50)}
	\chi_{L/R} = \frac{1\mp \gamma^5}{2} \:. 
\end{align*}
These are projection operators, i.e.~$\chi_{L/R}^2 = \chi_{L/R}$, satisfying the properties  
\begin{align*}
	\chi_{L}\chi_R = 0 \:, \qquad \chi_L + \chi_R = \Id  \:, \qquad \gamma^5 \: \chi_L = - \chi_L \:, \qquad \gamma^5 \chi_R = \chi_R \:, \qquad \chi_L^{\ast} = \chi_R
\end{align*}
(with respect to the spin scalar product~$\overline{\psi}\phi$).  
For any spinor~$\psi$, the projections~$\chi_L \psi$ and~$\chi_R \psi$ are referred to as \emph{left-} and \emph{right-handed} components of~$\psi$, respectively. We point out that a matrix is said to be \emph{even} or \emph{odd} if it commutes or anti-commutes with~$\gamma^5$, respectively. It is straightforward to verify that the Dirac matrices are odd, 
\begin{align*}
	\gamma^{\mu} \chi_{L/R} = \chi_{R/L} \gamma^{\mu} \qquad \text{for all~$\mu = 0, \ldots, 3$} \:. 
\end{align*}
Introducing~$\psi_{L/R} = \chi_{L/R}\psi$, this gives rise to the relation 
\begin{align}\label{c:Kane, (5.59)}
	\overline{\psi} \gamma^{\mu} \psi = \overline{\psi}_L \gamma^{\mu} \psi_L + \overline{\psi}_R \gamma^{\mu} \psi_R \:. 
\end{align}
The summands in~\eqref{c:Kane, (5.59)} are called \emph{left-} and \emph{right-handed currents}, respectively.\footnote{For the relation to weak isospin and hypercharge we refer to~\cite[Chapter~9]{griffiths}.}

Experiments have shown that neutrinos may convert from one flavor to another (for instance, $\nu_{\text{e}} \leftrightarrow \nu_{\mu}$).  
These phenomena, referred to as \emph{neutrino oscillations}, suggest that neutrinos have non-zero mass, which in turn motivates to also take \emph{right}-handed neutrinos into consideration~\cite{griffiths,quigg}.\footnote{These experimental observations are taken into account in the theory of causal fermion systems.} 
The neutrino oscillations can mathematically be described by the MNS matrix (or \emph{neutrino mixing matrix}, cf.~\cite[Section~11.5]{griffiths}). For quarks, a similar effect is described by the CKM matrix (or \emph{quark mixing matrix}, see~\cite[Section~9.5]{griffiths} or~\cite[Chapter~7]{quigg} for details).  
We point out that currently there is no experimental reason for not suspecting the quarks and leptons in Table~\eqref{c:table} to be the ultimate elementary particles~\cite{quigg}; in particular, it is unlikely to expect a fourth generation of fermions~\cite{beiser}.

\subsection{The Standard Model of Particle Physics}\label{c:S SM}
The quarks and leptons, which are the basic particles of matter, are subject to several interactions (or ``forces''). These forces between fundamental fermions are mediated via the exchange of bosons (so-called ``gauge bosons'').  
At present, there are four known fundamental interactions, comprising electromagnetic, weak and strong interaction as well as gravity. The first three of them are described by the  
\emph{Standard Model of particle physics}. More explicitly, electromagnetic and weak interaction are combined into electroweak theory, which is dealt with in quantum electrodynamics (QED), whereas the theory of strong interaction is quantum chromodynamics (QCD).  
The gauge bosons of electroweak theory are the photon and the~$W^{\pm}$ and~$Z$ bosons, while the gauge bosons of the strong force are (eight) gluons.  
In order to understand the number of gluons as well as its mathematical description in the Standard Model in some more detail, let us point out that modern theory of particle interactions is based on gauge theories~\cite{pal}. 
Without describing its precise structure we note that, 
in mathematical terms,  
gauge theory deals with principal bundles, connections on them, and the curvatures of these connections~\cite{morgan, zeidler-gauge}.

Before entering the basic idea of gauge theories, let us briefly outline some central structures of quantum field theory. In analogy to classical physics, it has become conventional to formulate particle physics in terms of a Lagrangian.  
More precisely, one usually introduces an \emph{action}~$S$ by
\begin{align*}
	S = \int L \: dt \qquad \text{with} \qquad L = \int \L \: d^3x \:, 
\end{align*}
where~$L$ is referred to as \emph{Lagrangian} and~$\L$ as \emph{Lagrangian density} (which often is also called ``Lagrangian'').  
In this formalism, the dynamics is determined by a single function~$\L$; whenever the Lagrangian~$\L$ is Lorentz invariant (that is, invariant under Lorentz transformations), the whole theory is Lorentz invariant~\cite{kane}.  
For instance, the Dirac equation can be derived from the Lagrangian  %(density)
\begin{align*}
	\L = i\overline{\psi}\slashed{\partial} \psi - m \overline{\psi}\psi \:,
\end{align*}
where~$m$ denotes mass and~$\overline{\psi}$ the adjoint spinor; this Lagrangian is invariant under the transformations~$\psi \mapsto e^{i\theta} \psi$ for~$\theta \in \R$ (cf.~\cite[Section~4.1]{folland-qft}). Varying the action with respect to~$\overline{\psi}$, treated as independent of~$\psi$, one obtains the Dirac equation~\eqref{c:Griffiths, (7.20)} (cf.~\cite[Section~1.5]{parker+toms}).

The modern viewpoint of gauge invariance is due to quantum theory~\cite{kane}. Since observables depend on~$|\psi|^2$, the structure of the theory is invariant under \emph{global gauge transformations} (cf.~\cite[eq.~(3.6)]{kane})
\begin{align*}%\label{c:Kane, (3.6)}
	\psi \to \psi' = e^{-i\theta} \psi \:, 
\end{align*}
where~$\theta \in \R$ is a constant. In case that~$\theta = \theta(t, \mathbf{x})$ depends on each spacetime point, the corresponding transformation (cf.~\cite[eq.~(3.7)]{kane})
\begin{align*}%\label{c:Kane, (3.7)}
	\psi(t, \mathbf{x}) \to \psi'(t, \mathbf{x}) = e^{-i\theta(t, \mathbf{x})} \psi(t, \mathbf{x})
\end{align*}
is called a \emph{local gauge transformation} (or, in this case, a \emph{local phase transformation}). 
Making use of the fact that group elements of the unitary group~$\U(1)$ are expressed by~$e^{i\theta}$ with~$\theta \in \R$, the above (local) gauge transformations give rise to the ``gauge'' group~$\U(1)$.

In order to prepare for Yang-Mills theory and to explain gauge theories in some more detail, let us return to electrodynamics, which is considered as both the simplest gauge theory and the most familiar~\cite{quigg}. 
In classical electrodynamics, the fields~$\mathbf{B}$ and~$\mathbf{E}$ are related to the vector potential~$\mathbf{A}$ by (see e.g.~\cite{kane})
\begin{align}\label{c:Kane, (3.1)}
	\mathbf{B} = \nabla \times \mathbf{A} \:, \qquad \mathbf{E} = - \nabla \phi - \partial_t \mathbf{A} \:. 
\end{align}
Whenever~$\Lambda$ is an arbitrary (smooth) real-valued function, 
it is evident that the equations~\eqref{c:Kane, (3.1)} remain unchanged with respect to the transformations
\begin{align}
	\mathbf{A} \to \mathbf{A}' = \mathbf{A} + \nabla \Lambda \:, \qquad %\label{c:Kane, (3.2)} \\
	\phi \to \phi' = \phi - \partial_t \Lambda \:. \label{c:Kane, (3.3)}
\end{align}
Combining~$\mathbf{A}$ and~$\phi$ into a four-vector~$A^{\mu} = (\phi, \mathbf{A})$, the transformations~\eqref{c:Kane, (3.3)} read 
\begin{align*}%\label{c:Kane, (3.5)}
	A^{\mu} \to {A^{\mu}}' = A^{\mu} - \partial^{\mu} \Lambda \:. 
\end{align*}
In other words, 
the electromagnetic potential~$A_{\mu}$ is not uniquely defined but only modulo the gauge transformations~$A_{\mu} \mapsto A_{\mu} + \partial_{\mu} \Lambda$, where~$\Lambda$ is an arbitrary (smooth) real-valued function. 
The resulting freedom to choose~$A_{\mu}$ is called \emph{gauge invariance} in electrodynamics (see~\cite[Section~4.2]{folland-qft} and~\cite[Chapter~3]{quigg}). Concerning the notion of ``gauge invariance'' we refer to~\cite{sakurai-qm} and~\cite{folland-qft}.

Let us also note the following invariance property: 
Whenever~$\psi$ and~$A_{\mu}$ satisfy the Dirac equation~$(i\slashed{\partial} - e \slashed{A}- m) \psi = 0$, then~$\psi' = \exp(ie\Lambda)\psi$ and~$A_{\mu}' = A_{\mu} + \partial_{\mu} \Lambda$ satisfy the equation~$(i\slashed{\partial} - e \slashed{A}' - m) \psi' = 0$.
In other words, the Dirac equation~$(i\slashed{\partial} - e \slashed{A}- m) \psi = 0$ is invariant under the simultaneous transformations~$\psi \to \exp(ie\Lambda) \psi$,~$A_{\mu} \to A_{\mu} + \partial_{\mu} \Lambda$ (cf.~\cite[Section~4.2 and Section~9.1]{folland-qft}), thus combining the two gauge transformations mentioned before.  
Accordingly, we obtain a gauge-invariant Lagrangian 
\begin{align*}
	\L = \overline{\psi}(i\partial_{\mu}\gamma^{\mu} - A_{\mu}\gamma^{\mu} -m) \psi \:;
\end{align*}
its physical interpretation is that the matter field~$\psi$ is coupled to a ``gauge field''~$A_{\mu}$. 

On the other hand, starting with the free Dirac Lagrangian~$\L_0 = \overline{\psi} (i\slashed{\partial} - m) \psi$, one observes that~$\L_0$ is invariant under the transformation~$\psi \mapsto e^{i\chi}\psi$ in case that~$\chi$ is a (real) constant. The natural question to ask is how the Lagrangian~$\L_0$ needs to be modified in order to be invariant under the transformation~$\psi \mapsto e^{i\chi}\psi$ for an \emph{arbitrary} smooth function~$\chi : \R^4 \to \R$. Namely, in the case that~$\chi$ is \emph{not} constant, the derivative~$\partial_{\mu}$ does not commute with multiplication by~$e^{i\chi}$. Indeed, one needs to replace the derivative~$\partial_{\mu}$ by a ``covariant derivative''~$D_{\mu} = \partial_{\mu} + i A_{\mu}$ for some gauge field~$A_{\mu}$, thus giving rise to the ``interacting'' Lagrangian~$\L = \overline{\psi}(i\slashed{\partial} - \slashed{A} -m) \psi$.\footnote{Replacing the derivative~$\partial$ by the covariant derivative~$D$ is known as ``minimal coupling''~\cite{peskin+schroeder}.}  
The important point to notice is that the gauge field~$A_{\mu}$ appearing in electrodynamics is closely related to the local gauge transformations~$\psi \to e^{i\chi}\psi$ with~$e^{i\chi}$ in the Lie group~$G = \U(1)$; more precisely, the gauge field~$A_{\mu}$ takes values in the corresponding Lie algebra~$\mathfrak{g} = \mathfrak{u}(1)$ (cf.~\cite[Section~9.1]{folland-qft}). 

Thus the transition from a global symmetry (constant~$\chi$) to a local symmetry which depends on spacetime coordinates (arbitrary~$\chi$) requires to introduce ``compensating gauge fields'' by means of the corresponding covariant derivative (for further details see~\cite[Section~18-6]{eisberg+resnick}); the gauge field is interpreted as interacting in a specific way, thus giving rise to a dynamical theory. These types of dynamical theories, based on local invariance principles, are called gauge theories~\cite{aitchison+hey-I}.

By contrast to electrodynamics, in which case the gauge group~$\U(1)$ is \emph{abelian}, 
quantum chromodynamics and the electroweak theory are built on a generalization of this gauge principle, in which case the ``phase factors'' become matrices, which in general do not commute with each other. Accordingly, the associated symmetry is called a \emph{non-abelian} one. The transition from a global symmetry to a local one in the non-abelian setting was first accomplished by Yang and Mills~\cite{yang+mills} in order to describe the weak interaction between quarks and leptons. More specifically, to obtain a local symmetry from the global symmetry of isospin invariance, Yang and Mills considered gauge transformations~$\psi \to U \psi$ with~$U \in \SU(2)$. The corresponding Lie algebra~$\mathfrak{su}(2)$ has three generators;  
the corresponding~$\mathfrak{su}(2)$-valued gauge fields~$A_{\mu}^{a}$ ($a=1,2,3$) 
are referred to as \emph{Yang-Mills fields}, and the resulting Euler-Lagrange equations are called \emph{Yang-Mills equations}~\cite{folland-qft}.  
In a similar fashion, the strong interaction between quarks is described by quantum chromodynamics, a non-abelian gauge theory of the Yang-Mills type corresponding to the gauge group~$\SU(3)$; the gauge potentials take values in the gauge algebra~$\mathfrak{su}(3)$ (cf.~\cite{aitchison+hey-II, folland-qft}). 
For a concise summary of Yang-Mills theory we refer to~\cite[Chapter~15]{peskin+schroeder}.

Thus in summary, the interactions in the Standard Model are described by Yang-Mills type equations~\cite{aitchison+hey-I}.  
The weak and electromagnetic theory usually are combined into \emph{electroweak theory} with gauge group~$\U(1) \times \SU(2)$; henceforth, the gauge group of the Standard Model is given by 
\begin{align*}
	\U(1) \times \SU(2) \times \SU(3) \:. 
\end{align*}

Let us finally establish the connection to the number of gluons involved in a gauge theory. For given~$n \in\N$, consider the Lie group~$\SU(n)$ and let~$\mathfrak{su}(n)$ be its Lie algebra. We point out that~$\mathfrak{su}(n)$ has~$n^2-1$ generators, denoted by~$T_a$ ($a=1, \ldots, n^2-1$).  
Considering gauge transformations~$\psi \to \psi' = U\psi$, in order to obtain local gauge invariance one is led to introduce the covariant derivative
\begin{align*}%\label{c:Pal, (11.12)}
	D_{\mu} = \partial_{\mu} + ig T_a A_{\mu}^a \:,
\end{align*}
where the ``coupling constant''~$g$ determines the strength of the interaction and~$A_{\mu}^{a}$ ($a= 1, \ldots, n^2-1$) are the \emph{gauge fields}, taking values in~$\mathfrak{su}(n)$.  
The number of these fields is equal to the number of generators, and the particles generated by the gauge fields are called \emph{gauge bosons} (for details see~\cite{pal}).  

\subsection{Quantum Field Theory}\label{c:S qft}
In short, quantum field theory (QFT) 
arose from the necessity to combine special relativity with quantum mechanics. More precisely, the world of everyday life is governed by classical mechanics. For classical objects that travel very fast compared to the speed of light, the laws of classical mechanics are modified by special relativity. For objects which are small (compared to the size of atoms, roughly speaking), classical mechanics is superseded by quantum mechanics.  
In order to describe objects that are both fast \emph{and} small, one requires a theory that incorporates relativity and quantum mechanics: quantum field theory~\cite{griffiths, zee}. It is essential for understanding the current state of elementary particle physics~\cite{peskin+schroeder}. An explanation for the necessity of the \emph{field} viewpoint  
can be found in~\cite[Section~2.1]{peskin+schroeder}. 

In order to deal with electroweak and strong interactions, quantum field theory is subdivided into quantum electrodynamics (QED) and quantum chromodynamics (QCD), where QED 
deals with the interaction of electrically charged particles and QCD 
describes  
the interaction of quarks by the exchange of gluons~\cite{gribbin}. 

In analogy to the Lagrangian formalism in classical mechanics (it is instructive to follow the explanations in~\cite{bjorken+drell-qf}),
quantum field theory is usually written in terms of an action~$S$, being the time integral over some Lagrangian~$L$, which in turn is given as the spatial integral over a Lagrangian density~$\L$ (where~$\L$ again is referred to as ``Lagrangian'').  
More specifically,  
``quantization'' in quantum mechanics amounts to replacing the (generalized) coordinate~$q$ of a classical particle by a Hermitian operator  
and its conjugate momentum~$p \equiv \partial L/\partial \dot{q}$ (where~$\dot{q} = \partial q/\partial t$) by~$-i\partial/\partial q$ in such a way that the commutator relation (cf.~\cite[eq.~(11.6)]{bjorken+drell-qf})
\begin{align*}%\label{c:BD2, (11.6)}
	[p,q] = - i 
\end{align*}
is satisfied. More generally, given a system with~$n$ degrees of freedom, the dynamics of a quantum mechanical system is determined by imposing the commutator relations
\begin{align*}%\label{c:BD, (11.24)}
	[p_i(0), q_j(0)] = -i\delta_{ij} \:, \qquad [p_i(0), p_j(0)] = 0 \:, \qquad [q_i(0), q_j(0)] = 0
\end{align*}
for all~$i,j \in \{1, \ldots, n \}$ at time~$t=0$ (cf.~\cite[Section~11.2]{bjorken+drell-qf}). This method, treating~$q_i$ and~$p_i$ as quantized variables, is referred to as \emph{first quantization}. In quantum field theory, on the other hand, one is mainly interested in quantizing \emph{fields} which have an \emph{infinite} number of degrees of freedom (corresponding to the limit~$n \to \infty$). This process is called \emph{second quantization}~\cite{kaku, peskin+schroeder}. For an example we refer to the quantization of the Klein-Gordon field~$\phi$ (see~\cite[Section~2.3]{peskin+schroeder}). In general, the idea is to start with a classical field theory  
and then ``quantize'' it, that is, reinterpret dynamical variables as operators which obey canonical commutation relations. This procedure, describing the transition from classical field theory to quantum field theory, is known as ``second quantization'' in order to distinguish the resulting field equation (in which~$\phi$ is an operator) from the old one-particle field equation (in which~$\phi$ was a wave function). 
In analogy to classical mechanics, the action in QFT 
is given as an integral over  
a Lagrangian~$\L(\phi, \partial_{\mu} \phi)$ depending on some field~$\phi$ and its derivative~$\partial_{\mu} \phi$,  
\begin{align*}%\label{c:Kaku, (1.41)}
	S = \int L \: dt \qquad \text{with} \qquad L = \int \L(\phi, \partial_{\mu} \phi) \: d^3x  \:,
\end{align*}
where~$L$ is 
integrated between initial and final times~\cite{kaku}.

For convenience, let us review some important QFT Lagrangians. To begin with, the Maxwell equations~$\partial^{\mu}F_{\mu\nu} = 0$ can be obtained from the Lagrangian
\begin{align*}%\label{c:PS, (3.7)}
	\LMaxwell = - \frac{1}{4}(F_{\mu\nu})^2 \:, 
\end{align*}
where~$F_{\mu\nu} = \partial_{\mu} A_{\nu} - \partial_{\nu} A_{\mu}$ is the field strength tensor and~$A^{\mu}$ is the vector potential (see~\cite[eq.~(3.7)]{peskin+schroeder}). Moreover, the Lorentz-invariant Dirac Lagrangian reads
\begin{align*}%\label{c:PS, (3.34)}
	\LDirac = \overline{\psi} (i\gamma^{\mu} \partial_{\mu} - m) \psi \:.
\end{align*}
Next, the QED Lagrangian~$\LQED$, given by 
%(cf.~\cite[eq.~(4.3)]{peskin+schroeder})
\begin{align*}%\label{c:PS, (4.3)}
	\LQED = \LDirac + \LMaxwell + \Lint
\end{align*}
(with~$\Lint$ describing the interaction), 
can be written in the simpler form 
\begin{align*}%\label{c:PS, (4.4)}
	\LQED = \overline{\psi} (i\slashed{D} - m) \psi - \frac{1}{4} (F_{\mu\nu})^2 \:,
\end{align*} 
where~$D_{\mu}$ is the \emph{gauge covariant derivative}, 
\begin{align*}%\label{c:PS, (4.5)}
	D_{\mu} \equiv \partial_{\mu} + i e A_{\mu}(x)
\end{align*}
with electron charge~$e$ (in order to describe a fermion of charge~$Q$, replace~$e$ by~$Q$). Varying~$\overline{\psi}$, the resulting Euler-Lagrange equations for~$\psi$ take the form
\begin{align*}%\label{c:PS, (4.7)}
	(i\slashed{\partial} + ie\slashed{A} - m) \psi (x) = (i\slashed{D} - m) \psi (x) = 0 \:,
\end{align*}
which is just the Dirac equation coupled to the electromagnetic field (with coupling constant~$e$). The Euler-Lagrange equations for~$A_{\nu}$ are given by
\begin{align*}%\label{c:PS, (4.8)}
	\partial_{\mu} F^{\mu\nu} = e\overline{\psi} \gamma^{\nu}\psi = ej^{\nu} \:,
\end{align*}
which are the inhomogeneous Maxwell equations with the current density~$j^{\nu} = \overline{\psi} \gamma^{\nu} \psi$ (see~\cite[eqs.~(4.3)--(4.8)]{peskin+schroeder}).  
The QCD Lagrangian, on the other hand, reads 
\begin{align*}
	\LQCD = \sum_f \overline{\psi}^f \big(i \slashed{\partial} - \alpha \slashed{A} - m^f \big)\psi^f - \frac{1}{4}  (F_{\mu\nu})^2 
\end{align*}
(cf.~\cite[Section~9.2]{folland-qft}), 
where~$f \in \{u,d,s,c,t,b \}$ are the quark flavors,~$m^f$ are the quark masses and~$\alpha$ is the strong coupling constant (the same for all quark flavors). For the Yang-Mills Lagrangian, which can be obtained by adding the gauge field Lagrangian to the usual Dirac Lagrangian (with the ordinary derivative replaced by the covariant derivative), we refer to~\cite[Section~15.2]{peskin+schroeder}. 

Note that the symmetry of a Lagrangian may be violated, in which case the field theory is said to have a \emph{hidden} or \emph{spontaneously broken} symmetry.  
One important consequence of spontaneous symmetry breaking is \emph{Goldstone's theorem} which states that for every spontaneously broken continuous theory there is a massless particle (known as \emph{Goldstone boson}, see~\cite[Section~11.1]{peskin+schroeder}). Applying Goldstone's theorem to gauge theories, however, one finds that these goldstone bosons convert massless gauge bosons into massive ones. This is the so-called ``Higgs mechanism'' (we refer the interested reader to~\cite[Section~10.2]{kaku} and~\cite[Section~20.1]{peskin+schroeder}). 

Applying the effect of spontaneous symmetry breaking to the gauge theory of weak interactions gives rise to the so-called \emph{GWS model}, introduced by Glashow, Weinberg and Salam (see~\cite[Section~10.4]{kaku} and~\cite[Section~20.2]{peskin+schroeder}). This model yields a unified description of weak and electromagnetic interactions. 
More explicitly, starting with a theory with~$\SU(2)$ gauge symmetry, one ends up with one massless gauge boson (photon), whereas the remaining three gauge bosons ($W^{\pm}, Z$) acquire masses from the Higgs mechanism.  
In the GWS model, only the left-handed components of the quark and lepton fields couple to the~$W$ bosons. 
As far as weak interactions are concerned, the resulting electroweak theory thus arranges the quarks and leptons as follows, 
\begin{align*}
	\begin{pmatrix}
		\nu_{\text{e}} \\ e
	\end{pmatrix}_L \:, \quad \begin{pmatrix}
		\nu_{\mu} \\ \mu
	\end{pmatrix}_L \:, \quad \begin{pmatrix}
		\nu_{\tau} \\ \tau
	\end{pmatrix}_L \qquad \text{and} \qquad \begin{pmatrix}
		u \\ d
	\end{pmatrix}_L \:, \quad \begin{pmatrix}
		c \\ s
	\end{pmatrix}_L \:, \quad \begin{pmatrix}
		t \\ b
	\end{pmatrix}_L \:.
\end{align*}
These are all weak isospin doublets; the right-handed components are weak isospin singlets (cf.~\cite[Section~18-8]{eisberg+resnick} and~\cite[eq.~(20.75)]{peskin+schroeder}). For this reason, physicists often denote the~$\SU(2)$ gauge group of weak interactions by~$\SU(2)_L$. 

The Standard Model of particle physics outlined above evolved as a consequence of experiments and advances in theoretical physics in the last century. Nevertheless, the predictions of a physical theory have to fit experimental results.  Concerning particle physics, so-called \emph{scattering experiments} are of crucial importance. Roughly speaking, in those experiments a beam of ``incoming'' particles is aimed at a target, and thereby ``scattered'' into ``outgoing'' particles. 
A typical quantity measured in the laboratory is the \emph{scattering cross section}, which can be thought of the effective ``size'' of each target particle  
as seen by an incoming beam. More precisely, the cross section is calculated in terms of the rate of collisions in a scattering experiment; it gives the probability  
that a collection of particles in some initial state will decay or scatter into another collection of particles in some final state.
For a mathematical description of such processes, the incoming particles are assigned an \emph{i}nitial state~$|i\rangle$ at initial time~$t = - \infty$, whereas the outgoing particles are in some \emph{f}inal state~$|f\rangle$ at time~$t = \infty$.   
In order to calculate the probability of the transition from the initial state to the final state, one introduces the so-called ``scattering matrix''~$S$ (cf.~\cite{kaku}). Then the transition from an initial state~$|i\rangle$ to a final state~$|f\rangle$ is given by the probability amplitude 
%(cf.~\cite[eq.~(16.57)]{bjorken+drell-qf}) 
\begin{align*}%\label{c:BD2, (16.57)}
	S_{fi} = \langle f \mid i \rangle \:,
\end{align*}
which can be thought of as one entry of the~$S$ matrix (see~\cite[Chapter~16]{bjorken+drell-qf}). 
Thus the task is to compute the~$S$ matrix for a scattering process. To this end, we recall that in quantum mechanics, a state~$\psi$ is a (normed) vector in a Hilbert space~$\mathcal{H}$.  
Following the axioms of quantum mechanics, the time evolution of a system is given by a one-parameter group of unitary operators~$U(t)$ on Hilbert space~$\mathcal{H}$ such that
\begin{align*}
	\psi(t) = U(t) \psi(0) \:, 
\end{align*}
where~$\psi(t)$ denotes the state at time~$t$. 
By Stone's theorem (see~\cite[Theorem~1.13]{dimock}), the time evolution~$U(t)$ is generated by a self-adjoint operator~$H$, 
\begin{align*}%\label{c:dimock, (3.5)}
	U(t) = e^{-itH} \:.
\end{align*}
The operator~$H$, called the \emph{Hamiltonian}, corresponds to the energy of the system (cf.~\cite[Section~3.1]{dimock}). In order to calculate the~$S$ matrix by means of perturbation theory (for details see~\cite[Section~3.5]{weinberg-I} and~\cite[Chapter~6]{folland-qft}),  
one assumes that the Hamiltonian~$H$ splits into a free-particle Hamiltonian~$H_0$ and an interaction term~$V$, 
\begin{align*}%\label{c:Weinberg-I, (3.1.8)}
	H = H_0 + V \:. 
\end{align*}
Introducing the operator
\begin{align*}%\label{c:Weinberg-I, (3.2.6)}
	U(\tau, \tau_0) \equiv \exp(iH_0 \tau) \exp(-iH(\tau - \tau_0)) \exp(-iH_0 \tau_0)
\end{align*}
for all~$\tau, \tau_0 \in \R$, 
the $S$ matrix is given by
\begin{align}\label{c:Weinberg-I, (3.2.5)}
	S = U(+ \infty, - \infty) \:. 
\end{align}
Starting from~\eqref{c:Weinberg-I, (3.2.5)} and defining 
\begin{align*}%\label{c:Weinberg-I, (3.5.5)}
	V(t) \equiv \exp(H_0t) V \exp(-iH_0t)
\end{align*}
for all~$t \in \R$, 
one obtains the following perturbation expansion for~$S$, 
\begin{align}\label{c:Weinberg-I, (3.5.8)}
	\begin{split}
		S = \Id &- i \int_{-\infty}^{\infty} V(t_1) \: dt_1 + (-i)^2 \int_{-\infty}^{\infty} \int_{-\infty}^{t_1} V(t_1) V(t_2) \: dt_2  \: dt_1 \\ 
		&+ (-i)^3 \int_{-\infty}^{\infty} \int_{-\infty}^{t_1} \int_{-\infty}^{t_2} V(t_1) V(t_2) V(t_3) \: dt_3 \: dt_2 \: dt_1 + \cdots \:.
	\end{split}
\end{align} 
In terms of the \emph{time-ordered product}~$T$ (cf.~\cite[Section~3.5]{weinberg-I}), expression~\eqref{c:Weinberg-I, (3.5.8)} reads 
\begin{align}\label{c:Weinberg-I, (3.5.10)}
	S = \Id + \sum_{n=1}^{\infty} \frac{(-i)^n}{n!} \int_{-\infty}^{\infty} T \left\{V(t_1) \cdots V(t_n) \right\} \: dt_1 \: dt_2 \cdots  \: dt_n\:.  
\end{align}
This expression, known as \emph{Dyson series}, is also written as \emph{time-ordered exponential}, 
\begin{align*}
	S = T\exp \left(-i\int_{\infty}^{\infty} V(t) \: dt \right) \:. 
\end{align*}
Making use of \emph{Wick's theorem}, each time-ordered summand of the perturbation expansion~\eqref{c:Weinberg-I, (3.5.10)} can be written as a sum of \emph{normal-ordered} summands (cf.~\cite[Section~4.3]{peskin+schroeder} or~\cite[Section~5.6]{kaku}).  
The resulting summands may be reexpressed diagrammatically in terms of so-called \emph{Feynman diagrams}. This yields a convenient method for evaluating the $S$ matrix to a given order by employing the corresponding \emph{Feynman rules}.  
In most calculations, it is preferable to express the Feynman rules in terms of momenta, thus giving rise to \emph{momentum-space Feynman rules} (for details see~\cite{peskin+schroeder}). 

One of the serious complications found in quantum field theory is that the theory is naively divergent. When higher-order corrections are calculated for QED, one finds that the integrals diverge in the ultraviolet region, that is, for large momentum~$p$. These divergences, which are related to the experimentally not accessible high-energy region, reflect our ignorance about the nature of physics at extremely small distances. In order to avoid these divergences, one ``cuts off'' momentum space at some large momentum~$\Lambda$, thereby obtaining finite quantities.
This manipulation is known as ``regularization.'' 
Afterwards one considers the limit~$\Lambda \to \infty$.  
A theory is said to be \emph{renormalizable} if the physical quantities turn out to be independent of~$\Lambda$ (cf.~\cite{peskin+schroeder}). For instance, it was proven by \textsc{Gerard 't Hooft} (1946--) that spontaneously broken Yang-Mills theory is renormalizable~\cite{kaku}. Following~\cite{zee}, quantum field theory thus should be regarded ``as an effective low energy theory. (...)  
It is thought that as we go to higher and higher energies the whole edifice of quantum field theory will turn out to be an approximation to a theory whose identity we don't know yet.''

Many different renormalization procedures have been proposed, but they all share some basic physical features. The essential idea is 
to assume 
a set of ``bare'' or ``naked'' parameters that are divergent, such as the coupling constants and masses.\footnote{Concerning the bare and effective masses of quarks we refer to~\cite[Table~4.4]{griffiths}.} By contrast to ``{physical}'' parameters, these {bare} parameters are unmeasurable.  
The divergences of these parameters are chosen so that they cancel against the ultraviolet infinities which probe the small-distance behavior of the theory. After these divergences have been absorbed by the bare parameters, we are left with the physical, renormalized, or ``dressed'' parameters that are indeed measurable~\cite{kaku}. We point out that, motivated by the renormalization program, the concept of bare parameters will show up in the theory of causal fermion systems. 

Note that most conventional regularization schemes are based on the perturbative Feynman expansion. 
A non-perturbative approach  
is to work on spacetime lattices instead of a spacetime continuum~\cite{aitchison+hey-II, creutz}. In this setting, one considers a discrete set of lattice points, separated by a minimum distance: the lattice spacing~$a$. Regarding the lattice merely as an ultraviolet cutoff, when removing the regularization observable quantities should approach their physical values in the limit~$a \searrow 0$, the so-called ``continuum limit''~\cite{gattringer+lang}. These explanations may serve as a motivation for the notion of ``continuum limit'' which we encounter in the theory of causal fermion systems.

As being of relevance for the theory of causal fermion systems, we finally outline the generalization of the Dirac equation to curved spacetime, modelled by a (globally hyperbolic) pseudo-Riemannian manifold~$(M, g)$  
(for details see~\cite{parker+toms}). In Minkowski spacetime~$\scrM$ (with Minkowski metric~$\eta_{\mu\nu}$), the Dirac equation
\begin{align*}%\label{c:Parker, (1.117)}
	(i\gamma^{\mu}\partial_{\mu} - m)\psi = 0
\end{align*}
is obtained from the Lagrangian  
\begin{align*}%\label{c:Parker, (1.115)}
	\L = \overline{\psi} (i \gamma^{\mu} \partial_{\mu} -m) \psi \:, 
\end{align*}
and the $\gamma$-matrices satisfy the anti-commutation relations 
\begin{align}\label{c:Parker, (1.116)}
	\{\gamma^{\mu}, \gamma^{\nu} \} = 2\eta^{\mu\nu} \:. 
\end{align}
In curved spacetime, on the other hand, formula~\eqref{c:Parker, (1.116)} is generalized to
\begin{align}\label{c:Parker, (3.206)}
	\underline{\gamma}^{\mu}(x) \underline{\gamma}^{\nu}(x) + \underline{\gamma}^{\nu}(x) \underline{\gamma}^{\mu}(x)  = 2g^{\mu\nu}(x) \:,
\end{align}
where~$g^{\mu\nu}$ is the inverse of~$g_{\mu\nu}$; the underline is used to distinguish the spacetime-dependent $\underline{\gamma}$-matrices from the constant~$\gamma$-matrices in~\eqref{c:Parker, (1.116)}.  
In order to represent the matrices~$\underline{\gamma}_{\mu}(x)$ in terms of the  
matrices~$\gamma_{\mu}$,  
one introduces a so-called \emph{vierbein}~${b^{\alpha}}_{\mu}(x)$ of vector fields (or \emph{tetrad}, see~\cite{defelice+clarke} for details), defined by
\begin{align*}%\label{c:Parker, (3.216)}
	g_{\mu\nu}(x) = \eta_{\alpha\beta} {b^{\alpha}}_{\mu}(x) {b^{\beta}}_{\nu}(x) \:. 
\end{align*}
Vierbein indices are lowered with~$\eta_{\alpha\beta}$, whereas spacetime indices are lowered with the metric~$g_{\mu\nu}$. 
In terms of the Dirac matrices~$\gamma^{\mu}$ in Minkowski space, 
the matrices~$\underline{\gamma}^{\mu}(x)$ in~\eqref{c:Parker, (3.206)} can be written in the form
\begin{align*}%\label{c:Parker, (3.217)}
	\underline{\gamma}^{\mu}(x) = {b_{\alpha}}^{\mu}(x) \gamma^{\alpha} \:. 
\end{align*}
Next, the spinorial affine connections~$\Gamma_{\mu}(x)$ (for details see~\cite{friedrich, kobayashi+Nomizu-I, lawson+michelson}) 
are defined by the vanishing of the covariant derivative of the~$\underline{\gamma}$-matrices, 
\begin{align*}%\label{c:Parker, (3.207)}
	\nabla_{\mu} \underline{\gamma}_{\nu} \equiv \partial_{\mu} \underline{\gamma}_{\nu} - {\Gamma^{\lambda}}_{\mu\nu}\underline{\gamma}_{\lambda} - \Gamma_{\mu} \underline{\gamma}_{\nu} + \underline{\gamma}_{\nu} \Gamma_{\mu} = 0 \:,
\end{align*}
where~$\underline{\gamma}_{\mu} = g_{\mu\nu}\underline{\gamma}^{\nu}$.
Introducing the covariant derivative acting on a spinor field~$\psi$ by
\begin{align*}%\label{c:Parker, (3.208)}
	\nabla_{\mu} \psi \equiv (\partial_{\mu} - \Gamma_{\mu}) \psi \:, 
\end{align*}
the generally covariant Dirac equation in curved spacetime reads
\begin{align*}%\label{c:Parker, (3.209)}
	(i \underline{\gamma}^{\mu}(x) \nabla_{\mu} - m) \psi(x) = 0 \:. 
\end{align*}

\section{The Principle of the Fermionic Projector}\label{c:Section Principle} 
The theory of causal fermion systems, as outlined in~\cite{cfs}, evolved from the principle of the fermionic projector~\cite{pfp}. For this reason, it seems a good starting point to first present the underlying ideas of the fermionic projector approach in more detail. To this end, we point out that the monograph~\cite{pfp} itself has its origins in~\cite{finster1996ableitung}, which in turn is based on previous considerations summarized in the articles~\cite{gauge, local}. In~\cite{gauge} it was suggested to link the physical gauge principle with non-relativistic quantum mechanical measurements of the position variable. The basic idea in~\cite{local} is to extend this concept to relativistic quantum mechanics and to explain local gauge freedom for Dirac spinors by a local~$\U(2,2)$ symmetry, which allows for a natural description of both electrodynamics and general relativity as a classical gauge theory~\cite{local}.  

Actually, the basic ideas in~\cite{gauge, local} may be considered as the starting point for the development of the theory of causal fermion systems. For this reason, let us outline the underlying concepts, which are essential in order to understand how the specific form of the causal action principle comes about, in more detail. This procedure eventually leads to the principle of the fermionic projector, which in turn allows us to formulate a variational principle in ``discrete'' spacetime. 

\subsection{Derivation of Local Gauge Freedom}\label{c:S gauge freedom}
To begin with, let us recall that in quantum mechanics one usually considers an abstract Hilbert space~$H$ endowed with a scalar product~$\langle \, . \, | \, . \, \rangle$. The  
physical observables in non-relativistic quantum mechanics correspond to self-adjoint operators on~$H$ with respect to the scalar product~$\langle \, . \, | \, . \, \rangle$, and measurements with respect to a state~$\psi \in H$ correspond to calculating the expectation value~$\langle \psi | \mathcal{O} | {\psi} \rangle$. 
Moreover, the wave function~$\psi(\vec{x})$ is introduced by
\begin{align}\label{c:pfp, (3.1.5)}
	\psi(\vec{x}) = \langle \vec{x} \mid \psi \rangle \:.  
\end{align}
It is of crucial importance to note that the observables for space are of particular interest as they determine the geometry of the physical system under consideration.  
Usually, they are given by mutually commuting operators~$(X^{i})_{i=1,\ldots,3}$ such that 
\begin{align*}
	X^{i} \psi(x) = x^{i}\psi(x) \qquad \text{for~$i=1,2,3$} \:. 
\end{align*}
Introducing the wave function~$\psi(\vec{x})$ by~\eqref{c:pfp, (3.1.5)} is known as ``position representation.''  
In bra/ket notation,\footnote{Given a Hilbert space~$H$ with scalar product~$\langle \, . \, | \, . \,\rangle$, in physics it is customary to ``split up'' the inner product~$\langle \, . \, | \, .\, \rangle$ into a \emph{ket}~$| \, . \, \rangle$ and a \emph{bra}~$\langle \, . \, |$ (this terminology is due to~\cite{dirac-1ed}). Note that the formal bra/ket notation can be made mathematically precise using spectral measures~\cite{gauge}.} 
this is done by choosing an ``eigenvector basis''~$|\vec{x} \rangle$ of the position operators in such a way that
\begin{align*}%\label{c:pfp, (3.1.4)}
	\vec{X} |\vec{x} \rangle = \vec{x} |\vec{x}\rangle \:, \qquad \langle \vec{x} \mid \vec{y} \rangle = \delta^{3}(\vec{x} - \vec{y}) \:. 
\end{align*}
At this point it is central to observe that the position representation is not unique, as the ``eigenvectors''~$|\vec{x}\rangle$ are only determined up to a phase; more precisely, they can be transformed according to
\begin{align}\label{c:local, (2)}
	|\vec{x}\rangle \to e^{ie\Lambda(\vec{x})} |\vec{x} \rangle
\end{align}
with a real function~$\Lambda(\vec{x})$. This corresponds to a local phase transformation
\begin{align}\label{c:local, (3)}
	\psi(\vec{x}) \to e^{-ie\Lambda(\vec{x})} \psi(\vec{x})
\end{align}
of the wave functions. This arbitrariness of the local phase of the wave functions can also be understood from the fact that the wave function itself is not an observable, but only its absolute square~$|\psi|^2$ has a physical interpretation as probability density. 
Starting point in~\cite{local} is to interpret  
the local phase transformations~\eqref{c:local, (2)}, \eqref{c:local, (3)}  
as~$\U(1)$ gauge transformations. 
Generalizing the previous considerations to wave functions with several components, for suitable unitary matrices~$U(\vec{x})$ one arrives at  
\begin{align*}%\label{c:local, (6)}
	\psi(\vec{x}) \to U(\vec{x}) \psi(\vec{x}) \:. 
\end{align*}
If these local transformations~$U(\vec{x})$ could be identified with physical gauge transformations, then the local gauge principle would no longer be an a-priori principle in physics. Instead, it would be a consequence of a quantum mechanical ``measurement principle,'' namely the description of space with observables~$X^{i}$ on an abstract Hilbert space. This idea might make it possible to describe additional interactions (like gravitation or the weak and strong forces). Furthermore, the local gauge group could no longer be chosen arbitrarily. These observations may be regarded as the starting point for the development of causal fermion systems. 

In order to generalize this ``measurement principle'' to the relativistic setting, it is convenient to introduce ``observables'' for space and time as multiplication operators with the coordinate functions by
\begin{align*}%\label{c:local, (11)}
	X^{i}\psi(x) = x^{i} \psi(x) \qquad \text{for all~$i=0, \ldots, 3$} \:,
\end{align*}
where~$\psi : \scrM \to \C^4$ is regarded as a Dirac spinor on Minkowski space~$\scrM$. Motivated by the ``spin scalar product''~$\overline{\psi}\phi$ of quantum mechanics, we let~$\C^4$ be endowed with an indefinite inner product~$\prec . \mid . \succ$ of signature~(2,2). Unfortunately, the usual positive scalar product~\eqref{c:pfp, (1.2.15)}, 
\begin{align}\label{c:local, (12)}
	(\psi \mid \phi) = \int_{\R^3} \psi^{\dagger}(t, \vec{x}) \, \phi(t, \vec{x}) \: d\vec{x} \:, 
\end{align}
where the spinors are integrated over a space-like hypersurface at a constant time~$t$, is not compatible with time measurement. More precisely, the expectation value~$$(\psi \mid X^0 \mid \psi) = t (\psi \mid \psi)$$ depends on the choice of the hypersurface, whereas for solutions of the Dirac equation, current conservation implies that~\eqref{c:local, (12)} is independent of~$t$. This motivates to introduce a different scalar product where the spinors are also integrated over the time variable, 
\begin{align}\label{c:local, (13)}
	{<\psi \mid \phi>} = \int_{\R^4} \prec \psi(x) \mid \phi(x) \succ \: d^4x \:. 
\end{align}
In contrast to~\eqref{c:local, (12)}, however, the inner product~\eqref{c:local, (13)} has no immediate physical interpretation.  
Introducing an ``eigenvector basis'' of the time and position operators with respect to~$< . \, | \, . >$,  
one again obtains local gauge freedom of the form
\begin{align}\label{c:pfp, (3.1.11')}
	\psi(x) \to U(x) \psi(x) \qquad \text{with~$U(x) \in \U(2,2)$} \:,
\end{align}
where~$\U(2,2)$ denotes the set of unitary matrices on~$\C^4$ with respect to~$\prec. \mid . \succ$. 

In order to clarify the connection of these ``gauge transformations'' to interactions between fermions, it is illustrative to consider the free Dirac equation (cf.~\eqref{c:Griffiths, (7.20)})
\begin{align*}
	(i\slashed{\partial} - m) \psi = 0 
\end{align*}
with~$\slashed{\partial} = \gamma^j \partial_j$,  
where~$\gamma^{j}$ ($j= 0, \ldots, 3$) are the usual Dirac matrices. 
Due to~\eqref{c:pfp, (3.1.11')}, we are given the freedom to modify~$\psi$ according to~$\psi \to U^{-1} \psi$ with~$U \in \U(2,2)$. Multiplying by~$U$ from the left, this leads us to consider the modified Dirac equation 
\begin{align*}
	(U(i\slashed{\partial})U^{-1} - m) \psi = 0 \:. 
\end{align*}
Thus
the gauge transformation~\eqref{c:pfp, (3.1.11')} yields a transformation of the Dirac operator, 
\begin{align*}
	i\slashed{\partial} \to G := U (i\slashed{\partial}) U^{-1} = i G^j \frac{\partial}{\partial x^j} + B 
\end{align*}
with
\begin{align*}
	G^j(x) = U(x) \gamma^j U(x)^{-1} \:, \qquad B(x) = i U(x) \gamma^j (\partial_j U(x)^{-1}) \:. 
\end{align*}
The resulting operator~$G$ is Hermitian with respect to the inner product~$< . \mid . >$ and Lorentz invariant. 
Furthermore, there exist gauge transformations such that~$G$ locally coincides with the original Dirac operator~$i\slashed{\partial}$ (see~\cite{local}). 

In order to motivate the generalization to curved spacetime, we point out that the Minkowski metric~$\eta$ can be derived from the Dirac matrices~$\gamma^j$ by~$2\eta^{ij} \Id = \{\gamma^{i}, \gamma^{j} \}$ for all~$i,j \in \{0, \ldots, 3 \}$.  
For the generalization to curved spacetime, one replaces Minkowski space~$\scrM$ by a four-dimensional smooth manifold~$M$. Given an abstract Hilbert space~$H$ endowed with an indefinite inner product~${<. \mid . >}$, the previous observations motivate to introduce the Dirac operator as a first order differential operator~$G$ on~$H$ by
\begin{align*}%\label{c:local, (20)}
	G = iG^j \frac{\partial}{\partial x^j} + B
\end{align*}
with~$(4 \times 4)$-matrices~$G^j(x), B(x)$ in such a way that~$G^j$ locally coincides with~$\gamma^j$ for a specific gauge in a specific chart. The time and position operators, which are defined by means of spectral measures~$(dE_x)_{x \in M}$, admit to introduce wave functions. Moreover, the Lorentzian metric~$g$ on~$M$ is given by
\begin{align*}%\label{c:local, (22)}
	g^{jk}(x) \Id = \frac{1}{2} \left\{G^j(x), G^k(x) \right\} \:.
\end{align*}
As a consequence, the Lorentzian metric allows us to construct further objects like the Levi-Civita connection or curvature tensors.  
It should be noticed that  
neither~$B(x)$ nor the~$\U(2,2)$ gauge symmetry enter these constructions. Indeed, the matrix~$B(x)$ can be used for describing additional interactions (like electromagnetism). In this way, the Dirac operator is considered as the basic object on the manifold; all required objects can be constructed from the Dirac operator~\cite{finster1996ableitung}. 
In particular, the gauge potentials are implicitly contained in the Dirac operator. 

In short, the above constructions can be summarized as follows~\cite{local}:  
Adapting the ideas in~\cite{gauge} to the relativistic context yields a local~$\U(2,2)$ gauge symmetry of the Dirac equation. In order to describe the physical interactions with this gauge symmetry, it is necessary to consider the Dirac operator as the basic object of the theory.  
By constructing the spin derivative (for details see~\cite{local}), the gauge potentials (which are implicitly contained in the Dirac operator) can be recovered as describing the electromagnetic and gravitational field. In this way, the concept of measurability of spacetime gives a fundamental explanation for local~$\U(2,2)$ gauge symmetry. In particular, this description has the advantage that both the Dirac theory and classical field theory are developed from few a-priori given objects: The fermionic particles correspond to vectors of an abstract Hilbert space~$H$ endowed with an indefinite scalar product~${<. \mid . >}$, and spacetime is described by spectral measures~$(dE_x)_{x \in M}$ on a manifold~$M$.  
The Dirac operator yields the gauge potentials and determines the interaction between the fermions and the gauge fields. This description, which is conceptually simple, is the starting point for further constructions which eventually lead to the ``principle of the fermionic projector'' as introduced in~\cite{pfp} and finally give rise to causal fermion systems~\cite{cfs}.

\subsection{The Principle of the Fermionic Projector}\label{c:S principle}
After these preliminaries, we are in the position to enter the principle of the fermionic projector and to clarify its relation to the causal action principle. 
For simplicity, let~$M= \scrM$ be Minkowski space. 
In generalization of~\cite{local}, the starting point for the ``discretization of spacetime'' is an abstract Hilbert space~$H$ together with an indefinite inner product~${<. \mid . >}$ of signature~$(2N,2N)$ for some~$N \in \N$.  
Following the explanations in~\cite[Section~3.1]{pfp}, each vector~$\psi \in H$ gives rise to a wave function~$\psi(x)$ with~$x \in M$, and in analogy to~\S \ref{c:S gauge freedom} one obtains local gauge freedom of the form
\begin{align*}%\label{c:pfp, (3.1.11)}
	\psi(x) \to U(x) \psi(x) \qquad \text{with~$U(x) \in \U(2N, 2N)$} \:. 
\end{align*}
Now every particle is described by its wave function~$\psi(x)$, or, in a gauge-independent way, by a vector~$\psi \in H$. For this reason, it seems convenient to consider the complex subspace~${<\psi>}$ in~$H$ spanned by~$\psi$ in order to describe the corresponding particle. Accordingly, a finite number of particles~$\psi_1, \ldots, \psi_f$ spans a finite-dimensional complex subspace~${Y := {<\psi_1, \ldots, \psi_f>}}$ in~$H$. One thus obtains an indefinite inner product space~$(Y, {<. \mid . >})$. Just as in positive definite scalar product spaces, every non-degenerate subspace~$Y \subset H$ uniquely determines a projector~$P_Y : H \to Y$ onto this subspace, characterized by the conditions~$P_Y^2 = P_Y = P_Y^{\ast}$ and~$P_Y(H) = Y$, where the star denotes the adjoint with respect to the inner product~${<. \mid. >}$ (for details see~\cite[Section~3.2]{pfp}). Instead of working with the subspace~$Y$, it is more convenient to consider the corresponding projector~$P = P_Y$ in order to describe the particles of the system.  
Having Dirac particles in mind, which are fermions, the projector~$P \in \LL(H)$ is called \emph{fermionic projector}, and 
\begin{align*}%\label{c:(number of particles)}
	f = \dim P(H)
\end{align*}
is referred to as the \emph{number of particles}. Since~$P$ is a projector, one has~$f = \tr(P)$, where~$\tr$ denotes the trace of a linear operator (cf.~\cite[Section~3]{continuum}).

Next, in order to present the underlying ideas  
for deriving a variational principle in discrete spacetime, we again let~$H$ be an abstract Hilbert space together with an indefinite inner product~${<. \mid . >}$ of signature~$(2N,2N)$.  
Then in analogy to~\cite{local}, 
one defines mutually commuting time and position operators~$X^{i}$ ($i=0, \ldots, 3$). In order to establish the connection to discrete spacetime, one replaces the time and position operators~$X^{i}$ by mutually commuting operators~$X^{i}$ with a \emph{purely discrete spectrum}. The joint spectrum of these operators is then regarded as discrete spacetime~$M$, and the joint eigenspaces~$(e_x)_{x \in M}$  
are $4N$-dimensional subspaces of~$H$. 
The corresponding projectors~$E_x$ on~$e_x$ for every~$x \in M$ are uniquely characterized as the spectral projectors of the operators~$X^{i}$. The projectors~$(E_x)_{x \in M}$ are called \emph{spacetime projectors}, and the resulting structure~$(H, {<. \mid . >}, (E_x)_{x \in M})$ is referred to as \emph{discrete spacetime} (for details see~\cite[Section~3.3]{pfp} and~\cite{discrete}). Roughly speaking, the underlying idea is to think of discrete spacetime as a ``loose set of points''~\cite{finster1996ableitung} of ``mean distance''~$\varepsilon \lesssim \ell_P$ (for some ``regularization length''~$\varepsilon > 0$). In order to take the principle of general coordinate invariance appropriately into account, spacetime~$M$ should merely be considered as an index set for the spectral projectors, whereas the projectors~$(E_x)_{x \in M}$ are regarded as the basic objects describing the geometry of spacetime~\cite{pfp}.

The \emph{principle of the fermionic projector} states that a physical system is completely described by the fermionic projector in discrete spacetime. In particular, the physical equations should be formulated exclusively with the operators~$P$ and~$(E_x)_{x \in M}$ on~$H$  
(see~\cite[Section 3.5]{pfp}). 
Thus
in order to formulate physical equations, the projectors~$P$ and~$(E_x)_{x \in M}$ need to be combined in a mathematically interesting way. Guided by the Lagrangian formalism of  
quantum field theory, a promising strategy is to define an ``action principle''  
in terms of a real-valued ``Lagrangian.'' A possible strategy towards this aim is to form endomorphisms and then to consider the eigenvalues thereof. 
To this end, one introduces the ``discrete kernel'' of the fermionic projector~$P(x,y)$ by~$P(x,y) \equiv E_x P E_y$. Following the reasoning in~\cite{finster1996ableitung, pfp}, it turns out that the so-called ``closed chain''~$A_{xy}$ is of particular interest, which for all~$x,y \in M$ is defined by
\begin{align*}
	A_{xy} \equiv P(x,y) P(y,x) : E_x(H) \to E_x(H) \:. 
\end{align*} 
Given a real-valued functional~$\L$ on the endomorphisms of~$E_x(H) \subset H$ (the so-called ``Lagrangian''), the expression~$\L[A_{xy}]$ depends on the two spacetime points~$x,y \in M$. Summing over~$x,y \in M$, the ansatz for the ``action'' is
\begin{align*}%\label{c:pfp, (3.5.3)}
	S = \sum_{x,y \in M} \L [P(x,y) P(y,x)] \:.
\end{align*} 
An obvious way to form a \emph{positive} functional is then to add up the absolute values of the eigenvalues of~$A_{xy}$ (counted with multiplicities). Denoting the eigenvalues of~$A_{xy}$ by~$\lambda_i$, the resulting quantity is referred to as ``spectral weight,'' 
\begin{align*}%\label{c:(3.5.8)}
	|A| = \sum_{i=1}^{2N} |\lambda_i| \:. 
\end{align*}
It turns out that it is preferable to formulate a variational principle which aspires to equalize the absolute values of all eigenvalues. This can be accomplished by combining the expressions~$|A^2|$ and~$|A|^2$. More precisely, it is reasonable to minimize~$|A^2|$, keeping~$|A|^2$ fixed. This is the motivation for considering the variational principle
\begin{align}\label{c:pfp, (3.5.9)}
	\text{minimize} \qquad S[P] = \sum_{x,y \in M} \left|A_{xy}^2 \right|
\end{align}
by varying~$P$ in the class of linear operators~$\LL(H)$  
under the constraint
\begin{align}\label{c:pfp, (3.5.10)}
	T[P] := \sum_{x,y \in M} \left| A_{xy}\right|^2 = C 
\end{align}
for some constant~$C > 0$. The corresponding \emph{Euler-Lagrange equations} read
\begin{align*}%\label{c:pfp, (3.5.20)}
	[P,Q] = 0 
\end{align*}
for some operator~$Q$ (cf.~\cite[eqs.~(3.5.9)--(3.5.10) and~(3.5.20)--(3.5.21)]{pfp}).  
The variational principle~\eqref{c:pfp, (3.5.9)}--\eqref{c:pfp, (3.5.10)} was first analyzed mathematically in~\cite{discrete}. In order to clarify its connection to the causal action principle in the theory of causal fermion systems, it is instructive to recall an existence result in~\cite{discrete}. More explicitly, in view of~\cite[Theorem~2.8]{discrete}  
there exists a minimizer
of the variational principle~\eqref{c:pfp, (3.5.9)}--\eqref{c:pfp, (3.5.10)} in the class of operators~$P \in \LL(H)$ with the property that~$\tr(P) = f$, $\rank(P) \le f$ and~$(-P)$ is positive in the sense that
\begin{align}\label{c:discrete, (11)}
	{<u \mid (-P) u>} \ge 0 \qquad \text{for all~$u \in H$} \:. 
\end{align}
The fact that~$P$ has a negative definite image makes it possible to introduce a Hilbert space~$(\H, \langle \, . \, | \, . \,\rangle_{\H})$ by setting~$\langle \, . \, | \, . \,\rangle_{\H} = {<. \mid (-P) \, . >}$ and dividing out the null space. This construction, which was first outlined in~\cite{rrev}, gives rise to an underlying Hilbert space structure. For similar constructions we refer to~\cite{langerma}.

\subsection{Connection to Causal Fermion Systems}\label{c:S connection CFS} 
According to~\cite[Preface to the second online edition]{pfp}, 
the step from indefinite inner product spaces to Hilbert spaces is regarded as ``maybe the most important change in the mathematical setup.''  
Actually, this transition from indefinite inner product spaces to Hilbert spaces is closely related to introducing a  ``regularization,'' one of the fundamental concepts of causal fermions systems. In order to outline this transition in a mathematical convincing way, let us explain the idea of introducing a ``regularization''  
in some more detail. The following explanations are due to~\cite[Section~1.2]{cfs} and~\cite[Section~1.5]{pfp}.  

We let~$\scrM$ be Minkowski space, endowed with the standard volume measure~$d\mu$. We define the \emph{spinor bundle}~$S\scrM$ as a vector bundle over~$\scrM$ with fiber~$\C^4$. 
Sections in the spinor bundle are called \emph{spinors} or wave functions. 
By~$C(\scrM, S\scrM)$ we denote the continuous Dirac wave functions (that is, the continuous sections of the spinor bundle, not necessarily solutions of the Dirac equation).  
The spinor space at a point~$x \in \scrM$ is denoted by~$S_x\scrM$, so that a wave function~$\psi$ takes values in
\begin{align*}
	\psi(x) \in S_x\scrM \simeq \C^4 \:. 
\end{align*}
The spinor space at~$x$ is endowed with an indefinite inner product of signature~$(2,2)$, which  
we denote by~$\prec . \mid . \succ$.  
We then consider solutions of the Dirac equation  
\begin{align*}%\label{c:cfs, (1.2.1)}
	(i\slashed{\partial} - m) \psi = 0 \:,
\end{align*} 
where~$\slashed{\partial} = \gamma^j \partial_j$. For a solution~$\psi$, the function~$(\overline{\psi}\gamma^0 \psi)(t, \vec{x})$ has the interpretation as the probability density of the Dirac particle at time~$t$ to be at the position~$\vec{x}$.  
The spatial integral of this probability density is time independent, and the corresponding bilinear form corresponding to this probability integral gives the scalar product (cf.~\eqref{c:pfp, (1.2.17)})
\begin{align}\label{c:cfs, (1.2.2)}
	(\psi \mid \phi) = \int_{\R^3} (\overline{\psi} \gamma^0 \phi)(t, \vec{x}) \: d^3x \:. 
\end{align} 
In order to ensure that the integral in~\eqref{c:cfs, (1.2.2)} is well-defined and finite, we first consider solutions which at time~$t$ are smooth and have compact support.  
Taking the completion, the solution space becomes a separable Hilbert space (see e.g.~\cite[Proposition~8.17]{folland}). Next, we choose~$\H$ as a closed subspace of this Hilbert space with the induced scalar product~$\langle \, . \, | \, . \,\rangle_{\H} := ( \, . \, | \, . \,)|_{\H \times \H}$. Then~$\H$ is again a separable Hilbert space. However, as Dirac solutions~$\psi, \phi \in \H$ are in general not continuous, 
the pointwise evaluation
\begin{align*}
	- \prec{\psi(x)} \mid \phi(x) \succ \qquad \text{for all~$\psi, \phi \in \H$}
\end{align*}
at~$x \in \scrM$ makes in general no mathematical sense. 
This is the reason for introducing an \emph{ultraviolet regularization}. Such regularizations are 
performed most conveniently with so-called ``regularization operators,''
which we now define. Consider a family of linear operators~$({\mathfrak{R}}_\varepsilon)_{\varepsilon \in (0, \varepsilon_{\max})}$ 
with~$\varepsilon_{\max} > 0$ 
which map~$\H$ to the continuous wave functions,
\begin{align}\label{c:(regularization function)}
	{\mathfrak{R}}_\varepsilon \::\: \H \rightarrow C(\scrM, S\scrM) \:.
\end{align} 
Whenever the technical conditions~\cite[eqs.~(1.2.6)--(1.2.8)]{cfs} in~\cite[Definition~1.2.3]{cfs} 
hold, 
the family~$({\mathfrak{R}}_\varepsilon)_{\varepsilon \in (0, \varepsilon_{\max})}$ is said to be a family of {\emph{regularization operators}}.  
A regularized wave function~$\mathfrak{R}_{\varepsilon}\psi$, however, need not again be a solution of the Dirac equation. 
Given a family of regularization operators~$({\mathfrak{R}}_\varepsilon)_{\varepsilon \in (0, \varepsilon_{\textup{max}})}$ for some~$\varepsilon_{\textup{max}} > 0$,  
the connection to causal fermion systems is accomplished as follows.  
Fixing~$\varepsilon \in (0, \varepsilon_{\max})$,  
for any~$x \in \scrM$ we consider the well-defined bilinear form
\begin{align*}%\label{c:cfs, (1.2.11)}
	b_x^{\varepsilon} \::\: \H \times \H \rightarrow \C\:, \qquad
	b_x^{\varepsilon}(u, v) = - \prec {({\mathfrak{R}}_\varepsilon \,u)(x)} \mid ({\mathfrak{R}}_\varepsilon \,v)(x) \succ \:.
\end{align*} 
Making use of the Fr{\'e}chet-Riesz theorem (see for instance~\cite[Section~6.3]{lax}),
there is a unique bounded linear operator~$F^{\varepsilon}(x) : \H \to \H$ such that
\begin{align*}
	\la u \mid F^{\varepsilon}(x) \, v \ra_\H = b_x^{\varepsilon}(u, v) = - \prec (\mathfrak{R}_{\varepsilon} \,u)(x) \mid  (\mathfrak{R}_{\varepsilon} \,v)(x) \succ \qquad \text{for all~$u,v \in \H$} \:. 
\end{align*}
Taking into account that the inner product~$\prec . \mid . \succ$ on the Dirac spinors at~$x$ has signature~$(2,2)$, 
we infer that~$F^{\varepsilon}(x)$ has at most~$2$ positive and at most~$2$ negative eigenvalues. 
Defining~$\F \subset \LL(\H)$ as the set of self-adjoint operators with at most~$2$ positive and at most~$2$ negative eigenvalues, we thus obtain a mapping
\begin{align*}
	F^{\varepsilon} : \scrM \to \F \:.
\end{align*}
The operators~$F^{\varepsilon}(x)$ are referred to as \emph{local correlation operators} (as they furnish a connection between the underlying measure space and the structure~$\F$ which plays a crucial role in the setting of causal fermion systems). It is important to observe that on~$\F$ we are given a measure~$d\rho$ (referred to as ``universal measure'') defined by
\begin{align*}
	d\rho^{\varepsilon} := F^{\varepsilon}_{\ast} \: d\mu
\end{align*}
as the push-forward measure of~$d\mu$ under~$F^{\varepsilon}$ (more precisely, for any Borel set~$\Omega \subset \F$ we have~$(F^{\varepsilon})_{\ast}\mu(\Omega) := \mu((F^{\varepsilon})^{-1}(\Omega))$).  
The resulting structure~$(\H, \F, d\rho^{\varepsilon})$ is a causal fermion system of spin dimension~$n=2$ (see Definition~\ref{c:defparticle}), and the support of the universal measure~$d\rho^{\varepsilon}$ is referred to as \emph{spacetime}. Note that in the context of causal fermion systems, the physical picture behind the ultraviolet regularization is that only the \emph{regularized} objects are regarded as the fundamental physical objects.  
In particular, the regularization has a physical significance as describing the microscopic structure of physical spacetime. Removing the regularization by taking the limit~$\varepsilon \searrow 0$ in a certain way (see~\cite[Section~3.5]{cfs}) is referred to as ``continuum limit.''

In this way, according to~\cite[Preface to the second online edition]{pfp}, 
when working out the existence theory of the variational principle~\eqref{c:pfp, (3.5.9)}--\eqref{c:pfp, (3.5.10)},  
``it became clear that instead of using the kernel of the fermionic projector, the causal action principle can be formulated equivalently in terms of the local correlation operators,'' thus giving rise to the causal action principle as stated in Section~\ref{c:Section intro-cfs}.

\section{The Theory of Causal Fermion Systems}\label{c:Section CFS}
Having derived the specific form of the causal action principle as outlined in Section~\ref{c:Section intro-cfs}, it remains to clarify the connection of causal fermion systems to the Standard Model of particle physics as well as general relativity. This is precisely the aim of this section. In order to clarify the underlying ideas, it seems best to first outline the general strategy (\S \ref{c:S general strategy}). 

\subsection{The General Strategy}\label{c:S general strategy}
The objective of this subsection is to present the general strategy behind the theory of causal fermion systems. 
It is worth noting that the basic procedure outlined in this subsection is closely related to the ideas presented in~\S \ref{c:S gauge freedom}.   
In short, the main ingredient of the theory can be regarded a class~$(M, g, \B)$ of four-dimensional, globally hyperbolic Lorentzian manifolds~$M$ (for details see~\cite{beem}) endowed with Lorentzian metric~$g$ and an external potential~$\B$ on~$M$ with appropriate properties. The underlying idea is that~$M$ models (curved) physical spacetime according to general relativity;  
the corresponding free Dirac equation reads (for details see~\S \ref{c:S qft})  
\begin{align}\label{c:(Dirac flat)}
	(i \underline{\gamma}^{\mu}(x) \nabla_{\mu} - m) \psi(x) = 0 
\end{align} 
with spacetime-dependent matrices~$\underline{\gamma}^{\mu}$ and covariant derivative~$\nabla_{\mu}$.  
Introducing an interaction~$\B$, the resulting Dirac equation takes the form
\begin{align}\label{c:(Dirac curved)}
	(i \underline{\gamma}^{\mu}(x) \nabla_{\mu} + \B(x) - m) \psi(x) = 0 \:.
\end{align}
Roughly speaking, the Dirac equation~\eqref{c:(Dirac curved)} can be regarded as the starting point for the theory of causal fermion systems. Albeit the theory is worked out only in Minkowski space, the general strategy is based on the following steps: The starting point of the theory is to consider a solution of the unperturbed Dirac equation~\eqref{c:(Dirac flat)}. The next step is to introduce a potential~$\B$ and to make use of the so-called ``causal perturbation expansion'' in order to construct a solution~$\psi$ of the Dirac equation~\eqref{c:(Dirac curved)}. Following Dirac's concept of describing the vacuum (in presence of an external potential~$\B$) by means of a ``completely filled Dirac sea,'' by introducing particles and anti-particles it is possible to modify the solution~$\psi$ of the Dirac equation~\eqref{c:(Dirac curved)}. As a consequence, the Dirac equation~\eqref{c:(Dirac curved)} will no longer be satisfied. Thus in order for~\eqref{c:(Dirac curved)} to hold again, it is necessary to also replace the potential~$\B$ by some other potential~$\tilde{\B}$. The resulting potential~$\tilde{\B}$ is then interpreted as ``interaction'' between fermions.

In order to establish a connection to the Standard Model of particle physics, it is reasonable to extend the previous considerations by introducing one Dirac sea for each of the fundamental fermions compiled in Table~\eqref{c:table}. This is done by introducing the so-called ``auxiliary fermionic projector,'' which is defined as the direct sum of the Dirac seas corresponding to the fundamental fermions. Employing the ansatz of ``massive neutrinos,''  
it is necessary to break the chiral symmetry; to this end,  
massless right-handed neutrinos are taken into account by adding one more direct summand to the auxiliary fermionic projector. In order to enter the setting of causal fermion systems, a ``regularization'' is inserted. Introducing particles and anti-particles, by forming the so-called ``sectorial projection'' one finally obtains the ``fermionic projector'' (for further details see~\cite[Remark~4.2.3]{cfs}).

Motivated by the variational principle~\eqref{c:pfp, (3.5.9)}--\eqref{c:pfp, (3.5.10)}, the underlying picture is that the fermionic projector  
is supposed to be of physical relevance if it is a minimizer of the causal action principle. In this way, the resulting Euler-Lagrange (EL) equations give rise to conditions for the fermionic projector.  
It is crucial to observe that the fermionic projector obtained by the just-mentioned method essentially depends on the external potential~$\B$ and the underlying spacetime~$(M,g)$. Accordingly, the EL equations pose conditions on ``admissible'' manifolds~$(M,g)$ and potentials~$\B$.  
In this way, only those external potentials~$\B$ and manifolds~$(M,g)$, for which the EL equations are satisfied, are considered as physically relevant; the resulting potentials are interpreted as ``bosonic interactions'' between fundamental fermions. In this way,  
the causal action principle  
allows to recover the interactions of the Standard Model and gives rise to the Einstein field equations of general relativity in a suitable limiting case (the so-called ``continuum limit''). More precisely, the basic idea to obtain a connection to the Standard Model and general relativity is to rewrite the resulting EL equations as corresponding to an ``effective action''~$\Sact_{\textup{eff}}$ of the form (cf.~\cite[eq.~(5.4.1)]{cfs})
\begin{align*}%\label{c:cfs, (5.4.1)}
	\Sact_\text{eff} = \int_\scrM \left( \LDirac + \LYM + \LEH \right) \sqrt{-\det g}\, d^4x 
\end{align*} 
for a suitable choice of the Dirac Lagrangian~$\LDirac$, the Yang-Mills Lagrangian~$\LYM$ and the Einstein-Hilbert Lagrangian~$\LEH$.  
Moreover, under appropriate assumptions one recovers the effective gauge group of the Standard Model,  
\begin{align*}%\label{c:cfs, (5.3.1)}
	\G = \U(1) \times \SU(2) \times \SU(3) \:.
\end{align*} 
The general strategy how to obtain these results shall be outlined in what follows.

\subsection{The Kernel of the Fermionic Projector}\label{c:S kernel}
Before entering the theory of causal fermion systems in more detail, let us briefly summarize some conceptual ideas which enter the theory from the very beginning. To begin with, taking into account that the causal action principle evolved from the principle of the fermionic projector~\cite{pfp}, the following physical principles implicitly enter the theory of causal fermion systems:  
Einstein's principle of relativity, the Pauli exclusion principle and, of course, the principle of least action. Moreover, the principle of causality is built in in such a way that spacelike separated points do not enter the Lagrangian and thus the causal action. Moreover, the local gauge principle arises as explained in~\S \ref{c:S gauge freedom}. 

Next, only the fermionic states are considered as the basic physical objects. The bosonic fields, on the other hand, are merely regarded as auxiliary objects used for describing the behavior of the fermionic states. For this reason, the fundamental fermions summarized in Table~\eqref{c:table} will play a central role in what follows. Motivated by the renormalization program, the concept of ``bare'' or ``naked'' masses and coupling constants is employed throughout; the ``physical'' masses which can be measured in experiments are obtained from the naked masses by self-interaction.

In order to take this fundamental character of fermions into account, the theory of causal fermion systems essentially employs the concept of a ``Dirac sea'' which  
was introduced by Paul Dirac in his paper~\cite{dirac1930theory} (cf.~\S \ref{c:S qm}). Based on the Pauli exclusion principle, Dirac assumed that 
\begin{quote}
	``(...) \emph{all the states of negative energy are occupied except perhaps a few of small velocity.} (...) \emph{Only the small departure from exact uniformity, brought about by some of the negative-energy states being unoccupied, can we hope to observe.} (...) We are therefore led to the assumption that \emph{the holes in the distribution of negative-energy electrons are the [positrons].}''
\end{quote}
Dirac made this picture precise in his paper~\cite{dirac1934discussion} by introducing a relativistic density matrix~$R(t, \vec{x}; t', \vec{x}')$ with $(t, \vec{x}), (t', \vec{x}') \in \R \times \R^3$ defined by
\begin{align*}
	R(t, \vec{x}; t', \vec{x}') = \sum_{\text{$l$ occupied}} \Psi_l(t, \vec{x}) \: \overline{\Psi_l(t', \vec{x}')} \:.
\end{align*}
In analogy to Dirac's original idea, in~\cite{external} the kernel of the fermionic projector is introduced as the sum over all occupied wave functions
\begin{align*}
	P(x,y) = - \sum_{\text{$l$ occupied}} \Psi_l(x) \: \overline{\Psi_l(y)} 
\end{align*}
for spacetime points~$x,y \in \scrM$ (also see~\cite{finster2011formulation}).  
A straightforward calculation shows that (cf.~\cite[\S 4.1]{finster+grotz} and~\cite[Section~3.8]{cfs}), up to an irrelevant constant, the relativistic density matrix~$R(t, \vec{x}; t', \vec{x}')$ coincides precisely with kernel of the fermionic projector~$P(x,y)$, 
\begin{align}\label{c:(completely filled sea)}
	P(x,y) = \int_{\hscrM} \frac{d^4k}{(2\pi)^4} \: (\slashed{k} + m) \: \delta(k^2-m^2) \: \Theta(-k^0) \: e^{-ik(x-y)} \:, 
\end{align}
where~$\delta$ is Dirac's delta distribution and~$\Theta$ denotes the Heaviside function, the slash in~$\slashed{k} = k^j\gamma_j$ denotes contraction with the Dirac matrices~$\gamma_j$ ($j=0, \ldots, 3$) and the expression~$k(x-y)$ is a short notation for the Minkowski inner product~$k_j(x-y)^j$.  
The quantity~$P(x,y)$ in~\eqref{c:(completely filled sea)} is called the (unregularized) \emph{kernel of the fermionic projector of the vacuum.}  
We also refer to the expression~\eqref{c:(completely filled sea)} as a \emph{completely filled Dirac sea}; it can be thought of as describing the ``vacuum'' with respect to \emph{one} kind of elementary particles of mass~$m$. 

Note that particles are obtained by adding states to a completely filled Dirac sea, whereas anti-particles are described by removing states therefrom.  
This picture of a Dirac sea is made precise by considering 
the free Dirac equation in Minkowski space, 
\begin{align}\label{c:cfs, (1.2.1')}
	(i\slashed{\partial} - m) \psi = 0 \:.
\end{align} 
Assume that~$\H$ is a separable Hilbert space of solutions of the Dirac equation~\eqref{c:cfs, (1.2.1')},  
and let~$(u_{\ell})_{\ell \in \N}$ be an orthonormal basis of~$\H$. 
Dirac's original idea is then specified  
with the help of regularization operators~\eqref{c:(regularization function)} by introducing    
the regularized kernel of the fermionic projector for all~$x,y \in \scrM$ as (cf.~\cite[eq.~(1.2.19)]{cfs}) 
\begin{align*}%\label{c:cfs, (1.2.19)}
	P^\varepsilon(x,y) = -\sum_\ell \big({\mathfrak{R}}_\varepsilon u_\ell \big)(x)\:
	\overline{\big({\mathfrak{R}}_\varepsilon u_\ell \big)(y)} = -\sum_\ell  |\big({\mathfrak{R}}_\varepsilon u_\ell \big)(x) \succ \:
	\prec {\big({\mathfrak{R}}_\varepsilon u_\ell \big)(y)} | \:. 
\end{align*} 
In the limit~$\varepsilon \searrow 0$, the regularized kernel of the fermionic projector~$P^\varepsilon(x,y)$ 
converges as a bi-distribution to the unregularized kernel~$P(x,y)$ defined by
\beq \label{c:cfs, (1.2.20)}
P(x,y) := -\sum_\ell u_\ell(x)\: \overline{u_\ell(y)} = -\sum_\ell | u_\ell(x) \succ \: \prec {u_\ell(y)} |  \:, 
\eeq  
as is made mathematically precise in~\cite[Proposition~1.2.7]{cfs}. Furthermore,  
if~$\H$ is the subspace of the solution space of the Dirac equation~\eqref{c:cfs, (1.2.1')}
which is
spanned by all negative-frequency solutions, then the unregularized kernel of the fermio\-nic projector
as defined by~\eqref{c:cfs, (1.2.20)} is the tempered bi-distribution (see~\cite[Lemma~1.2.8]{cfs}) 
\begin{align*}%\label{c:cfs, (1.2.23)}
	P(x,y) = \int \frac{d^4k}{(2 \pi)^4}\:(\slashed{k}+m)\: \delta(k^2-m^2)\: \Theta(-k_0)\: e^{-ik(x-y)} \:.
\end{align*}

In order to appreciate the relevance of the fermionic projector, it is instructive to outline the so-called ``external field problem'' which we recall now.

\subsection{The External Field Problem}\label{c:S external field problem} 
In order to understand the basic difficulty of introducing interactions into the Dirac equation in more detail,
let us consider an interaction described by an external field~$\B$ in Minkowski space. In this case, the Dirac wave functions are supposed to obey the modified Dirac equation
\begin{align*}%\label{c:pfp, (2.1.1)}
	(i\slashed{\partial} + \B - m)\psi = 0 \:, 
\end{align*}
where~$\B$ is assumed to be a smooth potential with suitable decay properties for large times and at spatial infinity (for details we refer to~\cite[Section~2.1]{pfp} and~\cite[\S 2.1.2]{cfs}). Whenever the potential~$\B$ is \emph{static} (that is, time-independent), one can separate the time-dependence of the wave function with a plane wave ansatz (cf.~\cite[eq.~(2.1.2)]{pfp}), 
\begin{align}\label{c:pfp, (2.1.2)}
	\psi(t, \vec{x}) = e^{-i\omega t} \psi_{\omega}(\vec{x}) \:, 
\end{align}
where~$\omega$ 
has the interpretation  
as the energy of the solution. In this way, the sign of~$\omega$ gives rise to a \emph{canonical splitting} of the  
solution space of the Dirac equation into subspaces of positive and negative energy, respectively. Unfortunately, the situation becomes much more difficult in the case that~$\B$ is \emph{time-dependent}. The basic difficulty in this setting is that the separation ansatz~\eqref{c:pfp, (2.1.2)} no longer works. More precisely, one is no longer given a canonical splitting of the solution space of the Dirac equation. Therefore it is not clear which solutions can be interpreted as ``negative-energy solutions'' and thus correspond to the anti-particle states. This difficulty is referred to as \emph{external field problem}. It is a common belief that in the presence of a general time-dependent external potential there exists no longer a \emph{canonical} decomposition of the solution space into two subspaces, 
and the resulting arbitrariness in decomposing the solution space into two subspaces is often associated to an ``observer.''  
In this way, the interpretation of particles and anti-particles becomes \emph{observer-dependent.} 
Nevertheless, one of the fundamental results in the theory of causal fermion systems is that \emph{even in the presence of a time-dependent external potential there is a canonical decomposition of the solution space into two subspaces}. This method allows us to construct solutions of the modified Dirac equation~\eqref{c:cfs, (2.1.1)} under suitable assumptions on~$\B$, as shall be outlined in the following. 

Starting point for the following constructions is a completely filled Dirac sea~\eqref{c:(completely filled sea)} of mass~$m$ (where the superscript ``vac'' clarifies Minkowski vacuum),    
\beq \label{c:cfs, (2.1.1)}
P^\text{vac}_m(x,y) = \int \frac{d^4k}{(2 \pi)^4}\: (\slashed{k}+m)\: \delta(k^2-m^2)\: \Theta(-k^0) \: e^{-ik(x-y)} \:, 
\eeq
which is referred to as \emph{kernel of the fermionic projector in the vacuum}. 
The basic idea in the theory of causal fermion systems in order to resolve the external field problem is to split up the
fermionic projector of the vacuum in such a way that
\beq
P^\text{vac}_m(x,y) = \frac{1}{2}\,\big(p_m(x,y)-k_m(x,y)\big) \:, \label{c:cfs, (2.1.6)}
\eeq
where~$p_m(x,y)$ and~$k_m(x,y)$ are the distributions
\begin{align*}
	p_m(x,y) &= \int \frac{d^4k}{(2\pi)^4} (\slashed{k} + m) \, \delta(k^2-m^2) \, e^{-ik(x-y)} \:, % \label{c:pfp, (2.2.4)} 
	\\
	k_m(x,y) &= \int \frac{d^4k}{(2\pi)^4} (\slashed{k} + m) \, \delta(k^2-m^2) \, \epsilon(k^0) \, e^{-ik(x-y)} \:, % \label{c:pfp, (2.2.5)}
\end{align*}
and~$\epsilon : \R \to \{-1, 1 \}$ is the sign function defined by~$\epsilon(x) = 1$ for~$x \ge 0$ and~$\epsilon(x) = -1$ otherwise. Remarkably, the distributions~$p_m$ and~$k_m$ can be rewritten in terms of Green's functions, as we now explain.  
For every~$x \in \scrM$ let~$J_x^{\lor}$ and~$J_x^{\land}$ denote the points in the \emph{causal future} respectively \emph{past} of~$x$, 
\begin{align*}
	J_x^\vee &= \{ y \in M \,|\, (y-x)^2 \geq 0,\;
	(y^0-x^0) \geq 0 \} \:, \\
	J_x^\wedge &= \{ y \in M \,|\, (y-x)^2 \geq 0, \;
	(y^0-x^0) \leq 0 \} \:. 
\end{align*}
Denoting Dirac's delta distribution in Minkowski space by~$\delta^4$, due to~\cite[Chapter~2]{cfs} the \emph{advanced} and \emph{retarded Green's functions} are uniquely defined by
\[ %\label{c:rGf}
(i \Pdd - m) \: {s}^\vee_m(x,y) = \delta^4(x-y) \qquad \text{with} \qquad
{\mbox{supp}}\: {s}^\vee_m(x,.) \subset J_x^\vee \]
and 
\[ %\label{c:rGf}
(i \Pdd - m) \: {s}^\wedge_m(x,y) = \delta^4(x-y) \qquad \text{with} \qquad
{\mbox{supp}}\: {s}^\wedge_m(x,.) \subset J_x^\wedge \:, \]
respectively. Then the \emph{causal fundamental solution}~$k_m$ can be recovered by 
\beq \label{c:cfs, (2.1.14)}
k_m(x,y) = \frac{1}{2\pi i} \left( s^{\vee}(x,y)-s^{\wedge}(x,y) \right) \:. 
\eeq
Introducing~$J_x := J_x^\vee \cup J_x^\wedge$, the distribution~$k_m$ is causal in the sense that
\begin{align*} %\label{c:cfs, (2.1.15)}
	\supp k_m(x,.) \subset J_x \:. 
\end{align*} 
Making use of the identity (see~\cite[eq.~(2.1.18) and~Lemma~2.1.3]{cfs})
\begin{align*} %\label{c:cfs, (2.1.18)}
	k_m\,k_{m'}=\delta(m-m')\:p_m \;, 
\end{align*} 
one can deduce~$p_m$ from~$k_m$.  
In this way, the Green's functions~$s_m^{\lor}$ and~$s_m^{\land}$ allow us to introduce the fermionic projector of the vacuum~$P_m^{\textup{vac}}$ by~\eqref{c:cfs, (2.1.6)}. 

The idea is to extend these constructions to the interacting situation, that is, to the Dirac equation in presence of an external potential~$\B$. According to~\cite[\S 2.1.4]{cfs}, the advanced and retarded Green's functions are uniquely defined even in the presence of an external potential~\eqref{c:cfs, (2.1.1)}, 
\begin{align}\label{c:cfs, (2.1.5)}
	(i\slashed{\partial} + \B - m) \psi(x) = 0 \:. 
\end{align}
More explicitly, the \emph{advanced} and \emph{retarded Green's functions}~$\tilde{s}_m^{\lor}$ and~$\tilde{s}_m^{\land}$ for the Dirac equation~\eqref{c:cfs, (2.1.5)} are characterized by
\begin{align*}
	(i \Pdd + \B - m) \: \tilde{s}^\vee_m(x,y) &= \delta^4(x-y) \qquad \text{with} \qquad
	{\mbox{supp}}\: \tilde{s}^\vee_m(x,.) \subset J_x^\vee \qquad \text{and}  \\
	(i \Pdd + \B - m) \: \tilde{s}^\wedge_m(x,y) &= \delta^4(x-y) \qquad \text{with} \qquad
	{\mbox{supp}}\: \tilde{s}^\wedge_m(x,.) \subset J_x^\wedge \:,  
\end{align*}
respectively. The crucial point in what follows is that the Green's functions~$\tilde{s}_m^{\lor}$ and~$\tilde{s}_m^{\land}$ can be uniquely expressed in terms of~${s}_m^{\lor}$ and~${s}_m^{\land}$. More precisely, one has the following unique perturbation series (cf.~\cite[eq.~(2.1.25)]{cfs})
\beq \tilde{s}_m^{\vee}=\sum_{n=0}^{\infty} \big( -s_m^{\vee}\, \mathscr{B} \big)^n \,s_m^{\vee}\;, \qquad
\tilde{s}_m^{\wedge}=\sum_{n=0}^{\infty} \big( -s_m^{\wedge}\, \mathscr{B} \big)^n \,s_m^{\wedge}\:.
\label{c:cfs, (2.1.25)}
\eeq
Its summands are referred to as ``Feynman diagrams'' (of~$n^{\textup{\tiny{th}}}$ order).\footnote{For the connection to Feynman diagrams in physics see~\cite[\S 3.8.4]{cfs}.} 

Having derived a perturbation series for the causal Green's functions,  
the \emph{causal fundamental solution} is defined in generalization of~\eqref{c:cfs, (2.1.14)} by
\begin{align*}%\label{c:cfs, (2.1.26)}
	\tilde{k}_m:=\frac{1}{2\pi i}(\tilde{s}_m^{\vee}-\tilde{s}_m^{\wedge}) \;.  
\end{align*} 
Omitting the mass indices,  
for any complex number~$\lambda$ in the resolvent set of~$k$ and~$\tilde{k}$, 
the unperturbed resolvent~$R_{\lambda}$  
and the (perturbed) resolvent~$\tilde{R}_{\lambda}$ are introduced by
\begin{align*} %\label{c:cfs, (2.1.56)}
	R_{\lambda} := (k-\lambda)^{-1} \qquad \text{and} \qquad \tilde{R}_\lambda = \big( \tilde{k} - \lambda \big)^{-1} \:,
\end{align*} 
respectively.
%(also see Appendix~\ref{i:Appendix Polish}).  
Rewriting 
the causal fundamental solution~$\tilde{k}$ as
\begin{align*} %\label{c:cfs, (2.1.55)}
	\tilde{k} = k + \Delta k \:,
\end{align*} 
the resolvent~$\tilde{R}_\lambda$
can be written as a Neumann series (see e.g.~\cite[Satz~II.1.11]{werner}),
\begin{align*} %\label{c:tR}
	\tilde{R}_\lambda = (k - \lambda + \Delta k)^{-1}
	= (1 + R_\lambda \cdot \Delta k)^{-1} \cdot R_\lambda 
	= \sum_{n=0}^\infty (-R_\lambda \cdot \Delta k)^n \cdot R_\lambda \:.
\end{align*} 
Furthermore, 
the unperturbed resolvent can be expressed in terms of~$p$ and~$k$, 
\begin{align*} %\label{c:cfs, (2.1.57)}
	R_\lambda = \frac{p+k}{2} \left( \frac{1}{1-\lambda} \right) + \frac{p-k}{2} \left( \frac{1}{-1-\lambda} \right)
	- \frac{\Id-p}{\lambda} \:.
\end{align*} 
Choosing a contour~$\Gamma_- \subset \C$ which encloses the point~$-1$ in counter-clockwise direction and does not enclose the points~$1$ and~$0$, the fermionic projector~$P^{\textup{sea}}$ is defined by
\begin{align*}%\label{c:cfs, (2.1.64)}
	P^{\sea} = -\frac{1}{2 \pi i} \ointctrclockwise_{\Gamma_-} (-\lambda)\: \tilde{R}_\lambda\: d\lambda \:. 
\end{align*} 
According to~\cite[Proposition~2.1.7]{cfs}, it has the desirable property that 
\begin{align} \label{c:cfs, (2.1.65)}  
	(i \Pdd+\B-m)\, P^\text{\rm{sea}} &= 0 \:.
\end{align}
Following the previous constructions, the fermionic projector~$P^{\textup{sea}}$ only depends on~$\B$ and the Green's functions~$s_m^{\lor}$, $s_m^{\land}$. Next, particles and anti-particles are introduced as follows.  
Taking the closure of the solution space of the Dirac equation~\eqref{c:cfs, (2.1.5)}  
with respect to the inner product~\eqref{c:cfs, (1.2.2)},  
one obtains a Hilbert space~$\H_m$ with the induced scalar product~$(. \, | \, .)_m$. Denoting by~$\scrN_t := \{t\} \times \R^3 \subset \scrM$ the spatial hyperplane at time~$t$, one can define the 
mapping~$\Pi^\sea \::\: C^\infty_0(\scrN_t, S\scrM) \rightarrow C^\infty(\scrM, S\scrM)$ by
\beq \label{c:cfs, (2.1.71)}
\begin{split} 
	&(\Pi^\sea \psi)(x) = -2 \pi \int_{\R^3} P^\sea \big( x, (t, \vec{y}) \big)\: \gamma^0\, \psi(\vec{y})\: d^3 y 
\end{split}
\eeq
(where by~$C^\infty_0(\scrN_t, S\scrM)$ and~$C^\infty(\scrM, S\scrM)$ we denote the set of smooth wave functions on~$\scrN_t$ with compact support and the set of smooth wave functions on~$\scrM$, respectively). 
Denoting the orthogonal projection onto a subspace~$A \subset \H_m$ by~$\Pi_A : \H_m \to \H_m$, the
operator~$\Pi^{\sea}$ 
can be extended by continuity to a projection operator on~$\H_m$, i.e. 
\[ \Pi^\sea : \H_m \rightarrow \H_m \qquad \text{with} \qquad (\Pi^\sea)^* = \Pi^\sea = (\Pi^\sea)^2 \:. \]
This allows us to introduce particles and anti-particles by 
\[ \Pi := \Pi^\sea + \Pi_{\text{span}(\psi_1,\ldots, \psi_{\np})} - \Pi_{\text{span}(\phi_1,\ldots, \phi_{\na})} \]
with~$\np, \na \in \N_0$, 
provided that the functions~$\phi_l \in \H_m$ ($l =1, \ldots, \np$) lie in the image of~$\Pi^\sea$,  
while the vectors~$\psi_k \in \H_m$ ($k=1, \ldots, \na$) are in the orthogonal
complement of the image of~$\Pi^\sea$. 
Employing the normalization conditions in~\cite[eq.~(2.1.72)]{cfs},  
the operator~$\Pi$ reads 
\[ \Pi \psi := \Pi^\sea \psi + \frac{1}{2 \pi} \sum_{k=1}^{\np} \psi_k \: (\psi_k | \psi)_m
- \frac{1}{2 \pi} \sum_{l=1}^{\na} \phi_l\: (\phi_l | \psi)_m \:. \]
The projection operator~$\Pi$ can be written in the form~\eqref{c:cfs, (2.1.71)} with 
the distribution\footnote{For the normalization of the particle and anti-particle states see~\cite[Section~2.8]{pfp} or~\cite[\S 3.4.3]{cfs} and~\cite{norm}.} 
\begin{align*}%\label{c:(Pm)}
	P_m(x,y) = P_m^\text{vac}(x,y)
	- \frac{1}{2 \pi} \sum_{k=1}^{\np} \psi_k(x) \overline{\psi_k(y)}
	+ \frac{1}{2 \pi} \sum_{l=1}^{\na} \phi_l(x) \overline{\phi_l(y)}\:,
\end{align*} 
thus allowing for~$\np$ particles and~$\na$ anti-particles of mass~$m$ (with~$\np, \na \in \N_0$). 

\subsection{The Auxiliary Fermionic Projector}\label{c:S auxiliary}
The previous considerations can be extended to several Dirac seas of different masses by forming the so-called ``auxiliary fermionic projector.'' In order to recover the Standard Model of elementary particle physics,   
one clearly needs to take into account all fundamental fermions as compiled in Table~\eqref{c:table}. 
As outlined in~\S \ref{c:S elementary particles}, the fundamental fermions consist of quarks and leptons, each of which obeys the corresponding Dirac equation. Quarks come in three colors (red, green, blue), each quark color as well as the leptons come in three generations, and each generation consists of a (weak) isospin doublet (isospin up and down). Accordingly, there are four ``blocks'' (three quark blocks and one lepton block), each of which consists of two ``sectors'' (corresponding to the isospin up and down component of the electroweak theory, respectively), and each sector comprises three generations. Furthermore, according to~\S \ref{c:S elementary particles}, the leptons split into a charged and a neutrino component, whereas all quarks carry an electric charge. In this way, all fundamental fermions either belong to a charged or neutrino component. We also point out that the theory of causal fermion systems does not distinguish the 
chirality of neutrinos in the lepton block and even allows for massive neutrinos.  
In order to take these facts into account, the (bare) mass of the~$i^{\textup{\tiny{th}}}$ neutrino generation will be denoted by~$\tilde{m}_i$, whereas the (bare) mass of the~$i^{\textup{\tiny{th}}}$ generation in the charged sector is denoted by~$m_i$ ($i=1,2,3$). Note that the bare mass of all charged particles in each generation is the same. Thus for describing the elementary particles of the Standard Model, one requires~24 Dirac seas in total.

Moreover, as mentioned in~\S \ref{c:S elementary particles}, the theory of causal fermion systems takes into account massless right-handed neutrinos as well as the fact that they are not observed for low energies. 
The mass of the right-handed neutrino is given by~$\tilde{m}_4 = 0$, and its chirality is distinguished by the chiral projector~$\chi_R$. It turns out that a suitable regularization of these right-handed states will break the chiral symmetry. This regularization will be chosen in such a way that it vanishes in the low-energy regime, thus taking into consideration that right-handed neutrinos have never been observed in nature.

In order to extend the constructions outlined in~\S \ref{c:S external field problem} to all fundamental fermions including right-handed massless neutrinos, it is useful to define the \emph{auxiliary fermionic projector of the vacuum}~$P^{\textup{aux}}$ by
\begin{align}\label{c:cfs, (4.2.40')}
	P^\text{aux} = P^N_\text{aux} \oplus P^C_\text{aux} \:,  
\end{align} 
where 
\beq \label{c:cfs, (5.2.4)} 
P^N_\text{aux} = \Big( \bigoplus_{\beta=1}^3 P^\text{vac}_{\tilde{m}_\beta} \Big) \oplus 0
\qquad \text{and} \qquad
P^C_\text{aux} = \bigoplus_{a=1}^7 \bigoplus_{\beta=1}^3 P^\text{vac}_{m_\beta}
\eeq
(where the Dirac seas~$P^\text{vac}_{\tilde{m}_\beta}$ and~$P^\text{vac}_{{m}_\beta}$ are given in analogy to~\eqref{c:cfs, (2.1.1)}). 
Thus~$P^\text{aux}$ is composed of~$25$ direct summands, four in the neutrino
and~$21$ in the charged sector.  
The fourth direct summand of~$P^N_\text{aux}$ has the purpose
of describing the right-handed massless neutrinos (or, in the terminology of~\cite{cfs}, the ``right-handed high-energy states''). 

Next, the \emph{chiral asymmetry matrix}~$X$ and the \emph{mass matrix}~$Y$ are introduced by 
\begin{align}
	X &= \left( \Id_{\C^3} \oplus \tau_\reg \,\chi_R \right) \oplus \bigoplus_{a=1}^7 \Id_{\C^3} \:, \label{c:cfs, (4.2.42')}\\
	m Y &= \text{diag} \big( \tilde{m}_1, \tilde{m}_2, \tilde{m}_3, 0 \big)
	\oplus \bigoplus_{a=1}^7 \text{diag} \big( m_1, m_2, m_3 \big) \:, \label{c:cfs, (4.2.43')}
\end{align}
where~$m$ is an arbitrary mass parameter and~$\tau_\reg \in (0,1]$ is a dimensionless parameter 
for which we always assume the scaling
\[ \tau_\reg = (m \varepsilon)^{p_\reg} \qquad \text{with} \qquad
0 < p_\reg < 2 \qquad \text{and} \qquad
0 < \varepsilon \lesssim \ell_P \:. \]
This allows us to rewrite  
the auxiliary fermionic projector as (cf.~\cite[eq.~(5.2.5)]{cfs}) 
\begin{align*} %\label{c:cfs, (5.2.5)}
	P^\text{aux} = X t %= t X^* 
	\qquad \text{with} \qquad t := \bigoplus_{\beta=1}^{25} P^\text{vac}_{m Y^\beta_\beta} \:, 
\end{align*} 
where~$t$ is a solution of the Dirac equation
\[ (i \Pdd - m Y) \,t = 0 \:. \]
In order to introduce an interaction, we insert an operator~$\B$ into the Dirac equation,
\beq \label{c:cfs, (5.2.6)}
(i \Pdd + \B - m Y) \,\tilde{t} = 0 \:.
\eeq 
The causal perturbation theory  
defines~$\tilde{t} = \tilde{t}[\B]$ in terms of a unique perturbation series (see~\S \ref{c:S external field problem} or~\cite[Section~2.1]{cfs}). 
Assuming the \emph{causality compatibility condition}\footnote{The causality compatibility condition is required in order to obtain \emph{bounded} line integrals, for details we refer to~\cite{finster1996ableitung}.} 
\beq \label{c:cfs, (4.2.48)}
(i \Pdd + \B - m Y)\, X = X^* \,(i \Pdd + \B - m Y) \qquad \text{for all~$\tau_\reg \in (0,1]$} \:, 
\eeq
the (unregularized) interacting auxiliary fermionic projector 
is then introduced  
by~$X\tilde{t}$.

The remainder of this section is devoted to establish the connection to physics. To this end, the procedure is as follows: The first step is to obtain a convenient form for~$\tilde{t}$ by performing the so-called ``light-cone expansion.'' The next step is to ``regularize'' the resulting expressions.  
Afterwards, one can introduce particles and anti-particles in the fashion of~\S \ref{c:S external field problem}, thus giving rise to an interacting auxiliary fermionic projector~$\tilde{P}^{\textup{aux}}$. Forming the so-called ``sectorial projection,'' one finally obtains a fermionic projector~$\tilde{P}$. The general procedure is summarized in~\cite[Remark~4.2.3]{cfs}. Assuming that the resulting operator~$\tilde{P}$ is a minimizer of the causal action, the corresponding Euler-Lagrange (EL) equations give rise to conditions on the interaction~$\B$ in a ``weak evaluation on the light cone.'' The underlying picture is to regard precisely those potentials~$\B$ as physically relevant for which the Euler-Lagrange equations are satisfied. 

Before entering the light-cone expansion in more detail, it is necessary to specify the admissible form of the external potential~$\B$ in the Dirac equation~\eqref{c:cfs, (5.2.6)}. To this end, we point out that the specific form of the interacting potential~$\B$ is determined by the causal action principle. For this reason, one should choose~$\B$ as general as possible, thereby even allowing for potentials which are usually not considered in physics. The following overview of potentials which are of relevance in the theory of causal fermion systems as worked out in~\cite{cfs} is due to~\cite[\S3.4.5]{cfs}.  
The most obvious choice is an electromagnetic potential,\footnote{Following the convention in~\cite{cfs}, the coupling constant~$e$ is omitted in the Dirac equation. This convention can be obtained from the usual choice~$\B = e\slashed{A}$ by the transformation~$A \to e^{-1}A$.}  
\begin{align*} %\label{c:cfs, (3.4.15)}
	\B = \slashed{A}\:.
\end{align*} 
More generally, one can choose {\em{chiral potentials}}, 
which may be non-diagonal in the
generations,
\begin{align*}\label{c:cfs, (3.4.16)}
	\B = \chi_L\: \slashed{A}_R + \chi_R\: \slashed{A}_L\:,
\end{align*} 
To describe a {\em{gravitational field}}, 
one needs to choose~$\B$ as a differential operator of
first order; more precisely,
\begin{align*}%\label{c:cfs, (3.4.17)}
	\B = \Dir - i \Pdd\:,
\end{align*} 
where~$\Dir$ is the Dirac operator in the presence of a gravitational field.

Apart from the above choices of~$\B$ motivated from physics, one can also
choose other physically less obvious operators, like  
{\em{scalar}} or {\em{pseudoscalar potentials}}, 
\begin{align*} % \label{c:cfs, (3.4.18)}
	\B = \Phi + i \pseudo \Xi
\end{align*} 
with~$\Phi = \Phi^\alpha_\beta$, $\Xi = \Xi^\alpha_\beta$ and~$\alpha, \beta=1,\ldots,g$ for some~$g \in \N$ (here~$\pseudo \equiv i \gamma^0 \gamma^1 \gamma^2 \gamma^3$ is the usual pseudoscalar matrix).
Next, one can consider a {\em{scalar differential operator}},
\[ \B = i \Phi^j \partial_j \:, \]
or a higher order differential operator. Furthermore, {\em{pseudoscalar
		differential potentials}} turn out to be useful, 
\[ \B = \pseudo \left( v^j \partial_j + \partial_j v^j \right) \]
(see e.g.~\cite[\S 3.7.4]{cfs}), as well as \emph{vector differential potentials} (cf.~\cite[\S 3.7.5]{cfs}). Further potentials and fields are discussed in~\cite[\S 3.9.3]{cfs}.

For clarity, we point out that the situation in which the external potential~$\B$ 
is composed of left- or right-handed potentials, i.e.\
\beq \label{c:cfs, (2.2.25)}
\B = \chi_L \:\slashed{A}_R + \chi_R \: \slashed{A}_L \:, 
\eeq
is of particular interest. The reason is that
potentials of the form~\eqref{c:cfs, (2.2.25)}, which are referred to as {\em{chiral potentials}},  
allow for the description of {\em{gauge fields}}. 
For instance, an electromagnetic field is described by choosing~$A_L=A_R=A$,
where~$A$ is the electromagnetic potential.
Furthermore, as weak potentials only couple to the left-handed component of fundamental fermions, 
a left-handed potential is  
required for describing the weak interaction in the Standard Model.  
Note that in the case of \emph{non-abelian} gauge fields, the potentials~$A_L$ and~$A_R$ take values in a Lie algebra (cf.~\cite[\S 2.2.3]{cfs}).

\subsection{The Light-Cone Expansion}\label{c:S light-cone expansion}
After these preliminaries, let us enter the light-cone expansion in more detail. 
For simplicity, we first restrict attention to a single Dirac sea. Thus the starting point is a Dirac sea~$P(x,y)$ of mass~$m$ as given by~\eqref{c:cfs, (2.1.1)}, 
\beq \label{c:cfs, (2.1.1')}
P(x,y) = \int \frac{d^4k}{(2 \pi)^4}\: (\slashed{k}+m)\: \delta(k^2-m^2)\: \Theta(-k^0) \: e^{-ik(x-y)} \:. 
\eeq
It is useful to observe that~$P(x,y)$ can be rewritten in the form 
\beq \label{c:cfs, (1.2.25)}
P(x,y) = (i \Pdd_x + m) \,T_{m^2}(x,y) \:,
\eeq
where~$T_{m^2}(x,y)$ is the scalar bi-distribution
\beq \label{c:cfs, (1.2.26)}
T_{m^2}(x,y) := \int \frac{d^4k}{(2 \pi)^4}\: \delta(k^2-m^2)\: \Theta(-k^0)\: e^{-ik(x-y)} \:.
\eeq
The important point in what follows is that~$T_{m^2}(x,y)$ can be reformulated as
\begin{align}
	T_{m^2}(x,y) &= -\frac{1}{8 \pi^3} \:\bigg( \frac{\PP}{(y-x)^2} + i \pi \delta \big( (y-x)^2 \big) \: \epsilon \big( (y-x)^0 \big)
	\bigg) \nonumber \\
	&\quad +\frac{m^2}{32 \pi^3}\sum_{j=0}^\infty \frac{(-1)^j}{j! \: (j+1)!} \: \frac{\big( m^2 (y-x)^2 \big)^j}{4^j} \notag \\
	&\qquad\qquad \times  \Big( \log \big|m^2 (y-x)^2 \big| + c_j
	+ i \pi \:\Theta \big( (y-x)^2 \big) \:\epsilon\big( (y-x)^0 \big) \Big) \label{c:cfs, (2.2.3)}
\end{align} 
with real coefficients~$c_j$ (here~$\Theta$ and~$\epsilon$ are again the Heaviside and the sign function,
respectively), where the distribution~$\PP/\xi^2$, denoted by {\em{principal value}}, is defined by
evaluating weakly with a test function $\eta \in C^\infty_0(\scrM)$
and by removing the positive and negative parts of the pole in a symmetric way, 
\begin{align*} %\label{c:PPdef}
	\begin{split}
		&\int \frac{\PP}{\xi^2} \: \eta(\xi)\: d^4\xi
		= \lim_{\nu \searrow 0} \int \Theta\big( |\xi^2| - \nu \big)\;
		\frac{1}{\xi^2} \: \eta(\xi)\: d^4\xi \\
		&\qquad = \lim_{\nu \searrow 0} \frac{1}{2} \sum_{\pm}
		\int \frac{1}{\xi^2 \pm i \nu} \: \eta(\xi)\: d^4\xi
		= \lim_{\nu \searrow 0} \frac{1}{2} \sum_{\pm}
		\int \frac{1}{\xi^2 \pm i \nu \xi^0} \: \eta(\xi)\: d^4\xi
	\end{split}
\end{align*} 
(where~$\xi^2 \equiv \xi^j \xi_j$ is the Minkowski inner product).  
It is important to observe that the distribution~$T_{m^2}(x,y)$ is smooth away from the light cone (i.e. for~$\xi^2 \not= 0$ with~$\xi = y-x$), whereas on the light cone
\begin{align*}
	L = \left\{ \xi \in \scrM : \xi^2 = 0 \right\} \:,
\end{align*} 
the distribution~$T_{m^2}(x,y)$ has singularities (see~\cite[Lemma~1.2.9]{cfs}). As a consequence, the main contributions to the fermionic projector~\eqref{c:cfs, (2.1.1')} arise from the contributions on the light cone (also cf.~\cite[\S 2.1.4]{finster1996ableitung} and~\cite[Section~4.2]{pfp}). A useful tool in order to analyze the singular structure of the fermionic projector on the light cone in more detail is the so-called ``light-cone expansion.'' In particular,  
the expression~\eqref{c:cfs, (2.2.3)} turns out to be a light-cone expansion, which is defined as follows (see~\cite[Definition~2.2.1]{cfs}): 
\begin{Def} \label{c:cfs, Definition 2.2.1}
	A distribution~$A(x,y)$  
	is \textbf{of the order~$\O((y-x)^{2p})$} for~$p \in \Z$ if  
	\[ (y-x)^{-2p} \: A(x,y) \] 
	is a regular distribution (i.e.\ a locally integrable function).
	An expansion of the form
	\beq
	A(x,y) = \sum_{j=g}^{\infty} A^{[j]}(x,y) \label{c:cfs, (2.2.1)}
	\eeq
	with $g \in \Z$ is called {\bf{light-cone expansion}} 
	if~$A^{[j]}(x,y) \in \O((y-x)^{2j})$ for all~$j \ge g$ (that is, the~$A^{[j]}(x,y)$ are distributions of the order
	$\O((y-x)^{2j})$ for every~$j \ge g$) and if~$A$ is approximated by the partial sums
	in the sense that for all~$p \geq g$,
	\beq \label{c:cfs, (2.2.2)}
	A(x,y) - \sum_{j=g}^p A^{[j]}(x,y) \qquad {\text{is of the order~$\O\big( (y-x)^{2p+2} \big)$}}\:.
	\eeq
\end{Def} \noindent
The parameter~$g$ gives the leading order of the singularity of~$A(x,y)$ on the light cone.
We point out that we do not demand that the infinite series in~\eqref{c:cfs, (2.2.1)} converges. Thus, similar
to a formal Taylor series, the series in~\eqref{c:cfs, (2.2.2)} is defined only via the approximation by the
partial sums~\eqref{c:cfs, (2.2.2)}. 

Due to the factors $(y-x)^{2j}$, the expression in~\eqref{c:cfs, (2.2.3)} is a light-cone expansion.
More precisely, the term with the leading singularity becomes integrable after multiplying by~$(y-x)^2$,
showing that~$g=-1$.
The light-cone expansion of the kernel of the fermionic projector~$P(x,y)$
in~\eqref{c:cfs, (2.1.1')} is readily obtained using the relation~\eqref{c:cfs, (1.2.25)}.
To this end, one simply applies the differential operator~$(i \Pdd+m)$ to the above series expansion of~$T_{m^2}(x,y)$
and computes the derivatives term by term. Since differentiation
increases the order of the singularity on the light cone by one, we thus obtain a light-cone expansion
of the form~\eqref{c:cfs, (2.2.1)} with~$g=-2$.

In order to study the effect of an interaction~$\B$ in more detail,  
our goal is to analyze the solution of the modified Dirac equation~\eqref{c:cfs, (2.1.5)}.  
Unfortunately, deriving a light-cone expansion of the fermionic projector~$P^{\textup{sea}}$ in~\eqref{c:cfs, (2.1.65)} is not quite straightforward (see formula~\eqref{c:cfs, (2.2.129)} below). Since the unique perturbation series in~\eqref{c:cfs, (2.1.25)} essentially enter the construction of~$P^{\textup{sea}}$, it is a promising strategy to    
develop a method
for performing the light-cone expansion of each summand of the perturbation series in~\eqref{c:cfs, (2.1.25)}. 
Let us begin by considering the 
free advanced Green's function $s^\vee_m$ of the Dirac equation of mass~$m$ in position space:
Similar to~\eqref{c:cfs, (1.2.25)}, it is again convenient to pull the Dirac matrices out of~$s^\vee_m$ by setting
\begin{align}\label{c:cfs, (2.2.4)}
	s^\vee_m(x,y) = (i \Pdd_x + m) \: S^\vee_{m^2}(x,y) \:,
\end{align} 
where $S^\vee_{m^2}(x,y)$ can be represented as (cf.~\cite[eq.~(2.2.7)]{cfs})
\begin{align}
	S^\vee_{m^2}(x,y) = &-\frac{1}{2 \pi} \:\delta \big( (y-x)^2 \big) \: \Theta\big(y^0 - x^0 \big) \notag \\
	&+ \frac{m^2}{8 \pi}
	\sum_{j=0}^\infty \frac{(-1)^j}{j! \:(j+1)!} \: \frac{\big( m^2 (y-x)^2 \big)^j}{4^j} \:
	\Theta \big( (y-x)^2 \big) \:\Theta \big(y^0 - x^0 \big) \:.
	\label{c:cfs, (2.2.7)}
\end{align}
This computation shows that $S^\vee_{m^2}(x,y)$ has a $\delta((y-x)^2)$-like
singularity on the light cone. Moreover, one sees that $S^\vee_{m^2}$
is a power series in $m^2$. The important point for what follows is
that higher order contributions in $m^2$ contain more 
factors $(y-x)^2$ and are thus of higher order on the light cone. 
More precisely,
\begin{align*}%\label{c:cfs, (2.2.8)}
	\left( \frac{d}{dm^2} \right)^n S^\vee_{m^2 }(x,y) \Big|_{m=0} \qquad
	\text{is of the order $\O\big((y-x)^{2n-2} \big)$}
\end{align*} 
(where~$m$ is treated as a variable parameter). 
According to \eqref{c:cfs, (2.2.4)}, the Dirac Green's function is obtained by 
computing the first partial derivatives of \eqref{c:cfs, (2.2.7)}. Therefore,
the Green's function~$s^\vee_m(x,y)$ has an even~$\sim \delta^\prime((y-x)^2)$-like singularity on the light cone, and
%which is even~$\sim \delta^\prime((y-x)^2)$, and
the higher order contributions in $m$ are again of increasing order on 
the light cone. This means that we can view the Taylor expansion of 
\eqref{c:cfs, (2.2.4)} in $m$,
\[ s^\vee_m(x,y) = \sum_{n=0}^\infty (i \Pdd + m) \;\frac{1}{n!}
\left( \frac{d}{dm^2} \right)^n S^\vee_{m^2}(x,y)  \Big|_{m=0} \: , \]
as a light-cone expansion of the free Green's function.  
These considerations motivate the strategy to first expand \eqref{c:cfs, (2.1.25)} with respect to
the mass (see~\cite[Section~2.2]{cfs}).

The expansion of \eqref{c:cfs, (2.1.25)} with respect to $m$ gives a double 
sum over the orders in the mass parameter and in the external 
potential. It is convenient to combine these two expansions in a single 
perturbation series. To this end, we rewrite the Dirac operator as
\beq \label{c:cfs, (2.2.10)}
i \Pdd + \B - m = i \Pdd + B \qquad {\mbox{with}} \qquad B:=\B-m \:.
\eeq 
For the light-cone expansion of the Green's functions, we will always
view $B$ as the perturbation of
the Dirac operator. This has the advantage that the unperturbed objects
are massless. Thus expanding in powers of~$B$
gives the mass expansion and the perturbation expansion in one step. 
In order to simplify the notation, for the massless objects we usually
omit the index~$m$. Thus we write the Green's function of the
massless Dirac equation in the Minkowski vacuum as
\begin{align*}%\label{c:cfs, (2.2.11)}
	s^\vee(x,y) = i \Pdd_x \:S^\vee_{m^2}(x,y) \big|_{m=0}\:,\qquad
	s^\wedge(x,y) = i \Pdd_x \:S^\wedge_{m^2}(x,y) \big|_{m=0}\:.
\end{align*} 
Then the interacting Green's functions are given by the perturbation series
\beq \tilde{s}^\vee = \sum_{k=0}^\infty (-s^\vee B)^k 
s^\vee \:,\qquad \tilde{s}^\wedge = \sum_{k=0}^\infty
(-s^\wedge B)^k s^\wedge \: .
\label{c:cfs, (2.2.12)}
\eeq 
The following constructions (for details see~\cite[Chapter~2]{cfs}) are exactly the same for
the advanced and retarded Green's functions. In order to treat both 
cases at once, in the remainder of this subsection we will omit all 
superscripts `$^\vee$', `$^\wedge$'. The formulas for the advanced and 
retarded Green's functions are obtained by either adding `$^\vee$' or
`$^\wedge$' to all factors $s$, $S$.
Moreover, in order to carry out the mass expansion of $S_{m^2}$, we set $a=m^2$ and
use the notation
\beq \label{c:cfs, (2.2.13)}
S^{(l)} = \left( \frac{d}{da} \right)^l S_a \big|_{a=0} \: .
\eeq
Furthermore, the combination~$(y-x)_k \:S^{(-1)}(x,y)$ is defined by (cf.~\cite[eq.~(2.2.24)]{cfs})
\begin{align*}%\label{c:cfs, (2.2.24} 
	(y-x)_k \:S^{(-1)}(x,y) := 2 \:\frac{\partial}{\partial x^k}
	S^{(0)}(x,y) \:.
\end{align*}

In the following, we restrict attention to chiral potentials~\eqref{c:cfs, (2.2.25)},
\beq \label{c:cfs, (2.2.25')}
\B = \chi_L \:\slashed{A}_R + \chi_R \: \slashed{A}_L \:.
\eeq
Considering Dirac particles with (bare) rest masses~$m_1, \ldots, m_g$ for some~$g \in \N$ and introducing the \emph{mass matrix}~$Y$ by
\[ Y = \frac{1}{m} \: \text{diag} \big(m_1, \ldots, m_g) \]
for some mass parameter~$m$, we combine the mass term with the potential in analogy to~\eqref{c:cfs, (2.2.10)} by setting
\beq \label{c:cfs, (2.2.29)}
B = \chi_L \:\slashed{A}_R + \chi_R \: \slashed{A}_L - m Y \:.
\eeq
Then the perturbation expansion of the causal Green's functions can again be written
in the form~\eqref{c:cfs, (2.2.12)}.  
In order to allow for scalar and pseudoscalar potentials it is most convenient to replace the mass matrix by a spacetime-dependent \emph{dynamical mass matrix}
\beq \label{c:cfs, (2.2.28)}
Y = Y(x) := \chi_L Y_L(x) + \chi_R Y_R(x) \:.
\eeq 

In what follows, we are mainly interested in analyzing the structure of solutions of the Dirac equation 
\beq \label{c:cfs, (2.2.27)}
(i \Pdd + \B - m Y) \,\psi(x) = 0 \:. 
\eeq
In order to recall the structural results on the contributions to the light-cone expansion of the Green's functions, let us state the following definitions: 
For the line integrals, we introduce the short notation
\begin{align*}%\label{c:l:29x}
	\int_x^y [l,r\:|\: n] \:dz \; f(z) := \int_0^1 d\alpha \;
	\alpha^{l}\:(1-\alpha)^{r} \:(\alpha-\alpha^2)^n \; f(\alpha y + (1-\alpha) x) \:.
\end{align*} 
Furthermore, we abbreviate the following products with multi-indices,
\[ \partial_z^J := \frac{\partial}{\partial z^{j_1}} \cdots
\frac{\partial}{\partial z^{j_l}} \:,\qquad
(y-x)^J := (y-x)^{j_1} \cdots (y-x)^{j_l} \:,\quad
\gamma^J \;:=\; \gamma^{j_1} \cdots \gamma^{j_l} \:, \]
where $J=(j_1, \ldots, j_l)$. 
Unfortunately, in order to clarify the detailed structure of the light-cone expansion it seems unavoidable to recall the following definitions and results. Let us begin with~\cite[Theorem~2.2.4]{cfs}: 

\begin{Thm} \label{c:cfs, Theorem 2.2.4}
	In the presence of chiral potentials~\eqref{c:cfs, (2.2.29)}, the light-cone expansion of the $k^{\mbox{\scriptsize{th}}}$ order 
	contribution $((-s B)^k \:s)(x,y)$ to the perturbation series 
	\eqref{c:cfs, (2.2.12)} can be written as an infinite sum of expressions, each of which 
	has the form 
	\begin{align}
		\chi_{c_0} \:C \:(y-x)^I &\int_x^y [l_1, r_1 \:|\: n_1] \:dz_1 \; 
		\partial_{z_1}^{I_1}\: \Box_{z_1}^{p_1} \:V^{(1)}_{J_1, c_1}(z_1)
		\int_{z_1}^y [l_2, r_2 \:|\: n_2] \:dz_2 \; \partial_{z_2}^{I_2}\:
		\Box_{z_2}^{p_2}\: V^{(2)}_{J_2, c_2}(z_2) \notag \\
		&\cdots \int_{z_{k-1}}^y [l_k, r_k \:|\: n_k] \:dz_k \; 
		\partial_{z_k}^{I_k}\: \Box_{z_k}^{p_k}\: V^{(k)}_{J_k, c_k}(z_k) \;
		\gamma^J \;S^{(h)}(x,y) \:. \label{c:cfs, (2.2.32)}
	\end{align}
	In this formula, $C$ is a complex number and the parameters
	$l_a$, $r_a$, $n_a$, and $p_a$ are non-negative integers; the indices
	$c$ and $c_a$ can take the two values $L$ or $R$.
	The functions $V^{(a)}_{J_a, c_a}$ (where~$J_a$ is a multi-index and~$c_a \in \{L,R\}$ is a chiral index)
	coincide with any of the individual potentials in~\eqref{c:cfs, (2.2.29)} 
	and~\eqref{c:cfs, (2.2.28)} with chirality $c_a$, i.e.
	\beq
	\begin{split}
		V^{(a)}_{c_a} &= A_{c_a} \qquad\;\;\,
		{\mbox{(in which case $|J_a|=1$)}} \qquad {\mbox{or}} \\
		V^{(a)}_{c_a} &= m Y_{c_a} \qquad {\mbox{(in which case 
				$|J_a|=0$)}} \:.
	\end{split}  \label{c:cfs, (2.2.33)}
	\eeq
	The chirality $c_a$ of the potentials is determined by the following rule:
	\begin{itemize}[leftmargin=2em]
		\item[(i)] The chirality is reversed precisely at every mass matrix, i.e.
		\[ {\mbox{$c_{a-1}$ and $c_a$}}\;\;\left\{ \begin{array}{cl} 
		\mbox{coincide} & \mbox{if $V^{(a)}_{c_a}=A_{c_a}$} \\
		\mbox{are opposite} & \mbox{if $V^{(a)}_{c_a}=m Y_{c_a}$}
		\end{array} \right. \]
		for all~$a=1,\ldots, k$.
	\end{itemize}
	The tensor indices of the multi-indices in \eqref{c:cfs, (2.2.32)} are all
	contracted with each other, according to the following rules:
	\begin{itemize}[leftmargin=2em]
		\item[(a)] No two tensor indices of the same multi-index are
		contracted with each other.
		\item[(b)] The tensor indices of the factor $\gamma^{J}$ are 
		all contracted with different multi-indices, in the order of their appearance in
		the product \eqref{c:cfs, (2.2.32)} (i.e., for $J=(j_1,\ldots,j_l)$ and $1 \leq a < b 
		\leq l$, the multi-index with which $j_a$ is contracted must stand to the 
		left of the multi-index corresponding to $j_b$).
	\end{itemize}
	The parameter $h$ is given by
	\begin{align*} %\label{c:l:l3}
		2h = k - 1 - |I| + \sum_{a=1}^k \Big( |I_a| + 2 p_a \Big) \:.
	\end{align*} 
	The number of factors $(y-x)$ is bounded by
	\begin{align*} %\label{c:l:l3a}
		|I| \leq k+1-\sum_{a=1}^k |I_a| \:.
	\end{align*} 
\end{Thm}
In a few words, 
Theorem~\ref{c:cfs, Theorem 2.2.4} states that the light-cone expansion of the 
$k^{\mbox{\scriptsize{th}}}$ order Feynman diagrams (cf.~\eqref{c:cfs, (2.1.25)}) can be written with $k$ 
nested line integrals. In particular, the number of summands \eqref{c:cfs, (2.2.32)} is finite to every 
order on the light cone. Therefore, the light-cone expansion of 
Theorem \ref{c:cfs, Theorem 2.2.4} makes mathematical sense in terms of Definition~\ref{c:cfs, Definition 2.2.1}. Moreover, the integer~$h$ is bounded from below by~$h \ge -1$. By neglecting all terms of the order~$\O((y-x)^{-2})$, to the leading singularity~$h=-1$ the sum over all Feynman diagrams~\eqref{c:cfs, (2.1.25)} takes the form (cf.~\cite[eq.~(2.2.57)]{cfs})
\begin{align*} %\label{c:cfs, (2.2.57)}
	\chi_c \:\tilde{s}(x,y) = \chi_c \:\Pexp \left( -i \int_x^y
	(y-x)_j \:A^j_c(z)\:dz \right) s(x,y) \:+\:
	\O((y-x)^{-2}) 
\end{align*} 
for~$c \in \{L, R\}$, 
where~$\Pexp$ is defined as follows (see~\cite[Definition~2.2.5]{cfs}).  

\begin{Def} %\label{c:cfs, Definition 2.2.5}
	For a smooth one-parameter family of matrices $F(\alpha)$, $\alpha
	\in \R$, the {\bf{ordered exponential}}~$\Pexp (\int F(\alpha) \:d\alpha)$
	is given by the Dyson series
	\begin{align*}
		\Pexp \bigg( \int_a^b F(\alpha) \:d\alpha \bigg)
		&= \Id + \int_a^b F(t_0) \:dt_0 \:+\: \int_a^b dt_0\:F(t_0) \int_{t_0}^b dt_1 \: F(t_1) \notag \\
		&\qquad \!+ \int_a^b dt_0\:F(t_0) \int_{t_0}^b dt_1 \: F(t_1) \int_{t_1}^b dt_2 \:F(t_2) + \cdots \:. 
	\end{align*}
	For ordered exponentials over the chiral potentials, we use the short notations
	\begin{align*}
		&\Pexp \bigg( -i \int_x^y (y-x)_j \:A^j_c(z)\:dz \bigg) =
		\Pexp \bigg(-i \int_x^y A^j_c \:(y-x)_j \bigg) = \Pe^{-i \int_x^y A^j_c \:(y-x)_j} \\
		&\qquad:= \Pexp \bigg( -i \int_0^1 A^j_c \big|_{\alpha y + (1-\alpha) x} \: (y-x)_j \: d\alpha \bigg) .
	\end{align*}        
\end{Def} 

\begin{Prp}  
	\label{c:cfs, Proposition 2.2.6}
	Every contribution~\eqref{c:cfs, (2.2.32)} to the light-cone expansion of Theorem~\ref{c:cfs, Theorem 2.2.4} can be written
	as a finite sum of expressions of the form
	\begin{align*}
		\chi_c\:C \:(y-x)^K \:W^{(0)}(x) &\int_x^y [l_1, r_1 \:|\: n_1] \:dz_1 \;
		W^{(1)}(z_1) \int_{z_1}^y [l_2, r_2 \:|\: n_2] \:dz_2 \; W^{(2)}(z_2) \notag \\
		& \cdots \int_{z_{\alpha-1}}^y [l_\alpha, r_\alpha \:|\: 
		n_\alpha] \:dz_\alpha \; W^{(\alpha)}(z_\alpha) \;
		\gamma^J \; S^{(h)}(x,y) %\label{c:l:70}
	\end{align*}
	with~$\alpha \leq k$, where the factors $W^{(\beta)}$ are composed of the potentials and 
	their partial derivatives,
	\begin{align*} %\label{c:l:71}
		W^{(\beta)} = (\partial^{K_{a_\beta}} \Box^{p_{a_\beta}} 
		V^{(a_\beta)}_{J_{a_\beta}, c_{a_\beta}}) \cdots (\partial^{K_{b_\beta}} \Box^{p_{b_\beta}} 
		V^{(b_\beta)}_{J_{b_\beta}, c_{b_\beta}})
	\end{align*} 
	with $a_1=1$, $a_{\beta+1}=b_\beta+1$, $b_\beta \geq a_\beta-1$ (in 
	the case $b_\beta=a_\beta-1$, $W^{(\beta)}$ is identically one),
	and $b_\alpha=k$.
	The parameters $l_a$, $r_a$, and $n_a$ are non-negative integers,
	$C$ is a complex number, and $c, c_a \in \{L, R \}$
%	=L\!/\!R$, $c_a=L\!/\!R$ 
	are chiral 
	indices. The potentials~$V^{(a)}$ are again given by~\eqref{c:cfs, (2.2.33)};
	their chirality is determined by the rule~(i) in Theorem \ref{c:cfs, Theorem 2.2.4}.
	The tensor indices of the multi-indices $J$, $K$, $J_a$, and 
	$K_a$ are all contracted with each other, according to the rules (a),(b) 
	of Theorem \ref{c:cfs, Theorem 2.2.4} and
	\begin{itemize}[leftmargin=2em]
		\item[(c)] The tensor indices of $(y-x)^K$ are all 
		contracted with the tensor indices of the factors $V^{(a)}_{J_a}$ or $\gamma^J$ 
		(but not with the factors $\partial^{K_a}$).
	\end{itemize}
	We have the relation
	\begin{align*}  %\label{c:cfs, (2.2.61)}
		2h = k - 1 - |K| + \sum_{a=1}^k \big( |K_a| + 2 p_a \big) \:.
	\end{align*} 
\end{Prp}

\begin{Def} %\label{c:cfs, Definition 2.2.7}
	A contribution of the form~\eqref{c:cfs, (2.2.32)} to the light-cone expansion of Proposition~\ref{c:cfs, Proposition 2.2.6}
	is called {\bf{phase-free}} (see~\cite[Definition~2.2.7]{cfs}) if  
	\[ |K_a|+2 p_a > 0 \qquad {\mbox{whenever}}\qquad {\mbox{$J_a$ is contracted with
			$(y-x)^K$.}} \]
	From every phase-free contribution the corresponding
	{\bf{phase-inserted}} contribution is obtained as follows: 
	We insert ordered exponentials according to the replacement rule
	\begin{align*} %\label{c:cfs, (2.2.62)}
		W^{(\beta)}(z_\beta) \;\longrightarrow\;
		W^{(\beta)}(z_\beta) \:\Pexp \left(-i \int_{z_\beta}^{z_{\beta+1}} A^{j_\beta}_{c_\beta}
		\: (z_{\beta+1} - z_\beta)_{j_\beta} \right),\quad \beta=0,\ldots,\alpha\:,
	\end{align*} 
	where we set~$z_0=x$ and~$z_{\alpha+1}=y$.
	The chiralities~$c_\beta$ ($\beta=0,\ldots,\alpha$) 
	are determined by the relations~$c_0 = c$ and
	\begin{align*} %\label{c:cfs, (2.2.63)}
		\begin{split}
			{\text{$c_{\beta-1}$ and $c_\beta$}} & \left\{ \!\!\!\begin{array}{c}
				{\text{coincide}} \\ {\text{are opposite}} \end{array} \!\!\right\} \\
			&{\text{if $W^{(\beta-1)}$ contains an}} \left\{ \!\!\!\begin{array}{c}
				{\text{even}} \\ {\text{odd}} \end{array} \!\!\right\}
			{\text{number of factors $Y_.$.}}
		\end{split}
	\end{align*} 
\end{Def}

\begin{Thm}
	%	\label{c:cfs, Theorem 2.2.8}
	To every order on the light cone, the number of phase-free contributions
	is finite.
	The light-cone expansion of the Green's function $\tilde{s}(x,y)$
	is given by the sum of the corresponding phase-inserted contributions. See~\cite[Theorem~2.2.8]{cfs}. 
\end{Thm}

After these technical preliminaries, we are in the position to outline the structure of the solution~$P^{\textup{sea}}$ of the Dirac equation~\eqref{c:cfs, (2.2.27)}, 
\beq \label{c:cfs, (2.2.27')}
(i \Pdd + \B - m Y) \,\psi(x) = 0 \:, 
\eeq
in a self-contained way.  
To this end, we consider the Green's functions
\begin{align*} %\label{c:cfs, (2.2.107)}
	s^\pm(p) = \slashed{p} \: S^\pm_{a \:|\: a=0}(p) \qquad {\mbox{with}} 
	\qquad S^\pm_a(p) = \lim_{\nu \searrow 0} \frac{1}{p^2-a \mp i \nu}
\end{align*} 
and again use the notation \eqref{c:cfs, (2.2.13)},
\[ S^{\pm \:(l)} = \left( \frac{d}{da} \right)^l S^\pm_{a \:|\: a=0} \qquad (l=0,1,2,\ldots) \: . \] 
The perturbation expansion of these Dirac Green's functions is,
in analogy to~\eqref{c:cfs, (2.2.12)} or~\eqref{c:cfs, (2.1.25)}, given by the formal series
\begin{align*} %\label{c:cfs, (2.2.108)}
	\tilde{s}^+ := \sum_{n=0}^\infty (-s^+ \:B)^n 
	s^+ \qquad \text{and} \qquad \tilde{s}^- := \sum_{n=0}^\infty (-s^- \:B)^n s^- \:.
\end{align*} 
Next, we introduce the
{\em{residual fundamental solution}}~$\tilde{p}^\res$ in analogy to~\eqref{c:cfs, (1.2.26)} by 
\begin{align*} %\label{c:l:E1}
	\tilde{p}^\res := \frac{1}{2 \pi i} \:\big( \tilde{s}^+ - \tilde{s}^- \big) \:, 
\end{align*} 
and define the {\em{residual fermionic projector}} 
$\tilde{P}^\res(x,y)$ by
\begin{align*} %\label{c:l:E0}
	\tilde{P}^\res(x,y) := \frac{1}{2}\,
	\big(\tilde{p}^\res - \tilde{k} \big)(x,y) \:. 
\end{align*} 
In analogy to the mass expansion of the Green's functions~\eqref{c:cfs, (2.2.13)}, we set
\begin{align} \label{c:cfs, (2.2.113)}
	T^{(l)}_\text{\tiny{formal}} = \left( \frac{d}{da} \right)^l T_a \big|_{a=0} \qquad (l=0,1,2,\ldots) \:.
\end{align} 
Then due to~\cite[Proposition~2.2.11]{cfs}, the formal light-cone expansion of the residual fermionic projector $\tilde{P}^\res(x,y)$
is obtained from that of the causal Green's functions 
by the replacement
\[ S^{(l)} \rightarrow T^{(l)}_\text{\tiny{\rm{formal}}} \:. \]

Unfortunately, the formal $a$-derivatives of~$T_a$ in~\eqref{c:cfs, (2.2.113)} contain logarithmic poles of the form~$\log |a|$ (cf.~\cite[eq.~(2.2.116)]{cfs}) which cause serious problems. These
difficulties, arising from the logarithm in \eqref{c:cfs, (2.2.3)}, are referred to as
{\em{logarithmic mass problem}}. In order to resolve the logarithmic mass problem, the first step is to
``regularize'' the formal light-cone expansion by subtracting the 
problematic~$\log |a|$ term, 
\begin{align*} %\label{c:cfs, (2.2.117)}
	T_a^\reg(x,y) := T_a(x,y) - \frac{a}{32 \pi^3} \:\log|a| \: \sum_{j=0}^\infty
	\frac{(-1)^j}{j! \: (j+1)!} \: \frac{(a \xi^2)^j}{4^j} 
\end{align*}
(where~$\xi^2 \equiv \xi^j \xi_j$ denotes again the Minkowski inner product). We then introduce
\begin{align} \label{c:cfs, (2.2.118)} 
	T^{(l)} := \left( \frac{d}{da} \right)^l
	T_{a \:|\: a=0}^\reg \qquad (l=0,1,2,\ldots) \:.  
\end{align}

By definition (see~\cite[Definition~2.2.12]{cfs}), the so-called {\em{causal contribution}} 
$\tilde{P}^{\mbox{\scriptsize{\rm{causal}}}}$ to the fermionic projector is obtained 
from the residual Dirac sea $\tilde{P}^\res$ by 
replacing all factors~$T^{(h)}_{\text{\tiny{\rm{formal}}}}$ in the formal light-cone expansion of~$\tilde{P}^\res(x,y)$ by~$T^{(h)}$. 
Next, the {\em{non-causal low energy contribution}}
$\tilde{P}^\lec$ to the fermionic projector is given by
\[ \tilde{P}^\lec(x,y) =
\tilde{P}^\res(x,y) -
\tilde{P}^{\mbox{\scriptsize{\rm{causal}}}}(x,y) \: . \]
Due to~\cite[Theorem~2.2.13]{cfs}, the light-cone expansion of the causal Green's functions also holds for the
causal contribution~$\tilde{P}^{\mbox{\scriptsize{\rm{causal}}}}$ to the 
fermionic projector if one simply replaces~$S^{(l)} \rightarrow T^{(l)}$
with~$T^{(l)}$ according to~\eqref{c:cfs, (2.2.118)}. 
Moreover, 
the {\em{non-causal high energy contribution}}
$\tilde{P}^\hec(x,y)$ to the fermionic projector is given by
\[ \tilde{P}^\hec(x,y) = P^\sea(x,y) - \tilde{P}^\res(x,y) \:. \] 
Note that to every order in the external potential~$\B$, both~$\tilde{P}^\lec(x,y)$ and~$\tilde{P}^\hec(x,y)$  
are smooth functions in $x$ and $y$ (see~\cite[Theorem~2.2.14 and Theorem~2.2.16]{cfs}).

Taken together, the above constructions show that in the presence of chiral and scalar/pseudoscalar potentials 
(see~\eqref{c:cfs, (2.2.25')}, \eqref{c:cfs, (2.2.27')}, \eqref{c:cfs, (2.2.28)}) 
the fermionic projector 
has a representation 
of the form 
\beq \label{c:cfs, (2.2.129)}
\begin{split}
	P^{\mbox{\scriptsize{sea}}}(x,y) &= \sum_{n=-1}^\infty
	{\mbox{(phase-inserted line integrals)}} \times  T^{(n)}(x,y) \\
	&\qquad + \tilde{P}^\lec(x,y) + \tilde{P}^\hec(x,y) \:,
\end{split}
\eeq
where~$(y-x)_k \: T^{(-1)}(x,y)$ is defined by the distributional equation (cf.~\cite[eq.~(3.4.13)]{cfs})
\begin{align*} %\label{c:cfs, (3.4.13}
	\frac{\partial}{\partial x^k} T^{(0)}(x,y) = \frac{1}{2} \: (y-x)_k \: T^{(-1)}(x,y) \:. 
\end{align*} 
The series in~\eqref{c:cfs, (2.2.129)} is a light-cone expansion which describes
the singular behavior of the fermionic projector on the light cone. 
It is obtained from the light-cone expansion of the
Green's functions by the simple replacement rule  
\[ S^{(n)} \longrightarrow T^{(n)} \]
with~$T^{(n)}$ as defined in~\eqref{c:cfs, (2.2.118)}. 
Note that the expression~\eqref{c:cfs, (2.2.129)} is ``causal'' in the sense that all phase-inserted line integrals are \emph{bounded}. 
The important point in what follows is that the higher $a$-derivatives of $T_a(x,y)$ 
contain more factors $(y-x)^2$ and are thus of higher order on the 
light cone. This makes it possible to make mathematical sense of the infinite 
series in~\eqref{c:cfs, (2.2.129)} as a light-cone expansion.

In order to put the following subsection into the context, we recall that composite expressions of the regularized fermionic projector enter the causal principle in form of the closed chain. In order to give such expressions a mathematical meaning, it is useful to evaluate composite expressions
in~$T^{(n)}$ and~$\overline{T^{(n)}}$ weakly on the light cone (where~$\overline{T^{(n)}}$ denotes the complex conjugate of~$T^{(n)}$), as we now explain.

\subsection{Weak Evaluation on the Light Cone}\label{c:S continuum limit}
The techniques presented in the previous subsection~\S \ref{c:S light-cone expansion} provide a method for analyzing the \emph{un}regularized kernel of the fermionic projector. The purpose of the formalism of the continuum limit is to extend these methods to the regularized setting. More precisely, 
the representation~\eqref{c:cfs, (2.2.129)}, which describes the singular behavior of the fermionic projector on the light cone, is 
the starting point for the formalism of the continuum limit. Introducing a ``regularization,'' the formalism of the continuum limit allows to evaluate composite expressions in the fermionic projector in a well-defined way.  
For details concerning the formalism of the continuum limit we refer to~\cite[Section~2.4 and Section~3.5]{cfs} as well as~\cite[Chapter~4]{pfp}. 

In order to explain the basic ideas of introducing a regularization in more detail,  
let~$P^{\textup{sea}}(x,y)$ be a solution of the interacting Dirac equation
\beq
(i \Pdd + \B - m Y) \, P^\text{sea}(x,y) = 0 \:. \label{c:cfs, (3.4.4)}
\eeq 
The solution~$P^{\textup{sea}}(x,y)$  
can be written as (cf.~\cite[Section~3.4]{cfs})  
\begin{align}
	P^{\text{sea}}(x,y) = & \sum_{n=-1}^\infty
	\sum_{k} m^{p_k} 
	{\text{(phase-inserted nested line integrals)}} \times  T^{(n)}(x,y) \nonumber \\
	&+ \tilde{P}^\lec(x,y) + \tilde{P}^\hec(x,y) \:.
	\label{c:cfs, (3.4.14)}
\end{align} 
Here the $n$-summands describe the different orders of the singularities on the light cone,
whereas the $k$-sum describes all contributions to a given order on the light cone.
The phase-inserted nested line integrals involve~$\B$ and its partial derivatives,
possibly sandwiched between time-ordered exponentials of chiral potentials.
Since these nested line integrals are smooth functions in~$x$ and~$y$,
the series in~\eqref{c:cfs, (3.4.14)} is a light-cone expansion in the sense of Definition~\ref{c:cfs, Definition 2.2.1},
provided that the $k$-sum is finite for every~$n$. This is indeed the case 
if~$\B$ is composed of scalar, pseudoscalar and chiral potentials~\cite{light}, 
whereas for a more general perturbation operator~$\B$ this condition still needs to be verified.
This expansion is {\em{causal}} in the sense that it depends on~$\B$ and its
partial derivatives only along the line segment~$\overline{xy}$.

Note that particles and anti-particles are introduced in analogy to~\eqref{c:cfs, (3.4.7)} by
\beq \label{c:cfs, (3.4.7)}
P^\text{aux}(x,y) = P^\sea(x,y)
-\frac{1}{2 \pi} \sum_{k=1}^{\np} \psi_k(x) \overline{\psi_k(y)}
+\frac{1}{2 \pi} \sum_{l=1}^{\na} \phi_l(x) \overline{\phi_l(y)} \:.
\eeq 
The formalism of the continuum limit is essentially based on the physically reasonable \emph{assumption of macroscopic potentials and wave functions} which states that both the bosonic potentials 
in~\eqref{c:cfs, (3.4.4)} and the fermionic wave functions in~\eqref{c:cfs, (3.4.7)} vary only on the ``macroscopic'' length scale~$\ell_{\text{macro}} \gg \ell_P$.  
In order to outline the basic ideas of the formalism of the continuum limit, it is necessary to first introduce a regularization of regularization length~$\varepsilon > 0$. As the regularized objects are thought of as describing the ``microscopic structure of spacetime,'' it is natural to choose~$ \varepsilon \lesssim \ell_P$. We are then led to ``evaluate weakly'' for  
spacetime points~$x,y \in \scrM$ in such a way that~$\varepsilon \ll |\vec{\xi}| \ll \ell_{\text{macro}}$, where~$\xi := y -x$ and~$\xi = (\xi^0, \vec{\xi})$ (cf.~\cite[eq.~(3.7.77)]{cfs}). 

In view of the causal action principle~\eqref{c:cfs, (1.1.2)},~\eqref{c:cfs, (1.1.14)} our main task is to evaluate composite expressions in the fermionic projector~$P(x,y)$ on the light cone. 
As smooth functions can be evaluated for any spacetime point, it clearly suffices to restrict attention to singular expressions on the light cone. In other words, we introduce a regularization in such a way that the smooth contributions are all left unchanged. 
Motivated by the special case of an~$i\varepsilon$-regularization (see~\cite[\S 2.4.1]{cfs}), 
introducing a regularization amounts in view of~\eqref{c:cfs, (3.4.14)} to regularizing the expressions~$T^{(n)}$ by the formal replacements   
\begin{align*} %\label{c:cfs, (2.4.42)}
	m^p \,T^{(n)} \rightarrow m^p \,T^{(n)}_{[p]} \:,
\end{align*} 
where the subscript~$[.]$ is added to the factors~$T^{(n)}$ in order to count the power in~$m$, 
and the factors~$T^{(n)}_{[p]}$ are smooth functions of~$\xi$. 
More precisely, one first rewrites every summand of the light-cone expansion~\eqref{c:cfs, (3.4.14)} in such a way that it involves at most one factor~$\slashed{\xi}$ (which can always be arranged), and then associates to every factor~$\slashed{\xi}$ the corresponding factor~$T^{(n)}_{[p]}$. This gives rise to the replacements (cf.~\cite[eq.~(2.4.43)]{cfs})
\begin{align*} %\label{c:cfs, (2.4.43)}
	m^p \,\slashed{\xi} T^{(n)} \rightarrow m^p \,\slashed{\xi}^{(n)}_{[p]}\, T^{(n)}_{[p]} \:.
\end{align*} 
Nevertheless, composite expressions diverge in the limit~$\varepsilon \searrow 0$.  
Setting~$r := |\vec{\xi}|$ for every~$\xi = (t, \vec{\xi})$,  
the proper method in order to analyze this singular behavior is 
to ``evaluate weakly'' in~$t$ for fixed~$r$. To this end, one considers integrals of the form
\begin{align*} %\label{c:cfs, (2.4.17)}
	\int_{-\infty}^\infty \eta(t) \: \big(\cdots\big) \: dt
\end{align*} 
for a smooth test function~$\eta$, where~``$\cdots$'' stands for a composite expression
in the~$T^{(n)}$ and~$\overline{T^{(n)}}$. 
In cases where the lower index does not need to be specified we write~$T^{(n)}_\circ$.
Next, to any factor~$T^{(n)}_\circ$ we associate the {\em{degree}} $\deg T^{(n)}_\circ$ by 
\[ \deg T^{(n)}_\circ = 1-n \:. \]
The degree is additive in products, whereas the degree of a quotient is defined as the
difference of the degrees of numerator and denominator. The degree of an expression
can be thought of as describing the order of its singularity on the light cone, in the sense that a
larger degree corresponds to a stronger singularity. Using formal Taylor series, we can expand in the degree. In all our applications, this will
give rise to terms of the form
\begin{align*} %\label{c:cfs, (2.4.47')}
	\eta(x,y) \:
	\frac{ T^{(a_1)}_\circ \cdots T^{(a_\alpha)}_\circ \:
		\overline{T^{(b_1)}_\circ \cdots T^{(b_\beta)}_\circ} }
	{ T^{(c_1)}_\circ \cdots T^{(c_\gamma)}_\circ \:
		\overline{T^{(d_1)}_\circ \cdots T^{(d_\delta)}_\circ} } \qquad \text{with~$\eta(x,y)$ smooth}\:.
\end{align*} 
The quotient of the two monomials in this equation is referred to as a {\em{simple fraction}}. 
A simple fraction can be given a quantitative meaning by considering one-dimensional integrals
along curves which cross the light cone transversely away from the origin~$\xi=0$.
This procedure is called {\em{weak evaluation on the light cone}}. 
For our purpose, it suffices to integrate over the time coordinate~$t=\xi^0$ for fixed~$\vec{\xi} \neq 0$.
light cone~$t \approx |\vec{\xi}|$. 
The resulting integrals, which in the leading degree are expressions of the form 
\beq
\int_{|\vec{\xi}|-\varepsilon}^{|\vec{\xi}|+\varepsilon} dt \; \eta(t,\vec{\xi}) \:
\frac{ T^{(a_1)}_\circ \cdots T^{(a_\alpha)}_\circ \:
	\overline{T^{(b_1)}_\circ \cdots T^{(b_\beta)}_\circ} }
{ T^{(c_1)}_\circ \cdots T^{(c_\gamma)}_\circ \:
	\overline{T^{(d_1)}_\circ \cdots T^{(d_\delta)}_\circ} } \:,  
\label{c:cfs, (2.4.48'')}
\eeq
diverge if the regularization
is removed. In view of the scalings~$\varepsilon \ll |\vec{\xi}| \ll \ell_{\text{macro}}$, it is natural to neglect in the resulting expressions error terms of the form 
\begin{align*} %\label{c:cfs, (2.4.18)}
	\cdots + \text{(higher orders in~$\varepsilon/|\vec{\xi}|$)} 
\end{align*} 
as well as 
\begin{align*} %\label{c:cfs, (2.4.19)}
	\cdots + \text{(higher orders in~$\varepsilon/\ell_\text{macro}$)} \:, 
\end{align*} 
where~$\ell_\text{macro}$ denotes the ``macroscopic'' length scale on which~$\eta$ varies.  
Moreover, for large values of~$|\vec{\xi}| \gg \ell_{\text{macro}}$, the resulting expansions are of ``rapid decay'' in the ``\emph{l}arge spacetime variable~$l$'' (for details see~\cite[Section~2.4]{cfs}), whereas expressions for small values~$|\vec{\xi}|$ describe pure ``regularization effects'' which should not be required in order to furnish the connection to conventional ``macroscopic'' physics. In this way, the weak evaluation on the light cone allows to extract information of composite expressions for values of~$|\vec{\xi}|$ in the scaling~$\varepsilon \ll |\vec{\xi}| \ll \ell_{\text{macro}}$ (see~\cite[eq.~(3.7.77)]{cfs}).

The method of weak evaluation will be useful in order to analyze the Euler-Lagrange equations corresponding to a minimizer of the causal action, which we present in the following subsection. A detailed analysis of these equations  
will eventually reveal the desired connection to physics.

\subsection{The Euler-Lagrange Equations} \label{c:S EL equations}
The derivation of the Euler-Lagrange (EL) equations outlined in this subsection (for details see~\cite[Section~3.2]{cfs}) is motivated by the variational principle~\eqref{c:pfp, (3.5.9)}--\eqref{c:pfp, (3.5.10)} (cf.~\cite{pfp}). For this reason, variations of a minimizing fermionic projector are considered.  
For simplicity, in analogy to~\cite[eq.~(2.1.3)]{cfs} let us  
consider an
operator~$P : C_0^{\infty}(\scrM, S\scrM) \to C(\scrM, S\scrM)$ of the form
\beq \label{c:cfs, (3.2.1)}
(P \psi)(x) = \int_\scrM P(x,y)\: \psi(y)\: d^4y 
\eeq
with an integral kernel~$P(x,y)$, where~$\scrM$ denotes Minkowski space and~$C_0^{\infty}(\scrM, S\scrM)$ is the set of smooth wave functions with compact support. Furthermore, we shall assume that~$P$ is symmetric with 
respect to the Lorentz invariant inner product
\beq \label{c:cfs, (3.2.2)}
\bra \psi \mid \phi \ket = \int_\scrM \overline{\psi(x)} \: \phi(x)\: d^4x 
\eeq 
(where~$\overline{\psi} \equiv \psi^\dagger \gamma^0$ is the usual adjoint spinor, and~$\psi^\dagger$ the complex conjugate spinor).  
We then refer to~$P$ as the {\em{fermionic projector}},\footnote{Note that, with a slight abuse of language, sometimes the kernel~$P(x,y)$ is also referred to as ``fermionic projector.''}  
and the vectors in the image of~$P$
are called {\em{physical wave functions}}. Moreover, we introduce the {\em{closed chain}} $A_{xy} =A_{xy}[P]$ by 
\begin{align*} %\label{c:cfs,(3.2.5)}
	A_{xy} = P(x,y)\, P(y,x) \qquad \text{for all~$x,y \in \scrM$} 
\end{align*} 
and define the {\em{spectral weight}} $|A|$ by 
\begin{align*} %\label{c:cfs, (3.2.6)}
	|A| = \sum_{i=1}^4 |\lambda_i| \:,
\end{align*} 
where~$\lambda_1, \ldots, \lambda_4$ are the eigenvalues of~$A$ counted with algebraic
multiplicities. Then for any~$x,y \in \scrM$ the \emph{Lagrangian}~$\L$ is given by 
\begin{align*} %\label{c:cfs, (3.2.7)}
	\L_{xy}[P] = |A_{xy}^2| - \frac{1}{4}\: |A_{xy}|^2 \:,
\end{align*} 
and the functionals~\eqref{c:cfs, (1.1.2)}, \eqref{c:cfs, (1.1.5)} formally read (cf.~\cite[Section~3.2]{cfs})
\beq  
\begin{split}
		\Sact[P] \;&\stackrel{\text{formally}}{=}\; \iint_{\scrM \times \scrM} \L_{xy}[P] \:d^4 x\: d^4y \:, \label{c:cfs, (3.2.8)} \\	\T[P] \;&\stackrel{\text{formally}}{=}\; \iint_{\scrM \times \scrM} |A_{xy}|^2 \:d^4 x\: d^4y \:. 
	\end{split}  
\eeq 
These expressions are only formal because the integrands need not decay for large~$x$ or~$y$,
and thus the integrals may be infinite.
The functional~$\Sact$ is called the {\em{causal action}}.

In order to clarify the class in which to minimize the causal action, let us point out that 
in~\cite[Section~3.5]{pfp} it is argued to only consider continuous variations of~$P$ of the form~$UPU^{-1}$ with unitary transformations~$U$. In order to avoid divergences in~\eqref{c:cfs, (3.2.2)} and~\eqref{c:cfs, (3.2.8)}, 
we shall only consider such unitary variations~$U$ with respect to the inner product~\eqref{c:cfs, (3.2.2)} which outside of a compact set coincide with the identity. Variations of this kind are referred to as \emph{unitary in a compact region} (see~\cite[Definition~3.2.1]{cfs}). According to~\cite[Definition~3.2.2]{cfs}, a fermionic projector~$P$ of the form~\eqref{c:cfs, (3.2.1)} is said to be a \emph{minimizer} of the variational principle 
\beq \label{c:cfs, (3.2.9)}
\text{minimize $\Sact$ for fixed~$\T$}
\eeq
if for any operator~$U$ which is unitary in a compact region and satisfies the constraint
\beq \label{c:cfs, (3.2.10)}
\int_\scrM d^4x \int_\scrM d^4y \:\Big( \big|A_{xy}[P] \big|^2 - \big| A_{xy}[U P U^{-1}] \big|^2 \Big)
= 0\:,
\eeq
the functional~$\Sact$ satisfies the inequality
\begin{align*} %\label{c:cfs, (3.2.11)}
	\int_\scrM d^4x \int_\scrM d^4y \:\Big(\L_{xy}[U P U^{-1}] -  \L_{xy}[P] \Big)
	\;\geq\; 0 \:.
\end{align*}

In~\cite{cfs}, the connection to conventional physics is established essentially by means of the Euler-Lagrange (EL) equations corresponding to a minimizer of the causal action principle~\eqref{c:cfs, (3.2.9)}. To this end, the existence of a family~$(P^{\varepsilon})_{\varepsilon > 0}$ of regularized vacuum minimizers is taken for granted (see~\cite[Assumption~3.3.1]{cfs}).  
In order to derive the corresponding Euler-Lagrange (EL) equations, it is convenient to treat the side condition~\eqref{c:cfs, (3.2.10)} with a Lagrange multiplier~$\mu$ (for details and a justification of this procedure we refer to~\cite[\S 3.5.2]{cfs} and~\cite{bernard+finster}). More precisely, introducing the regularized closed chain~$A_{xy}^{\varepsilon} = P^{\varepsilon}(x,y) P^{\varepsilon}(y,x)$ as well as the functional 
\begin{align*} % \label{c:cfs, (3.5.13)}
	\Sact_\mu[P^{\varepsilon}] \;\stackrel{\text{formally}}{=}\; \iint_{\scrM \times \scrM}\L_\mu[A^{\varepsilon}_{xy}]\:d^4x\: d^4y
	\qquad \text{with} \qquad
	\L_\mu[A_{xy}^{\varepsilon}] = |(A_{xy}^{\varepsilon})^2| - \mu |A_{xy}^{\varepsilon}|^2 \:, 
\end{align*} 
unitary variations of~$P$ in a compact region according to~\cite[Definition~3.2.1]{cfs} give rise to 
the Euler-Lagrange (EL) equations
\beq \label{c:cfs, (3.5.20)}
\boxed{ \quad [P^{\varepsilon}, Q^{\varepsilon}] = 0 \:, \quad }
\eeq
where~$Q^{\varepsilon}$ is an operator with integral kernel
\begin{align}\label{c:cfs, (3.5.17)}
	Q^{\varepsilon}(x,y) = \frac{1}{4} \big(R^{\varepsilon}(x,y) + R^{\varepsilon}(y,x)^{\ast} \big) \:, 
\end{align}
and~$R^{\varepsilon}(y,x)$ is given by~$R^{\varepsilon}(y,x) := \nabla \L_{\mu}[A^{\varepsilon}_{xy}]$ with (cf.~\cite[eq.~(3.5.15)]{cfs})
\begin{align*} %\label{c:cfs, (3.5.15)}
	(\nabla \L_{\mu})^\alpha_\beta := \frac{\partial \L_{\mu}}{\partial \re P(x,y)^\beta_\alpha} - i
	\frac{\partial \L_{\mu}}{\partial \im P(x,y)^\beta_\alpha} \qquad \text{for~$\alpha, \beta \in \{1, \ldots, 4 \}$} \:. 
\end{align*} 
Employing the method of \emph{testing on null lines} (for details see~\cite[Appendix~A]{cfs}), the EL equations~\eqref{c:cfs, (3.5.20)} reduce to (cf.~\cite[eq.~(3.5.29)]{cfs})
\beq \label{c:cfs, (3.5.29)}
\boxed{ \quad Q^{\varepsilon}(x,y) = 0 \quad \text{if evaluated weakly on the light cone}\:. \quad }
\eeq
We refer to~\eqref{c:cfs, (3.5.29)} as the \emph{Euler-Lagrange equations in the continuum limit}. 
The crucial point in what follows is to observe that the EL equations~\eqref{c:cfs, (3.5.20)} pose conditions on the regularized fermionic projector~$P^{\varepsilon}(x,y)$ and in this way on the interaction~$\B$ entering the fermionic projector. In a sufficiently general model, the connection to conventional physics will be obtained by means of EL equations of this form. For a generalization to several generations we refer to~\cite[Section~3.4]{cfs}. 

In order to analyze the EL equations~\eqref{c:cfs, (3.5.29)}, 
it seems reasonable and sufficiently general to assume that the regularized fermionic projector of the vacuum~$P^{\varepsilon}(x,y)$ is \emph{homogeneous} in the sense that~$P^{\varepsilon}(x,y) = P^{\varepsilon}(y-x)$ for all~$x,y \in\scrM$ and that~$P^{\varepsilon}$ has a \emph{vector-scalar structure}, meaning that it is of the form (cf.~\cite[eq.~(2.6.2)]{cfs}) 
\begin{align*} %\label{c:cfs, (2.6.2)}
	P^\varepsilon(x,y) = g_j(x,y)\: \gamma^j + h(x,y) \: \Id 
\end{align*} 
with appropriate smooth functions~$g_j$ and~$h$. 
We then introduce the closed chain by
\begin{align*} %\label{c:cfs, (2.6.4)}
	A^\varepsilon_{xy} = P^\varepsilon(x,y)\: P^\varepsilon(y,x) 
\end{align*} 
in view of~\eqref{c:cfs, (1.1.4)}. 
Omitting 
the superscript~``$\varepsilon$'' and considering~$g$ generations, in the formalism of the continuum limit the quantities (cf.~\cite[eqs.~(2.6.15) and~(3.6.5)]{cfs})
\beq \label{c:cfs, (2.6.15)}
\lambda^{xy}_+ = g^2 T^{(0)}_{[0]} \,\overline{T^{(-1)}_{[0]}} + (\deg < 3) \:,\qquad
\lambda^{xy}_- = g^2 T^{(-1)}_{[0]} \,\overline{T^{(0)}_{[0]}} + (\deg < 3)\:.
\eeq
can be interpreted as the eigenvalues of the closed chain, 
where the bracket~($\deg < 3$) stands for terms of degree at most two. The corresponding spectral projectors read (see~\cite[eqs.~(2.6.16) and~(3.6.6)]{cfs} as well as~\cite[eq.~(5.3.21)]{pfp})
\beq \label{c:cfs, (2.6.16)}
F^{xy}_\pm = \frac{1}{2} \Big( \Id \pm \frac{[\slashed{\xi}, \overline{\slashed{\xi}}]}{z-\overline{z}} \Big)
+ \slashed{\xi} (\deg \leq 0) + (\deg < 0) 
\eeq
(where the functions~$z$ and~$\overline{z}$ are given in terms of the so-called ``contraction rules,'' see~\cite[eqs.~(2.4.44)--(2.4.46)]{cfs}). 
The corresponding regularized fermionic projector is given by (cf.~\cite[eq.~(3.6.2)]{cfs}) 
\begin{align*} %\label{c:cfs, (3.6.2)}
	P(x,y) = \frac{ig}{2}\: \slashed{\xi}\: T^{(-1)}_{[0]} + (\deg < 2 )\:,
\end{align*} 
where the indices~$^{(-1)}_{[0]}$ of the factor~$\slashed{\xi}$ are omitted for notational convenience. 
In view of~\cite[Proposition~3.6.1]{cfs},
the operator~$Q$
as defined by~\eqref{c:cfs, (3.5.17)} takes the form
\beq \label{c:cfs, (3.6.8)}
Q(x,y) = i \slashed{\xi}  \, g^3\,(1-4 \mu)\:
T^{(0)}_{[0]} T^{(-1)}_{[0]}\,\overline{T^{(-1)}_{[0]}} + (\deg < 5)\:. 
\eeq
Although only formal, the underlying calculations have a well-defined meaning
in the formalism of the continuum limit, because to the resulting formulas one can apply
the weak evaluation formula~\eqref{c:cfs, (2.4.48'')}. 
In view of~\eqref{c:cfs, (3.6.8)} we conclude that the degree of the leading singularity of~$Q(x,y)$ is five. Analyzing the EL equations~\eqref{c:cfs, (3.5.29)} degree by degree, the connection to conventional ``macroscopic'' physics can be accomplished as we now explain. 

\subsection{Derivation of Classical Field Equations}\label{c:S field equations}
In a few words, the result of the detailed analysis carried out in~\cite[Chapter~3]{cfs} shows that the EL equations~\eqref{c:cfs, (3.5.29)} can be satisfied to degree five on the light cone in the case of a single charged sector. More precisely, it is useful to rewrite a chiral potential~$\B$ in the form
\begin{align*} %\label{c:cfs, (3.6.15)}
	\B = \slashed{A}_\text{\rm{v}} + \pseudo \slashed{A}_\text{\rm{a}}
\end{align*} 
with a {\em{vector potential}}~$A_\text{\rm{v}}$ and an {\em{axial potential}}~$A_\text{\rm{a}}$ defined by 
\begin{align*} %\label{c:cfs, (3.6.16)}
	A_\text{\rm{v}}=(A_L+A_R)/2 \qquad \text{and} \qquad A_\text{\rm{a}}=(A_L-A_R)/2 \:. 
\end{align*}  
It is found that the eigenvalues~$(\lambda^c_s)$ of the closed chain and the corresponding spectral projectors~$F^c_s$ with~$c \in \{L,R\}$ and~$s \in \{+,-\}$ are given by (cf.~\cite[eq.~(3.6.21)]{cfs})
\begin{align*} %\label{c:cfs, (3.6.21)} 
	\lambda^{L\!/\!R}_\pm = \nu_{L\!/\!R}\: \lambda_\pm \qquad \text{and} \qquad
	F^{L\!/\!R}_\pm = \chi_{L\!/\!R} \:F_\pm \quad 
\end{align*}
with~$\lambda_s$ and~$F_s$ as in~\eqref{c:cfs, (2.6.15)} and~\eqref{c:cfs, (2.6.16)}, where (cf.~\cite[eq.~(3.6.20)]{cfs})
\begin{align*} %\label{c:cfs, (3.6.20)}
	\nu_L = \overline{\nu_R} %= e^{-i (\Lambda_L^{xy} - \Lambda_R^{xy})} 
	= \exp \Big(-2i \int_x^y A_\text{\rm{a}}^j \,\xi_j \Big) \:. 
\end{align*}
The fact that all eigenvalues have the same absolute value (cf.~\cite[eq.~(3.6.22)]{cfs}) implies that the EL equations in the continuum limit~\eqref{c:cfs, (3.5.29)} are satisfied to degree five on the light cone 
(cf.~\cite[Section~3.6]{cfs}).

Thus the task is to deal with the EL equations~\eqref{c:cfs, (3.5.29)} to the next lower degree four on the light cone in the case of a single sector (cf.~\cite[Section~3.6]{cfs}). 
Introducing the \emph{axial (bosonic) current} by (cf.~\cite[eq.~(3.7.4)]{cfs})
\begin{align*} %\label{c:cfs, (3.7.4)}
	j_\text{\rm{a}}^k = \partial^k_{\;j}A^j_\text{\rm{a}} - \Box A_\text{\rm{a}}^k
\end{align*}
as well as the \emph{axial Dirac current} by (cf.~\cite[eq.~(3.7.8)]{cfs})
\begin{align*} %\label{c:cfs, (3.7.8)}
	J^i_\text{\rm{a}} = \sum_{k=1}^{\np} \overline{\psi_k} \pseudo \gamma^i \psi_k
	- \sum_{l=1}^{\na} \overline{\phi_l} \pseudo \gamma^i \phi_l \:, 
\end{align*}
it turns out that the Euler-Lagrange equations~\eqref{c:cfs, (3.5.29)} give rise to the equation
\[ \xi_k \left( j^k_\text{\rm{a}} \: N_1- m^2 A_\textup{a}^k\:N_2 - J^k_\textup{a}\: N_3 \right) = 0 \:, \]
where~$N_1$, $N_2$ and~$N_3$ are simple fractions which are defined by~\cite[eqs.~(3.7.5), (3.7.6) and~(3.7.9)]{cfs}, respectively. 
Evaluating weakly on the light cone, by contrast to~$N_3$ the simple fractions~$N_1$ and~$N_2$ give rise to \emph{logarithmic poles on the light cone}.  
In order to obtain physically reasonable equations, one must get rid of these logarithmic poles. In short, this is achieved by means of a so-called \emph{microlocal chiral transformation} (we refer the interested reader to~\cite[Section~3.7]{cfs}). After this has been accomplished, the EL equations to degree four give rise to the desired field equations, including several ``correction terms'' (see~\cite[Theorem~3.8.2]{cfs}).  Dropping one of these correction terms, one obtains the Dirac-Maxwell equations in~\cite[eq.~(3.8.23)]{cfs}.

Unfortunately, the situation is more complicated for a system involving neutrinos (see~\cite[Section~4.3]{cfs}). 
Nevertheless, employing the ansatz of ``massive neutrinos''~\cite[eq.~(4.1.8)]{cfs}, in the setting of one charged and one neutrino sector one gets the following results: the chiral gauge potentials satisfy a classical Yang-Mills-type equation, coupled to the fermions (cf.~\cite[eq.~(4.1.10) and Theorem~4.8.1]{cfs}). The chiral potentials are of the form (cf.~\cite[eq.~(4.8.2)]{cfs})
\begin{align*} %\label{c:cfs, (4.8.2)}
	\B &= \chi_R \begin{pmatrix} \slashed{A}_L^{11} & \slashed{A}_L^{12}\, \UMNS^* \\[0.2em]
		\slashed{A}_L^{21}\, \UMNS & -\slashed{A}_L^{11}  \end{pmatrix}
	+ \chi_L \begin{pmatrix} 0 & 0 \\
		0 & 0 \end{pmatrix} 
\end{align*}
with left-handed~$\mathfrak{su}(2)$-valued gauge potentials and 
a unitary matrix~$\UMNS \in \U(3)$, which in analogy to the Standard Model is referred to as MNS matrix. Moreover, the setting allows for describing a gravitational field by the Einstein equations in~\cite[eq.~(4.1.11)]{cfs}.

These results are obtained for a system of one charged and one neutrino sector (including right-handed neutrinos). In this case, the auxiliary fermionic projector is given in analogy to~\eqref{c:cfs, (4.2.40')}--\eqref{c:cfs, (5.2.4)} with \emph{one} charged sector instead of seven and comprises seven direct summands. 
In analogy to~\eqref{c:cfs, (4.2.42')} and~\eqref{c:cfs, (4.2.43')},
the chiral asymmetry matrix~$X$ and the mass matrix~$Y$ are given with \emph{one} charged sector instead of seven as well. The regularization is built in by the formal replacements (cf.~\cite[eqs.~(4.2.49)--(4.2.50)]{cfs})
\begin{align*}
	m^p \,T^{(n)} &\rightarrow m^p \,T^{(n)}_{[p]}\:, %\label{c:cfs, (4.2.49)} 
	\\
	\tau_\reg \,T^{(n)} &\rightarrow \tau_\reg \sum_{k=0}^\infty \frac{1}{k!}\: \frac{1}{\delta^{2k}}
	T^{(k+n)}_{[R,2n]} \:, %\label{c:cfs, (4.2.50)}
\end{align*}
where the factor~$\delta^{-2n}$ is required in order to get the scaling dimensions right, while the subscript~``$R$'' indicates the regularization of the right-handed component (for details see~\cite[Section~4.2]{cfs}). The scaling of the length scale~$\delta \gg \varepsilon$ will be specified in~\eqref{c:cfs, (4.5.5)} below, whereas the parameter~$\tau_{\textup{reg}}$ is chosen in accordance with~\cite[eq.~(4.3.34)]{cfs}, 
\begin{align*}\label{c:cfs, (4.3.34)}
	\tau_\reg = (m \varepsilon)^{p_\reg} \qquad \text{with} \qquad
	0 < p_\reg < 2 \:.  % \label{c:l:preg}
\end{align*} 
Next, particles and anti-particles are introduced by occupying additional states or by removing
states from the sea in the fashion of~\eqref{c:cfs, (3.4.7)}. The regularized interacting auxiliary fermionic projector~$\tilde{P}^{\textup{aux}}$ is then obtained in the fashion of~\S \ref{c:S auxiliary}. 
Finally, the regularized fermionic projector~$\tilde{P}$  
is introduced by forming the {\em{sectorial projection}} (cf.~\cite[eq.~(4.2.51)]{cfs}) 
\begin{align*} %\label{c:cfs, (4.2.51)}
	(\tilde{P})^i_j = \sum_{\alpha, \beta} (\tilde{P}^\text{aux})^{(i,\alpha)}_{(j, \beta)} \:, 
\end{align*}
where~$i,j \in \{1, 2\}$ is the sector index, and the generation indices~$\alpha$ and~$\beta$ take the 
values~$\alpha, \beta \in \{1, \ldots 4\}$ if~$i=1$ and~$\alpha, \beta \in \{1, 2, 3 \}$
if~$i=2$.\footnote{Forming the sectorial projection is necessary in order to obtain the correct size of the effective gauge group; concerning further details see \S \ref{c:S connection SM} below.}

The basic procedure is to first prove that in the presence of a chiral potential~$\B$ the EL equations in the continuum limit to degree five on the light cone can be satisfied. To this end, 
one needs to pose certain conditions on the regularization. More precisely, introducing
\[ L^{(n)}_{[p]} = T^{(n)}_{[p]} + \frac{1}{3}\:\tau_\reg\, T^{(n)}_{[R,p]} \:, \]
the regularization is required to satisfy the condition (cf.~\cite[eq.~(4.3.35)]{cfs}) 
\begin{align*} %\label{c:cfs, (4.3.35)}
	T^{(n)}_{[p]} = L^{(n)}_{[p]} \:\big(1+\O \big((m \varepsilon)^{p_\reg} \big)\big) \qquad \text{pointwise} \:. 
\end{align*}
Following the arguments in~\cite[\S 4.3.2]{cfs}, 
one obtains the following two cases:
\beq \label{c:cfs, (4.3.36)}
\text{\bf{(i)}} \quad |\vec{\xi}| \gg \frac{(m \varepsilon)^{p_\reg}}{\|A_L^{12}\|+\|A_L^{21}\|}
\:,\qquad \qquad
\text{\bf{(ii)}} \quad |\vec{\xi}| \ll \frac{(m \varepsilon)^{p_\reg}}{\|A_L^{12}\|+\|A_L^{21}\|} \:.
\eeq
It is found that in both cases, the EL equations can be satisfied to degree five on the light cone (see~\cite[Section~4.3]{cfs}). However, the analysis in~\cite[Chapter~5]{cfs} indicates that Case~\textbf{(ii)} in~\eqref{c:cfs, (4.3.36)} is of physical relevance, which is therefore assumed in what follows.

In order to allow for left-handed gauge fields in the neutrino sector, 
the simplest method is to pose the following additional condition on the regularization functions, 
\beq \label{c:cfs, (4.6.38)}
\big| L^{(0)}_{[0]} \big| = \big| T^{(0)}_{[0]} \big| \left( 1 + \O \big( (m \varepsilon)^{2 p_\reg} \big) \right)
\qquad \text{pointwise} \:. 
\eeq
Then according to~\cite[Proposition~4.6.8]{cfs} the EL equations to degree four on the light cone can be satisfied only if~\eqref{c:cfs, (4.3.36)} and~\eqref{c:cfs, (4.6.38)} are satisfied. Furthermore, in order to compensate the logarithmic poles on the light cone, one again employs a microlocal chiral transformation. It turns out that the logarithmic poles can be compensated if the condition in~\cite[eq.~(4.4.64)]{cfs} holds (see~\cite[Proposition~4.4.6]{cfs}). For the relation of this condition to the neutrino masses we refer to~\cite[Remark~4.4.9]{cfs}. 

Finally, in order to derive the Einstein equations, in the case~$n=-1$ and~$p=0$ one needs to pose the following condition on the regularization, 
\beq \label{c:cfs, (4.9.2)}
\big| L^{(-1)}_{[0]} \big| = \big| T^{(-1)}_{[0]} \big| \left( 1 + \O \big( (m \varepsilon)^{2 p_\reg} \big) \right)
\qquad \text{pointwise} \:.
\eeq
The connection to the Einstein field equations is then obtained in~\cite[Theorem~4.9.3]{cfs} 
provided that~\eqref{c:cfs, (4.6.38)} and~\eqref{c:cfs, (4.9.2)} hold. Nevertheless, this connection becomes more clear in~\cite[Chapter~5]{cfs}.  
Introducing the Einstein-Hilbert Lagrangian~$\LEH$ according to~\cite[Theorem~5.4.4]{cfs}, 
then under the assumptions~\eqref{c:cfs, (4.3.36)}, \eqref{c:cfs, (4.6.38)} and
\begin{align}\label{c:cfs, (4.5.5)}
	\varepsilon \ll \delta \ll \frac{1}{m} (m\varepsilon)^{p_{\reg}/2} \:, 
\end{align}
the EL equations to degree four
can be expressed in terms of the effective action 
\beq \label{c:cfs, (5.4.1')} 
\Sact_\text{eff} = \int_\scrM \left( \LDirac + \LYM + \LEH \right) \sqrt{-\det g} \: d^4x 
\eeq 
by suitably choosing the parameter~$\tau$ in the Dirac Lagrangian~$\LDirac$ in~\cite[eq.~(5.4.3)]{cfs} (and a Yang-Mills-type Lagrangian~$\LYM$ of the form~\cite[eqs.~(5.4.8)--(5.4.10)]{cfs}). Then variations of the effective action~\eqref{c:cfs, (5.4.1')} with respect to the Lorentzian metric~$g$ 
yield the Einstein field equations (also cf.~\cite[eqs.~(4.7.16) and~(4.7.17)]{cfs}), where the energy-momentum tensor is taken into account according to~\cite[\S 5.4.3]{cfs}. On the other hand, varying the effective action~\eqref{c:cfs, (5.4.1')} with respect to the gauge potential, one obtains the bosonic field equations (see~\cite[\S 5.4.1]{cfs}). Note that all variations are carried out under the constraint that the Dirac equation~\cite[eq.~(5.4.2)]{cfs} is satisfied. 

Following the arguments in~\cite[\S 3.9.4]{cfs}, it is expected that the EL equations to degree three or lower on the light cone do \emph{not} give rise to dynamical field equations. For a discussion of \emph{nonlocal} potentials we refer to~\cite[Section~3.10]{cfs}.

\subsection{Connection to the Standard Model} \label{c:S connection SM}
In order to furnish the connection to the Standard Model, we are led to consider the auxiliary fermionic projector as outlined in~\S \ref{c:S auxiliary}. 
The interaction by chiral potentials is introduced by inserting the following operator into the Dirac equation (for details see~\cite[Section~5.2]{cfs}), 
\begin{align*} %\label{c:q:chiral}
	\B = \chi_L\: \slashed{A}_R + \chi_R\: \slashed{A}_L \:,
\end{align*}
where~$A^j_L$ and~$A^j_R$ are Hermitian $25 \times 25$-matrices acting on the sectors (implying that~$\B$ is a symmetric operator). However, in order to obtain unitary phase transformations, one needs to assume that~$\B$ is a unitary operator. Bearing this in mind, 
the chiral gauge potentials can a-priori be chosen according to the gauge group
\[ \U(25)_L \times \U(25)_R\:. \]
This gauge group, however, is too large for mathematical and physical reasons. 
On the one hand, 
the causality compatibility condition~\eqref{c:cfs, (4.2.48)} 
gives rise to the smaller gauge group
\begin{align*} %\label{c:q:U24}
	\U(24)_L \times \U(24)_R \times \U(1)_R \:,
\end{align*}
where the group~$\U(24)$ acts on the first three direct summands of~$P^N_\text{aux}$ 
and on the 21 direct summands in~$P^C_\text{aux}$ in~\eqref{c:cfs, (5.2.4)}. On the other hand, taking into consideration the corresponding regularized interacting 
auxiliary fermionic projector~$\tilde{P}^{\textup{aux}}$  
(for details we refer to~\cite[Appendix~F]{cfs} and~\S \ref{c:S auxiliary}), in order to build in the sectorial projection 
\begin{align*} %\label{c:cfs, (5.2.8)}
	(\tilde{P})^i_j = \sum_{\alpha, \beta} (\tilde{P}^\text{aux})^{(i,\alpha)}_{(j, \beta)}
\end{align*}
in a compatible way  
(where~$i,j \in \{1, \ldots, 8\}$ is the sector index, and the generation indices~$\alpha$,~$\beta$ take values
$\alpha, \beta \in \{1, \ldots 4\}$ if~$i=1$ and~$\alpha, \beta \in \{1, 2, 3 \}$
if~$i=2, \ldots, 8$),
one is led to consider the ansatz~\cite[eq.~(5.2.17)]{cfs}. This gives rise to the gauge group
\beq \label{c:cfs, (5.2.16)}
\U(8)_L \times \U(1)_R \times \U(7)_R \:. 
\eeq
As a consequence, 
whenever the assumptions in~\cite[Theorem~5.3.2]{cfs} are satisfied, one obtains precisely the effective gauge group of the Standard Model, 
\begin{align}\label{c:cfs, (5.3.1')}
	\mathcal{G} = \U(1) \times \SU(2) \times \SU(3) \:. 
\end{align}
For clarity, however, we remark that the assumptions of~\cite[Theorem~5.3.2]{cfs} can\emph{not} be derived from the causal action principle. 
Nevertheless, under these assumptions, one recovers the mixing matrix~$\UCKM$ (which is regarded as the CKM matrix), 
and in view of~\cite[Proposition~5.4.3]{cfs} one obtains precise agreement with electroweak theory (for details see~\cite[Theorem~5.4.2]{cfs}). 
Let us finally note that
the reduction from the large gauge group~\eqref{c:cfs, (5.2.16)} to the effective gauge group~\eqref{c:cfs, (5.3.1')} and to gauge potentials of the specific form given in~\cite[Theorem~5.3.2]{cfs} can be regarded as spontaneous breaking of the gauge symmetry; this effect is referred to as ``spontaneous block formation.''

\Thanks {{\em{Acknowledgments:}}
	I would like to thank Felix Finster for many discussions and Marco Oppio for valuable comments on the manuscript. 
	%Moreover, 
	I gratefully acknowledge financial support by the ``Studienstiftung des deutschen Volkes.''

\end{document}